\documentclass[aps,prb,amsmath,amssymb,floatfix,twocolumn,amsmath,superscriptaddress,
twocolumn,nofootinbib,tighten,letterpaper]{revtex4-2}
\usepackage{lmodern}
\usepackage[T1]{fontenc}
\usepackage[latin9]{inputenc}
\usepackage{float}
\usepackage[normalem]{ulem}
\usepackage{amsmath, dsfont}
\usepackage{amssymb}
\usepackage{mathtools}
\usepackage{mathdots}
\usepackage{physics}

\usepackage{graphicx}
\usepackage{wasysym}
\usepackage{color}
\usepackage[table]{xcolor}
\usepackage{bm}
\usepackage{hyperref}
\hypersetup{pdfpagemode=UseNone}
\usepackage[many]{tcolorbox}
\newsavebox\MBox

\allowdisplaybreaks

\newcommand{\bsub}{\begin{subequations}}
	\newcommand{\esub}{\end{subequations}}

\newcommand{\bg}{ \begin{gather} }
	\newcommand{\eg}{\end{gather}}
\newcommand{\be}{ \begin{equation} }
	\newcommand{\ee}{\end{equation}}
\newcommand{\bea}{ \begin{eqnarray} }
	\newcommand{\eea}{\end{eqnarray}}

\newcommand{\fullstop}{\text{\,.}}                    
\newcommand{\comma}{\text{\,,}}                       

\renewcommand{\d}{\ensuremath{\text{d}}}               

\graphicspath{{Figures/}}

\begin{document}
	
	\title{Ergodicity-to-localization transition on random regular graphs with large connectivity and in many-body quantum dots}
	
	\author{Jan-Niklas Herre}
	\affiliation{{Institut f\"ur Theorie der Kondensierten Materie, Karlsruhe Institute of Technology, 76128 Karlsruhe, Germany}}
	
	\author{Jonas F.~Karcher}
	\affiliation{{Pennsylvania State University, Department of Physics, University Park, Pennsylvania 16802, USA}}
	\affiliation{{Institute for Quantum Materials and Technologies, Karlsruhe Institute of Technology, 76021 Karlsruhe, Germany}}
	\affiliation{{Institut f\"ur Theorie der Kondensierten Materie, Karlsruhe Institute of Technology, 76128 Karlsruhe, Germany}}
	
	\author{Konstantin S.~Tikhonov}
	\affiliation{{Capital Fund Management, 23 rue de l'Universit\'e, 75007 Paris, France}}
	
	\author{Alexander D.~Mirlin}
	\affiliation{{Institute for Quantum Materials and Technologies, Karlsruhe Institute of Technology, 76021 Karlsruhe, Germany}}
	\affiliation{{Institut f\"ur Theorie der Kondensierten Materie, Karlsruhe Institute of Technology, 76128 Karlsruhe, Germany}}
	
	\date{\today}
	
	\begin{abstract}
		Anderson localization on random regular graphs (RRG) serves as a toy-model of many-body localization (MBL). We explore the transition from ergodicity to localization on RRG with large connectivity $m$. In the analytical part, we focus on the inverse participation ratio of eigenstates and identify several regimes on the ergodic side of the transition---self-consistent golden rule, golden rule, precritical, and critical---that the system consecutively goes through when the disorder increases towards the point $W_c$ of the localization transition. We also perform exact-diagonalization numerics as well as a population-dynamics analysis combining analytical and numerical techniques. Results of all the approaches are in excellent mutual agreement. We further explore the evolution from ergodicity to localization  in two models of Fock-space MBL: fermionic and spin quantum dots. Large-connectivity RRG models serve as an approximation for this class of many-body problems; one of our central goals is to better understand the status of this approximation. Our numerical simulations support the conjecture that, in the limit of a large system, there is a sharp ergodicity-to-localization transition in the quantum-dot models. Furthermore, our results are consistent the RRG-like scaling of the critical disorder, $W_c \sim m \ln m$. While in the golden-rule range the behavior of quantum-dot models is in agreement with analytical predictions based on RRG model, substantial deviations occur in the pre-critical regime. Our results indicate that the
		pre-critical and critical behavior (as well as the numerical coefficient in the formula for $W_c$) in the quantum-dot models may be different from what one would expect from the RRG-like approximation. 
		
	\end{abstract}
	
	\maketitle
	
	\section{Introduction}
	\label{sec:intro}
	
	\subsection{Anderson and many-body localization}
	
	Anderson localization \cite{anderson58} is one of the most fundamental phenomena in condensed matter physics. Anderson transitions between localized and delocalized phases (and between topologically distinct localized phases) exhibit a rich fascinating physics that depends on symmetry class and spatial dimensionality of the system and of the underlying topology \cite{evers08}. While Anderson localization in its conventional setting refers to a single-particle problem, recent years have witnessed a great interest to the problem of many-body localization (MBL) in highly excited states (i.e., those with a finite energy density) of interacting disordered systems \cite{gornyi2005interacting,basko2006metal}. The reader is referred to Refs.~\cite{nandkishore15,Alet2018a,abanin2019colloquium,gopalakrishnan2020dynamics,tikhonov2021from} for reviews of various facets of MBL. 
	
	A conceptual importance of MBL may be additionally emphasized by noticing that the MBL implies a breakdown of a basic postulate of the statistical mechanics: the ergodicity. 
	One way to probe this is to study properties of eigenstates and eigenvalues of the many-body Hamiltonian. On the ergodic side of the MBL transition, many-body states strongly  hybridize within an energy shell including a very large number of levels (increasing with increasing system size). This manifests itself in the Wigner-Dyson statistics of energy levels and in the scaling of the inverse participation ratio (IPR) describing an efficient spreading of eigenstates over basis states within the energy shell. On the other hand, in the MBL phase, even adjacent-in-energy states typically do not hybridize, which implies the Poisson energy-level statistics and is obviously also reflected in the behavior of the IPR. 
	In this general form, a notion of the ergodicity-to-MBL transition applies also to models without a real-space structure, like quantum dot models considered in the present work.  
	
	A controllable analytical solution of the MBL problem has turned out to be an extremely challenging task. Original approaches to the MBL transition in Refs.~\cite{gornyi2005interacting,basko2006metal} (as well as in a related later work \cite{Ros2015420}) were based on the analysis of the perturbative series.  Later, it was demonstrated \cite{gornyi2017spectral} that Hartree-Fock matrix elements (disregarded in Refs.~\cite{gornyi2005interacting,basko2006metal}) lead to spectral diffusion and, as a result, parametrically enhance delocalization and shift the transition point  as given by the perturbative treatment.
	Furthermore, recent works have shown that exponentially rare regions of anomalously strong or weak disorder may essentially affect the scaling of the MBL transition and the corresponding critical behavior in the limit of a large system \cite{Agarwal2016a,Thiery2017a}. These ideas served as a basis for several phenomenological renormalization-group schemes \cite{Dumitrescu2019a,morningstar2020a} that were conjectured to describe scaling properties around the MBL transition. 
	
	Since a complete analysis of genuine MBL models turns out to be so difficult, simplified models amenable to a controllable analytical study are of great use. In this context, the problem of Anderson localization on random regular graphs (RRG) has emerged as a toy-model for the MBL problem, see Ref.~\cite{tikhonov2021from} for a review. An RRG is a graph of finite size that has locally a tree-like structure with a fixed coordination number (to be denoted $m+1$) for all vertices. An RRG is thus locally similar to Bethe lattice. However, an RRG has a finite size (at variance with an infinite Bethe lattice) and is also crucially different from a finite portion of Bethe lattice: it does not have a boundary but instead possesses large-scale loops. The structure of an RRG approximately mimics that of the Hamiltonian of an interacting systems in the associated many-body Hilbert space. Specifically, vertices correspond to eigenstates of the non-interacting part of the Hamiltonian, while the links correspond to matrix elements of the interaction. 	
	
	\subsection{Goals and structure of the paper}
	\label{sec:goals}
	
	
	Our central goal in this work is to explore the transition from ergodicity to localization in two closely related classes of models---Anderson models on RRG with large values of connectivity $m$ and many-body quantum-dot models in fermionic and spin versions---by a combination of analytical and computational approaches. For the RRG problem, we extend previous analytical studies to large-$m$ models and show that it leads to emergence of broad regimes that do not exist at $m \sim 1$. We verify the analytical results by a numerical study, which also allows us to study which of the analytically predicted regimes can be observed in systems accessible to exact diagonalization. In particular, we demonstrate that the ultimate critical behavior becomes out of reach for exact diagonalization at large $m$ but, at the same time, a broad pre-critical regime opens that is numerically accessible. For the quantum-dot models, we develop approximate analytical treatment based on a mapping on RRG (an ``RRG-like approximation''). The status of this approximation becomes unclear for sufficiently strong disorder, making a comparison between the analytical and numerical results particularly important. More generally, the numerical study of quantum-dot models addresses the following fundamental questions: What is the scaling of the critical disorder $W_c$ with the number $n$ of orbitals (respectively, spins) in these models? Does a crossover from ergodicity to MBL in this class of models become a sharp MBL transition in the large-$n$ limit?

The structure of the article is as follows. In Sec.~\ref{sec:rrg}, we explore the RRG model with large connectivity $m$ by using three approaches: (i) purely analytical, (ii) first-principle numerics (exact diagonalization), and (iii) numerical solution (by means of population dynamics) of analytically derived self-consistency equation. Excellent agreement between all three approaches is found. In Sec.~\ref{sec:quantum-dots}, we study the fermionic and spin quantum dots by exact diagonalization. Our results support analytical predictions for $W_c$ and provide evidence of a sharp MBL transition in the thermodynamic limit.   Finally, in Sec.~\ref{sec:quantum-dot-rrg}, we extend analytical results of Sec.~\ref{sec:rrg} to derive an RRG-like approximation for the spin quantum dot model. We then compare the exact-diagonalization numerics from Sec.~\ref{sec:quantum-dots} to this approximation.  Our findings in Sec.~\ref{sec:quantum-dot-rrg} confirm the validity of the RRG approximation in a large part of the ergodic phase (golden-rule regime) but also demonstrate substantial deviations for stronger disorder, when $W$ approaches $W_c$. Section \ref{sec:summary} contains a summary of our results as well as a discussion of their implications and of perspectives for future research.
	
\section{Localization on RRG and in various many-body models}

In this section, we briefly discuss previous works on the Anderson localization on RRG and its relations to various MBL problems, with a particular focus on quantum-dot models. These results serve as a starting point for the present paper. 
	
	\subsection{MBL and Anderson localization on RRG}
		 
	 An idea to think about a many-body problem in terms of a tree (Bethe lattice) in Fock space was put forward in Ref.~\cite{altshuler1997quasiparticle} in the context of a decay of a hot quasiparticle in a zero-temperature quantum dot. It was subsequently understood that this connection is much more general and that the appropriate graphs are not strictly trees but rather have the RRG structure. (The importance of this distinction was demonstrated in Refs.~\cite{tikhonov2016fractality,sonner17}.)
	
	The Anderson localization on RRG (including some variations of this model) was addressed in a considerable number of recent works \cite{biroli2012difference,de2014anderson,tikhonov2016anderson,garcia-mata17,metz2017level,Biroli2017,kravtsov2018non,biroli2018,PhysRevB.98.134205,tikhonov19statistics,tikhonov19critical,PhysRevResearch.2.012020,PhysRevLett.125.250402,tikhonov2021eigenstate,garcia-mata2022critical,sierant2023universality,valba2022mobility}, 
	see also earlier papers \cite{mirlin1991universality,fyodorov1991localization,fyodorov1992novel} on a closely related sparse random matrix model. 
	We refer the reader to Ref.~\cite{tikhonov2021from} for a recent review of the RRG model and its relations to the MBL. These works have yielded a rather detailed understanding of various key properties of the RRG model. These include, in particular, the thermodynamic-limit value of the critical disorder $W_c$, values of critical exponents, and many observables characterizing statistics of eigenstates and of energy levels. On a qualitative level, the most important features of the localization transition on RRG are as follows. First, the critical point has a localized character. Second, the delocalized phase is ergodic. Third, the ``correlation volume'' in the Hilbert space (beyond which the ergodicity is reached) grows exponentially when the  transition point  is approached from the delocalized side. Fourth, there are strong finite-size effects that induce a drift of the apparent transition point towards stronger disorder with increasing system size as well as a non-monotonic flow of various observables on the delocalized side. Analytical and numerical studies of the MBL transitions in genuine many-body problems indicate that they share similar properties, although analytical results for the MBL problem are much less rigorous than for RRG. 
	
Despite these analogies between the MBL and RRG problems, the structure of the Hamiltonian of a genuine many-body problem in the corresponding Fock space is of course more complex than that of the RRG model. In particular, the role of Fock-space correlations in MBL problems was emphasized (for models with a real-space structure) in Ref.~\cite{roy2020fock-space}. We will return to this issue  (in the context of Fock-space MBL in quantum dots) in Sec.~\ref{sec:intro-Fock-MBL}.
	
	
	The link to Anderson localization on RRG can be made particularly explicit in the case of MBL models with long-range interaction decaying as $1/r^\alpha$ with distance $r$. 
	An important class of such models include models with random $1/r^\alpha$  interaction of spins (or pseudospins),
with $d \le \alpha \le 2d$, where $d$ is the spatial dimensionality \cite{burin2006energy,Demler14,gutman2015energy,burin2015many,tikhonov18}. 
It was found \cite{tikhonov18} that the MBL transition in such a model 
can be approximately mapped to Anderson transition on RRG with a large coordination number $m$ (increasing as a power of the system size $L$).  A relation to the RRG with a large 
$m$ emerges also for ``most-long-range'' models characterized by $\alpha=0$; we will term them ``quantum dots'', see Sec.~\ref{sec:intro-Fock-MBL}.
It is worth emphasizing that the critical disorder $W_c$ is a non-trivial function of system size $L$ in models with long-range interaction: it increases with $L$ according to a power law, with an additional logarithmic factor. (For quantum-dot models, the role of $L$ is played by the number of orbitals $n$.)  Thus, to study the MBL transition, one should investigate observables as functions of $W/W_c(L)$. This situation was termed a ``non--standard thermodynamic limit'' in Ref.~\cite{gopalakrishnan2019instability}.

	
	\subsection{Fock-space MBL transition in quantum dots}
	\label{sec:intro-Fock-MBL}
		
	The localization physics in a model of fermionic quantum dot model was studied analytically in Refs.~\cite{altshuler1997quasiparticle,jacquod1997emergence,mirlin1997localization,silvestrov1997decay,silvestrov1998chaos,gornyi2016many,gornyi2017spectral}
	and numerically in Refs.~\cite{jacquod1997emergence, georgeot1997breit, leyronas2000scaling, shepelyansky2001quantum,  PhysRevE.62.R7575, jacquod2001duality, rivas2002numerical,bulchandani2022onset}. Recently, an analogous Majorana quantum-dot model (expected to have essentially the same properties) was considered in Refs.~\cite{garcia-garcia2018chaotic,micklitz2019nonergodic,monteiro2020minimal,monteiro2020quantum,nandy2022delayed,larzul2022quenches}. Related questions were earlier addressed in computer simulations of a similar model in the nuclear-physics context \cite{aaberg1990onset, aaberg1992quantum}.
	
	The model (see Sec.~\ref{sec:quantum-dots} for a formal definition of the Hamiltonian) involves $n$ single-particle fermionic orbitals ($j=1, \ldots, n$), all coupled by a random two-body interaction, with characteristic magnitude $V$ of the interaction matrix elements $V_{ijkl}$. In addition, single-particle states $j$ have random energies $\varepsilon_j$ with a distribution of width $W$.  
	Setting $V=1$, we are left with two parameters, $n$ and $W$.  Eigenstates of the non-interacting part of the Hamiltonian serve as a basis in the Fock space. If one thinks about them as vertices and associates links connecting them with interaction matrix elements, one gets a graph. For the case of half-filling, each vertex is connected to  
	$m \approx n^4 / 64$ vertices, whose energies are distributed in the interval of width $ \sim W$.  We thus get a structure reminding an RRG with a coordination number $m \sim n^4$ 
	and disorder $W$.  The critical disorder $W_c^{\rm RRG}$ of this RRG model scales as $m \ln m$. However, such an approximation by an RRG discards important 
	correlations in the quantum-dot model \cite{gornyi2017spectral}. Indeed, consider, starting from a given basis state, two second-order processes: in one of them we first move two particles $i,j \mapsto k,l$ and then another pair $i', j' \mapsto k',l'$, while in the second one we do the same in the opposite order. Clearly, we get the same final state, which is at variance with a loopless structure of RRG at short distances. Furthermore, the amplitudes of these two processes are correlated since they involve the same matrix elements. These correlations tend to enhance localization. While these arguments are not mathematically rigorous, they strongly suggest that the RRG critical disorder serves as an upper boundary for the transition point $W_c$ in the quantum-dot model:
	\be
	W_c \le W_c^{\rm RRG} \,, \qquad  W_c^{\rm RRG} \sim n^4 \ln n \,.
	\label{Wc-RRG}
	\ee
	At first sight, correlations can modify even the leading, power-law factor $n^4$ (i.e., can lower the exponent 4) in the scaling of $W_c$. It turns out, however, that interaction-induced shifts of single-particle energies (known as spectral diffusion) efficiently counteract these correlations, thus enhancing the many-body delocalization and making the quantum-dot model much closer to the RRG approximation. Specifically, it was argued in Ref.~\cite{gornyi2017spectral} that the quantum-dot critical disorder $W_c$ satisfies
	\be
	\label{Wc-fermi-quantum-dot}
	W_c \sim n^4\ln^\mu n \,,
	\ee
	with $-\frac{3}{4} \le  \mu \le 1$, i.e., differs from $W_c^{\rm RRG}$ at most by a power of the subleading logarithmic factor. 
	
	Another related model---that of a spin quantum dot---was introduced in Ref.~\cite{gornyi2017spectral}. It involves $n$ spins subjected to a random Zeeman field along $z$ axis, with a distribution of a width $\sim W$. The interacting part of the Hamiltonian (see Sec.~\ref{sec:quantum-dots} for a precise definition) includes all pairwise interactions between spin components, with random amplitudes $\sim V$. Setting $V = 1$, we are again left with two parameters, $n$ and $W$. The relation to an RRG model is obtained in essentially the same way as for a fermionic dot, with the main difference that the coordination number $m$ now scales as $m \approx n^2/2$. This yields \cite{gornyi2017spectral}, in analogy with Eqs.~\eqref{Wc-RRG} and \eqref{Wc-fermi-quantum-dot},
	\be
	W_c \le W_c^{\rm RRG} \,, \qquad  W_c^{\rm RRG} \sim n^2 \ln n \,,
	\label{Wc-spin-RRG}
	\ee
	and
	\be
	\label{Wc-spin-quantum-dot}
	W_c \sim n^2\ln^\mu n \,, \qquad \mu \le 1 \,.
	\ee
The status of Eqs.~\eqref{Wc-spin-RRG} and \eqref{Wc-spin-quantum-dot} is the same as that of Eqs.~\eqref{Wc-RRG} and \eqref{Wc-fermi-quantum-dot}:  there is a strong analytical evidence in favor of them but a rigorous proof is missing.

	\section{Observables}
	\label{sec:observables}
	
As a central observable, we use in this paper the (averaged) inverse participation ratio (IPR) defined as 
\begin{equation*}
P_2 = \left\langle \sum_r |\psi_\alpha(r)|^4 \right\rangle,
\end{equation*}
where $\psi_\alpha(r)$ are eigenfunctions of the Hamiltonian. The averaging $\langle \ldots \rangle$ is performed over realizations of disorder and over eigenstates $\psi_\alpha$ in a window of energies around the band center. (The exact parameters of averaging are specified below for each of the models). The basis $r$ corresponds to eigenstates of the Hamiltonian in the limit of infinite disorder, $W \to \infty$ (or, equivalently, zero hopping in the RRG model and zero interaction in the quantum-dot models), so that $P_2= 1$ in this limit. 

The IPR characterizes spreading of eigenstates $\psi_\alpha$ over the basis states $r$ and thus serves as a very useful characteristics of the evolution from ergodicity to localization. In particular, in the RRG model, the IPR scales as $P_2 \propto 1/N$ with the number of sites $N$  (in the large-$N$ limit) in the ergodic phase and is $P_2 \sim 1$ in the localized phase. Furthermore, the coefficient in front of $1/N$ provides an important information about different regimes in the ergodic phase (in particular, about the correlation volume in a vicinity of the transition). Importantly, the IPR is a suitable observable both for numerical and analytical investigations.

It is worth mentioning that, in the MBL phase of models with real-space localization of single-particle states, the IPR shows a fractal scaling $P_2 \propto N^{-\tau}$ with the Hilbert-space volume. A numerical evidence of this behavior was presented in Ref.~\cite{luitz2015many}. As was shown in Refs.~\cite{gornyi2017spectral,tikhonov18}, such a  fractal scaling
with $\tau(W)  \propto 1/W$ originates from short-scale resonances. A detailed analysis of the multifractality of eigenstates in the MBL phase of a Heisenberg chain was carried out
in Ref.~\cite{mace19multifractal}. Like in the RRG model, the delocalized phase was found to be ergodic (with $P_2 \propto 1/N$), and the critical point essentially sharing properties of the localized phase. In view of a close relation between the quantum-dot and RRG models, one may expect  that $P_2 \sim 1$ in the MBL phase of quantum-dot models. 

For quantum-dot models, we complement the investigation of IPR of eigenstates by a numerical study of the level statistics. It is well known that the level statistics has the Wigner-Dyson form in the ergodic phase and the Poisson form in the localized phase, thus serving as an indicator of the transition. 
Specifically, we consider the mean adjacent gap ratio $r$, which is a very convenient and broadly employed spectral observable for locating an ergodicity-to-localization transition \cite{oganesyan2007localization,atas2013distribution,giraud2022probing}. It is defined as a mean ratio of the smaller and the larger of two consecutive level spacings $s_i$ and $s_{i+1}$, 
\begin{equation*}
r =  \left \langle \frac{\min(s_i,s_{i+1}) }{\max(s_i,s_{i+1})} \right\rangle.
\end{equation*}
In the localized phase, one has Poisson statistics, with $r=0.386$. On the other hand, in the ergodic phase, the level statistics is as in the Gaussian Orthogonal Ensemble (GOE),
Gaussian Unitary Ensemble (GUE), or Gaussian Symplectic Ensemble (GSE) of the random matrix theory, depending on the symmetry of the system; the corresponding values of $r$ are $r = 0.536$ (GOE), $r=0.603$ (GUE), and  $r=0.676$  (GSE).

	\section{Random regular graphs with large connectivity}
	\label{sec:rrg}
	
	\subsection{Model and regimes}
	\label{sec:rrg-model}
	
	We consider a tight-binding model for a particle hopping over an RRG with $N \gg 1$ sites, coordination number $m+1 \gg 1$, and a potential disorder,
	\begin{equation}
		\label{H}
		\hat H= V \sum_{\left<i, j\right>}\left(\hat c_i^\dagger \hat c_j + \hat c_j^\dagger \hat c_i\right)+\sum_{i=1}^N \varepsilon_i \hat c_i^\dagger \hat c_i\,,
	\end{equation}
	where $V$ is the hopping and the summation in the first term goes over pairs of nearest-neighbor sites. The random potentials $\varepsilon_i$ are independent random variables sampled from a distribution $\gamma(\varepsilon)$.  
	In Sec.~\ref{sec:rrg}  we will mainly focus on the case of  a box distribution $\gamma(\varepsilon)$ (i.e., a uniform distribution on $[-W/2,W/2]$) but the analysis is straightforwardly extended to a generic distribution (as will be done later in Sec.~\ref{sec:quantum-dot-rrg} where we will use the RRG model for an approximate treatment of a quantum-dot model). On a qualitative level, the results are independent on the form of $\gamma(\varepsilon)$ as long it is characterized by a single parameter (width) $\sim W$ and vanishes (or decays fast) for larger energies $\varepsilon$.
	Further, we will focus on zero energy, $E=0$ (i.e., the middle of the band).
	
	Before embarking in more accurate and technical analytical calculations, it is useful to perform estimates that permit to understand what regimes are expected and what are corresponding disorder ranges. For this purpose, let us consider states with energies $\varepsilon_i$ localized on individual sites $i$ as a starting point and analyze how they are mixed (and thus broadened in energy) by hopping. Each site is coupled to $m$ sites with energies distributed in the range $W$, implying a level spacing $W/m$ of directly coupled states.
	The self-consistent Golden-rule formula for the broadening $\Gamma$ reads
	\begin{equation}
		\Gamma \sim \frac{mV^2}{{\rm max} \{W, \Gamma\} }.
		\label{eq:RRG-Gamma}
	\end{equation}
	Here, and in analogous estimates below, we keep only the parametric dependence, discarding numerical factors of order unity. The self-consistency is operative when $\Gamma$ exceeds $W$, so that the bare density $m/W$ of coupled states is replaced by $m/\Gamma$.
	From Eq.~\eqref{eq:RRG-Gamma}, we infer the following regimes:
	\begin{itemize}
		\item Random-matrix theory (RMT) regime,
		\begin{equation}
			W/V \ll m^{1/2} \,.
			\label{eq:RRG-RMT}
		\end{equation}
		In this regime, the broadening is $\Gamma \sim m^{1/2}V$ and satisfies $\Gamma \gg W$, which implies that all states are mixed. We thus expect the IPR to be $P_2 \simeq 3/N$ as in the Gaussian ensemble of RMT, which explains the name that we give to this regime. 
		
		\item Golden-rule regime,
		\begin{equation}
			m^{1/2}  \ll W/V \ll m\,.
			\label{eq:RRG-Golden-rule}
		\end{equation}
		The broadening in this regime is given by the conventional Golden-rule formula, $\Gamma \sim mV^2/W$, and is smaller than the total disorder-induced band width, $\Gamma \ll W$. Thus, only states within the energy window $\Gamma$ get strongly mixed, implying $N P_2 \sim W / \Gamma \sim W^2 / m V^2$. 
		
		\item Pre-critical and critical regimes,
		\begin{equation}
			m \ll W/V < W_c / V \,,
			\label{eq:RRG-precrit-crit}
		\end{equation}
		where $W_c / V \sim m \ln m$.  
		
		Once $W$ increases above the upper border $mV$ of the Golden-rule regime, the broadening $\Gamma$ as calculated by the Golden-rule formula becomes smaller than the spacing $W/m$ of states coupled by hopping to a given one. This indicates that the Golden-rule calculation  is not valid any more and suggests that the system approaches the localization transition. However, the transition point $W_c$ is larger by a factor $\sim \ln m \gg 1$ than $mV$, which implies that the regime $mV < W < W_c$ is parametrically broad for large $m$. At the same time, one can expect that the actual critical regime, with the power-law divergence of the correlation length, takes place sufficiently close to the critical point, $(W_c - W) / W_c \ll 1$. It follows that there should be an intermediate regime (that we call pre-critical) between the Golden-rule and critical regimes. 
		
		Note that, when one speaks about the critical regime, one should distinguish two situations, depending on the relation between the system size $N$ and the correlation volume $N_\xi$ \cite{tikhonov19statistics,tikhonov19critical}.  The correlation volume increases when W approaches the transition point, diverging in the limit $W \to W_c$. 
		If $N$ is larger than $N_\xi$, we have an ergodic regime of ``critical metal'', with IPR $P_2$ proportional to $1/N$. On the other hand, if $N$ is smaller than $N_\xi$, the system is essentially at the critical point and, in particular, $P_2 \sim 1$. 
		
		\item Localized regime,
		\begin{equation}
			W  > W_c \,, 
		\end{equation}
		for which $P_2 \approx 1$. 
		
	\end{itemize}
	
	In Sec.~\ref{sec:rrg-analytical} we carry out a detailed analysis that fully confirms these estimates and provides accurate results for $P_2$ in all the regimes.

	\subsection{Analytical treatment}
	\label{sec:rrg-analytical}
	
	\subsubsection{Generalities}
	\label{sec:RRG-analytical-general}
	
	The general analytical approach to statistics of eigenfunctions on RRG is presented in detail in Ref.~\cite{tikhonov19statistics}  (see also a subsequent work \cite{tikhonov19critical} and a review  \cite{tikhonov2021from}). Using a supersymmetric field-theory formalism for the partition function of the problem, one derives a saddle-point equation (justified for a large system size $N$), which is a non-linear integral equation for a function $g(\Psi)$ of a supervector $\Psi$ [Eq.~(24) of Ref.~\cite{tikhonov19statistics}]. For symmetry reasons, the solution $g(\Psi)$ depends on two scalar variables. This equation turns out to be identical to the self-consistency equation as obtained by supersymmetry approach for a model of an infinite Bethe lattice \cite{mirlin1991localization}, which establishes a relation between the two problems.  Upon Fourier-Laplace transformation, the equation takes a form of a self-consistency equation for distribution of local Green functions $G$ as originally derived in the seminal paper by Abou-Chacra, Thouless, and Anderson \cite{abou1973selfconsistent} that pioneered the investigation of the Anderson transition on Bethe lattice. 
	
	A compact form of the saddle-point (self-consistency) equation is 
	\be
	\label{pool_sc}  
	G^{(m)}\stackrel{d}{=}\frac{1}{E+ i\eta-\varepsilon- V^2 \sum_{i=1}^{m} G_i^{(m)}},
	\ee
	where $\eta$ is infinitesimal positive and  the symbol $\stackrel{d}{=}$ denotes the equality in distribution for random variables. Here
	$G^{(m)}$ is the (retarded) Green function with equal spatial arguments, $G^{(m)} = G_{\rm R}(0,0;E) = \langle 0| (E- {\cal H} + i \eta)^{-1}|0\rangle$, defined on a modified lattice, with the site 0 having only $m$ neighbors.  On the right-hand-side of Eq.~(\ref{pool_sc}), $G_i^{(m)}$ represent independent copies of the random variable $G^{(m)}$, while $\varepsilon$ is a random variable with distribution  $\gamma(\varepsilon)$.  Clearly, Eq.~(\ref{pool_sc}) can be equivalently written as a non-linear integral equation for the distribution of $G^{(m)}$. 
	Equation (\ref{pool_sc}) contains  all the information about the position of the localization transition and the critical volume $N_\xi$. Furthermore,
	with a solution of this equation at hand, one can determine the distribution of the Green function $G^{(m+1)}$ on an original lattice (where all sites have $m+1$ neighbors), which is given by
	\be
	\label{pool_simple}
	G^{(m+1)}\stackrel{d}{=}\frac{1}{E+ i\eta-\varepsilon- V^2 \sum_{i=1}^{m+1} G_i^{(m)}}.
	\ee
	From this distribution, one can in particular calculate the IPR $P_2$ in the RRG model. The distributions of $G^{(m)}$ and $G^{(m+1)}$ are very similar. 
	Furthermore, since we are interested in the behavior of the model at large $m$ in this paper, the difference between $G^{(m)}$ and $G^{(m+1)}$  is negligible. 
	
	\subsubsection{RMT and golden rule regimes}
	\label{sec:RRG-RMT-GR}
	
	We begin the analysis of the behavior of average IPR of eigenstates from the regime of relatively weak disorder, $W / V \ll m$ (that can be termed ``good metal''). The corresponding calculation is closely related to that in Ref.~\cite{mirlin1997localization} where fluctuations of the local density of states  at a Bethe lattice with large connectivity were considered. 
	In the regime under consideration, the solution $g(\Psi)$ of the supersymmetric self-consistency equation is Gaussian (up to small corrections) \cite{mirlin1997localization}, which leads to a major simplification of the self-consistency equation. The result can be obtained by averaging Eq.~\eqref{pool_sc} over disorder realizations and assuming that summation over $m$ terms in the denominator justifies a replacement of $G^{(m)}$ by its average. Defining $g_0 = (i/2) V^2 \langle G^{(m)} \rangle $, we reduce this equation to the form
	\begin{equation}
		g_0 = \frac{V^2}{4} \int \d \varepsilon \gamma
		(\varepsilon) \frac{1}{mg_0 - \frac{i}{2}(E - \varepsilon)} \comma 
		\label{eq:generalg0sce}
	\end{equation}
	which is Eq.~(16) of Ref.~\cite{mirlin1997localization}. It is easy to see that Eq.~\eqref{eq:generalg0sce} is nothing but the self-consistent Born approximation (or, a weak-disorder limit of the coherent potential approximation).
	We are interested in the solution with $\Re \, g_0 > 0$ and discard the infinitesimal $\eta$ (as it is negligible in comparison with  $\Re \, g_0$). 
	Choosing the energy in the band center, $E = 0$, as well as a box distribution $\gamma(\varepsilon)$, 
	\begin{equation}
		\gamma(\varepsilon) = \frac{1}{W} \Theta\left(\frac{W}{2} - |\varepsilon|\right) \,,
		\label{gamma-epsilon-box}
	\end{equation}
	and also setting the hopping to be $V=1$, we get
	\begin{equation}
		g_0 = \frac{1}{4W} \int_{-W/2}^{W/2} \d \varepsilon \frac{1}{mg_0 + \frac{i}{2}\varepsilon} \fullstop \label{eq:g0_sce}
	\end{equation} 
	The integral is straightforwardly calculated, and the equation takes the form
	\begin{equation}
		g_0 = \frac{1}{W} \arctan \frac{W}{4mg_0} \fullstop 
		\label{eq:g0_arctan}
	\end{equation}
	Let us consider two limiting cases---of small and large arguments of the arctangent---which correspond to two distinct regimes:
	\begin{enumerate}
		
		\item[(i) ] \  $mg_0 \gg W$, in which case the self-consistency---i.e., the term proportional to $g_0$ in the denominator of Eq.~\eqref{eq:generalg0sce}---is crucial.  The equation takes the form    $ g_0 \simeq 1/4mg_0$, yielding the solution
		\begin{equation}
			g_0 \simeq \frac{1}{2\sqrt{m}}\fullstop
			\label{eq:RRG-g0-RMT}  
		\end{equation}
		This is applicable in the disorder range 
		\begin{equation}
			W \ll m^{1/2} \,,
			\label{eq:RRG-RMT-regime}
		\end{equation}
		which is exactly the RMT regime identified above [Eq.~\eqref{eq:RRG-RMT}].  As discussed below in more detail, properties of the system in this regime are largely equivalent to those of a Gaussian ensemble of RMT.  In particular, the density of states $\rho(E)$ has a semicircular form and the IPR is given by $P_2 \simeq 3/N$. 
		
		\item[(ii) ] \ $mg_0 \ll W$. In this case (when $g_0$ in the denominator of Eq.~\eqref{eq:generalg0sce} can be neglected), the solution reads
		\begin{equation}
			g_0 \simeq \frac{\pi}{2W}\fullstop 
			\label{eq:RRG-g0-GR}  
		\end{equation}
		This is the Golden-rule regime, which is realized in the disorder range
		\begin{equation}
			m^{1/2} \ll W \ll m \,,
			\label{eq:RRG-GR-regime}
		\end{equation}
		in agreement with Eq.~\eqref{eq:RRG-Golden-rule}. 
		
	\end{enumerate}
	
	With a solution for $g_0$ at hand, we can evaluate averaged moments of the local density of states (LDOS) on an infinite Bethe lattice, which is what we need to find to calculate the average IPR of the RRG model. Let $\nu(E,j)$ be the LDOS on a site $j$ at energy $E$ for a given disorder realization. We define $\rho_1(E, \varepsilon) = \langle \nu(E,1) \rangle_{\varepsilon_1 = \varepsilon}$, with averaging going over disorder realizations with a condition $\varepsilon_1 = \varepsilon$, where 1 is a certain fixed lattice site. Further, let $\rho(E)$ be the global density of states. Clearly, 
	\begin{equation}
		\rho(E) = \int \d \varepsilon \gamma(\varepsilon) \rho_1(E,\varepsilon) \fullstop
		\label{eq:RRG-rho}
	\end{equation}
	We have 
	\begin{eqnarray}
		\rho_1 (E,\varepsilon) &=& - \frac{1}{\pi} \Im G^{(m+1)}  \nonumber \\
		& = &  \frac{1}{2\pi} \Re \frac{1}{(m+1)g_0 - \frac{i}{2}(E-\varepsilon)} \,,
		\label{eq:RRG-rho1}
	\end{eqnarray}
	which yields, after neglecting the difference between $m+1$ and $m$ and setting $E=0$ (in which case $g_0$ is real), 
	\begin{equation}
		\rho_1 (0,\varepsilon) = \frac{1}{\pi} \frac{2mg_0}{(2mg_0)^2 + \varepsilon^2} \fullstop 
		\label{eq:RRG-rho1-result}
	\end{equation}
	Importantly, the spectral function $\rho_1 (E,\varepsilon)$ is self-averaging in the RMT and Golden-rule regimes: when calculated without disorder averaging (with the only condition $\varepsilon_1 = \varepsilon$), it is found to fluctuate only weakly from one disorder realization to another one \cite{mirlin1997localization}. 
	Substituting Eq.~\eqref{eq:RRG-rho1-result} in Eq.~\eqref{eq:RRG-rho}, we get the global density of states,
	\begin{equation}
		\rho(0) = \frac{1}{\pi W}\int_{-W/2}^{W/2} \d\varepsilon \frac{2mg_0}{(2mg_0)^2 + \varepsilon^2} = \frac{2}{\pi} g_0 \comma
		\label{eq:RRG-rho-result}
	\end{equation}
	where we used on the last step the self-consistency equation \eqref{eq:g0_sce}. Equation \eqref{eq:RRG-rho-result}   is of course  in full agreement with $\rho = - (1/\pi) \Im \langle G^{(m+1)} \rangle$ and $g_0 = (i/2) \langle G^{(m)} \rangle $.
	
Up to this point [i.e., in the derivation of Eqs.~\eqref{eq:RRG-rho1-result}, \eqref{eq:RRG-rho-result}], we followed Ref.~\cite{mirlin1997localization}, adjusting the calculation performed there to the box distribution \eqref{gamma-epsilon-box} chosen in the present work.  Now we can use these results for the LDOS on an infinite Bethe lattice to calculate the IPR the average IPR $P_2$ of the RRG model. Indeed, the sought IPR is given by  \cite{tikhonov19statistics} 
	\begin{equation}
		P_2 = \frac{3}{N} \frac{ \langle \nu^2(E,1) \rangle }{ \langle \nu(E,1) \rangle^2 },
		\label{eq:RRG-P2-general}
	\end{equation}
	where the averages in the right-hand-side are calculated on an infinite Bethe lattice (or, equivalently, from the self-consistency equation). Note that this formula holds everywhere in the metallic phase, $W < W_c$, as long as the RRG system is sufficiently large, $N \gg N_\xi$. 
	Using the self-averaging of $\rho_1(E, \varepsilon)$ in the considered range of disorder, $W \ll m$, we obtain  
	\begin{align}
		NP_2 &= 3 \int \d \varepsilon \gamma(\varepsilon) \left[\frac{\rho_1(0,\varepsilon)}{\rho(0)}\right]^2  \nonumber \\
		&= \frac{3}{W}\int_{-W/2}^{W/2} \d\varepsilon \left[\frac{m}{(2mg_0)^2 + \varepsilon^2} \right]^2 \fullstop 
		\label{eq:NP2_Integral}
	\end{align}
	In the second line of Eq.~\eqref{eq:NP2_Integral}, we used Eqs.~\eqref{eq:RRG-rho1-result} and \eqref{eq:RRG-rho-result}. 
	
	In the RMT regime \eqref{eq:RRG-RMT-regime}, $g_0$ is given by Eq.~\eqref{eq:RRG-g0-RMT}. Substituting it in  Eq.~\eqref{eq:NP2_Integral}, one gets
	\begin{equation}
		NP_2 \simeq 3 \,,
	\end{equation}
	as expected. In the Golden-rule regime \eqref{eq:RRG-GR-regime}, $g_0$ is given by Eq.~\eqref{eq:RRG-g0-GR}. The integral in Eq.~\eqref{eq:NP2_Integral} can be then extended to $\pm \infty$, yielding
	\begin{equation}
		NP_2 \simeq \frac{3m^2}{W} \int_{-\infty}^{\infty} \d \varepsilon \frac{1}{\left[\left(\frac{\pi m}{W}\right)^2 + \varepsilon^2\right]^2} 
		= \frac{3}{2\pi^2} \frac{W^2}{m} \fullstop
		\label{eq:RRG-GR-P2-final}
	\end{equation}
	To describe accurately both (RMT and Golden-rule) regimes, as well as a crossover between them, we calculate the integral in Eq.~\eqref{eq:NP2_Integral} exactly,
	\begin{equation}
		NP_2 = \frac{3}{8g_0^3} \left[ \frac{4mg_0W}{16g_0^2m^2 + W^2} + \frac{\mathrm{arccot}\left(\frac{4mg_0}{W} \right)}{mW} \right] \fullstop \label{eq:NP2_fullSolution}
	\end{equation}
	Substituting here $g_0$ as obtained by a numerical solution of Eq.~\eqref{eq:g0_arctan}, we obtain $P_2$ in the whole range $W \ll m$. 
	
	For stronger disorder, $W \gtrsim m$, the local spectral function acquires strong fluctuations signalling an approach of the system to the critical point $W_c$, and the above approximation ceases to be applicable. This regime can be analysed by a different approach, as we are now going to discuss. 
	
	\subsubsection{Pre-critical and critical regimes} 
	\label{sec:RRG-critical}   
	
	To calculate analytically the scaling of IPR in the pre-critical and critical regimes, $m \ll W < W_c$, we extend the analysis presented in Ref.~\cite{tikhonov19critical}. While in that work 
	(where the focus was on the case of coordination number $m=2$) only the critical regime was considered, here we are interested in the limit of large $m$, which implies the emergence of a broad pre-critical regime, as discussed above. The starting formulas (summarized below) are, however, the same as in Ref.~\cite{tikhonov19critical}. 
	
	For $m \ll W < W_c$, a large correlation volume $N_\xi (W)$ emerges, which governs a broad distribution of the local density of states and thus all associated observables. In particular, under the condition of a sufficiently large system, $N \gg N_\xi$, the IPR scales as 
	\begin{equation} 
		P_2 \sim N_\xi / N \,.
		\label{eq:RRG-P2-Nxi}
	\end{equation}
	To calculate $N_\xi$, one starts with the self-consistency integral equation (that is equivalent to Eq.~\eqref{pool_sc}) and performs its linearization, which yields
	\begin{equation}
		m\lambda_{\beta} = 1 \comma 
		\label{eq:mlambda=1}
	\end{equation}
	where $\lambda_{\beta}$ is the largest eigenvalue of a certain integral operator. For large $W$, the eigenvalue $\lambda_{\beta}$ is found to be
	\begin{equation}
		\lambda_{\beta} \simeq \frac{1}{\beta -1/2} \frac{1}{W - 4/W} \left[\left(\frac{W}{2}\right)^{2\beta - 1} - \left(\frac{W}{2}\right)^{-2\beta + 1}\right] \fullstop  
		\label{eq:lambdalargem}
	\end{equation}
	
	The value $\beta = 1/2$ corresponds to the transition point $W_c$. Taking the limit $\beta \to 1/2$ in Eq.~\eqref{eq:lambdalargem}, one gets the large-$m$ equation for $W_c$, 
	\begin{equation}
		W_c = 4m\ln \frac{W_c}{2} \fullstop 
		\label{eq:Critical_Disorder}
	\end{equation}
	In the leading order, this yields the large-$m$ asymptotics of the critical disorder of the form 
	\begin{equation}
		W_c \simeq 4 m \ln m \,.
		\label{eq:RRG-Wc-asymptotic}
	\end{equation} 
	It is worth mentioning that $W_c$ as obtained from Eq.~\eqref{eq:Critical_Disorder} is in excellent agreement with the exact value of $W_c$ (found from $m \lambda_{1/2} = 1$ with $\lambda_{1/2}$ computed numerically by determining the largest eigenvalue of the corresponding integral operator) already for rather small $m$. In particular, for $m=2$, 
	Eq.~\eqref{eq:Critical_Disorder} gives $W_c = 17.23$, while the exact value is $W_c = 18.17$ \cite{tikhonov19critical}. Further, for $m=3$, 
	Eq.~\eqref{eq:Critical_Disorder} yields $W_c = 34.00$, whereas the exact value reported in Ref.~\cite{sierant2023universality}  is $W_c = 34.95$. Therefore, for large $m$, the value of $W_c$ predicted by Eq.~\eqref{eq:Critical_Disorder} can be viewed as exact. At the same time, the asymptotic formula \eqref{eq:RRG-Wc-asymptotic} discards terms that have a relative smallness $\sim  \ln \ln m / \ln m$ and thus decay rather slowly. As a result, even for a large value of connectivity such as $m=200$, the value of $W_c$ given by \eqref{eq:RRG-Wc-asymptotic} turns out to be smaller than the exact one by a factor $\approx 1.5$. 
	
	The eigenvalue $\lambda_\beta(W)$, when considered around the point $W= W_c$ and $\beta=1/2$, decreases as a function of $W$ and has a minimum (on the real axis of $\beta$) at $\beta = 1/2$. Thus, for $W < W_c$, the solution of Eq.~\eqref{eq:mlambda=1} for $\beta$ has the form 
	\begin{equation}
		\beta = 1/2 \pm i \sigma \,,
		\label{eq:RRG-beta-sigma}
	\end{equation} 
	with a real $\sigma$.  Once $\beta$ (and thus $\sigma$) is determined, as a function of $W$, from Eq.~\eqref{eq:mlambda=1}, the correlation volume $N_\xi$ is obtained via 
	\begin{equation}
		N_\xi \sim \exp \left( \frac{\pi}{\sigma} \right) \,.
		\label{eq:RRG-Nxi-sigma}
	\end{equation}
	
Up to here [i.e., in Eqs.~\eqref{eq:RRG-P2-Nxi}--\eqref{eq:RRG-Nxi-sigma}] we followed Ref.~\cite{tikhonov19critical}. 
The key idea underlying our analysis starting from this point is that, for $m \gg 1$, this formalism can be used to determine $N_\xi$ (and thus $P_2$ and other related observables) not only in the critical regime, $W_c - W \ll W_c$ (as in Ref.~\cite{tikhonov19critical}), but in a much broader disorder range, $m \lesssim W < W_c$, which includes also the (parametrically broad) pre-critical regime. It is worth noting at this point that, as our numerical simulations below show, the critical regime becomes in fact unaccessible to exact diagonalization for $m \gg 1$, whereas the pre-critical regime can be observed numerically. 

	Substituting Eq.~\eqref{eq:RRG-beta-sigma}
	in Eq.~\eqref{eq:lambdalargem}, we rewrite Eq.~\eqref{eq:mlambda=1} as
	\begin{equation}
		\frac{2m}{W \sigma} \sin \left(2\sigma \ln \frac{W}{2} \right) = 1 \,.
		\label{eq:RRG-crit-precrit}
	\end{equation} 
	Further, this equation can be cast into the form
	\begin{equation}
		\frac{\sin x}{x} = \frac{f(W)}{f(W_c)} \comma \qquad f(W) = \frac{W}{\ln \frac{W}{2}} \,,
		\label{eq:Sinc}
	\end{equation}
	where we have introduced
	\begin{equation}
		x= 2\sigma \ln \frac{W}{2} 
		\label{eq:RRG-crit-x}
	\end{equation}
	and used Eq.~\eqref{eq:Critical_Disorder} for the critical disorder $W_c$. 
	
	For a given disorder $W$, Eq.~\eqref{eq:Sinc} can be easily solved numerically for $x$.  By virtue of Eq.~\eqref{eq:RRG-crit-x}, this yields $\sigma$, which in turns allows us to determine $N_\xi$ and $P_2$ by using  Eqs.~\eqref{eq:RRG-Nxi-sigma}  and \eqref{eq:RRG-P2-Nxi}. In this way, we can obtain $NP_2 \sim N_\xi$ as a function of $W$ in the whole range $m \lesssim W < W_c$, i.e., in the pre-critical and critical regimes and in the crossover between them.  Let us emphasize again the role of the condition $m \gg 1$:  it ensures that $\ln m$ in Eq.~\eqref{eq:RRG-Wc-asymptotic} is large and thus that a parametrically broad pre-critical regime $m \lesssim W \ll W_c$ emerges.
	
	We can also proceed analytically for each of the two regimes. Consider first the critical regime, for which the system is close to the transition point,
	\be
	W < W_c \,, \qquad W_c - W \ll W \,.
	\ee
	In this case, $f(W)/ f(W_c)$ in Eq.~\eqref{eq:Sinc} is close to unity, implying that $x \ll 1$. 
	Expanding $\sin x \approx x - x^3 / 6$ in Eq.~\eqref{eq:RRG-crit-precrit} and solving the resulting equation for $\sigma$, we obtain
	\begin{equation}
		\sigma \simeq \frac{1}{\ln \frac{W}{2}} \left[\frac{3}{2} \left(1 - \frac{W}{4m \ln \frac{W}{2}}\right) \right]^{1/2} \fullstop
	\end{equation}
	Equation \eqref{eq:RRG-Nxi-sigma} yields now the correlation volume
	\begin{equation}
		\ln N_{\xi} = \frac{\pi}{\sigma} = \pi \ln \frac{W}{2} \left[ \frac{3}{2} \left( 1 - \frac{W}{4m\ln \frac{W}{2}} \right)\right]^{-1/2} \,.
	\end{equation}
	In the leading approximation, this becomes
	\begin{equation}
		\ln N_{\xi} \simeq \pi \ln m \left[ \frac{3}{2} \left( 1 - \frac{W}{W_c}\right) \right]^{-1/2} \fullstop 
		\label{eq:AsymptoticCriticalBehavior}
	\end{equation}
	Equation \eqref{eq:AsymptoticCriticalBehavior} is the known critical behavior of the correlation volume $N_\xi$ and of the associated correlation length $\xi = \ln N_\xi / \ln m$ on the delocalized side of the transition, with the critical exponent $\nu_{\mathrm{del}} = 1/2$ \cite{tikhonov19critical}.
	
The following comment is in order here. As has been pointed out above, the localization transition on RRG is strongly ``asymmetric'': the critical point shares many key properties of the localized phase. In particular, $P_2 \sim 1$ at $W=W_c$, and the levels statistics has (in the large-$N$ limit) the Poisson form, as in the localized phase. For this reason, we largely focus in this work on the disorder range $W < W_c$, where the evolution from ergodicity to localization take place. This asymmetry is also reflected in values of critical exponents. While the delocalized side, $W<W_c$ is described by the above exponent $\nu_{\mathrm{del}} = 1/2$, on the localized side, $W>W_c$, average and typical observables can be characterized by different exponents $\nu_{\mathrm{loc}} = 1$  and $\tilde{\nu}_{\mathrm{loc}} = 1/2$ \cite{garcia-mata2022critical}. The physics on the localized side is governed by rare resonances \cite{tikhonov2021eigenstate}. While this is an interesting physics that has its direct counterpart in many-body problems, it is beyond the scope of this paper.
	
	Let us now turn to the analysis of the pre-critical regime, which is located between the Golden-rule and critical regimes,
	\begin{equation}
		m \lesssim W \ll W_c \,.
	\end{equation}
	Here, $f(W) / f(W_c) \simeq W/W_c \ll 1$, which implies, according to Eq.~\eqref{eq:Sinc}, that 
	$x = \pi - \delta$ with $\delta \ll 1$. Equation~\eqref{eq:Sinc} then yields
	$\delta \simeq \pi W / W_c$ and thus $x = \pi (1- W/ W_c)$ and, consequently,
	\begin{equation}
		\sigma \simeq \frac{\pi \left(1 - \frac{W}{W_c} \right)}{2 \ln \frac{ W}{2}} \fullstop
	\end{equation}
	According to Eq.~\eqref{eq:RRG-Nxi-sigma}, this gives the correlation volume
	\begin{equation}
		N_{\xi} \sim e^{\pi/\sigma} \sim W^2 e^{\frac{W}{2m}} \fullstop
		\label{eq:RRG-Nxi-precrit}
	\end{equation}
	We see that the correlation volume $N_\xi$ (as thus the IPR $P_2$) grows exponentially with $W$ in the pre-critical regime. Furthermore, we observe that $W^2$ factor in Eq.~\eqref{eq:RRG-Nxi-precrit} is in perfect correspondence with the $W$ dependence of IPR in the Golden-rule regime, Eq.~\eqref{eq:RRG-GR-P2-final}. The only mismatch between the two formulas at the boundary between the two regimes, $W \sim m$, is due to a factor $\sim 1/m$ in Eq.~\eqref{eq:RRG-GR-P2-final}. This is not surprising as 
	such a pre-exponential factor is beyond the accuracy of Eq.~\eqref{eq:RRG-Nxi-sigma} that was used in course of the analysis of the pre-critical regime. To have a full matching, we include the pre-exponential factor $\sim 1/m$ in Eq.~\eqref{eq:RRG-Nxi-sigma}, $N_\xi \sim m^{-1} e^{\pi/\sigma}$,  which yields the final result for the pre-critical regime,
	\begin{equation}
		N_{\xi} \sim \frac{1}{m}e^{\pi/\sigma} \sim \frac{W^2}{m} \exp \left( \frac{W}{2m}\right), \ \ \  m\lesssim W \ll W_c \fullstop
		\label{RRG-Nxi-precrit-final}
	\end{equation}
	This formula works reasonably well up to $W \approx W_c / 2$, where it crosses over to the critical behavior, Eq.~\eqref{eq:AsymptoticCriticalBehavior}. 
	To estimate the value of $NP_2 \sim N_\xi$ at the crossover between the pre-critical and critical regimes, we substitute $W = W_c/2$ into
	Eqs.~\eqref{eq:Sinc} and \eqref{eq:RRG-crit-x}, and the resulting $\sigma$ into $N_\xi \sim m^{-1} e^{\pi/\sigma}$. This yields an estimate $N_\xi^{\rm cr}$ of the correlation volume $N_\xi$ at the beginning of the critical regime:
	\begin{equation}
		N^{\rm cr}_\xi (m) \sim \frac{1}{m} \left(\frac{W_c(m)}{2} \right)^\gamma \sim \frac{1}{m} (2m \ln m)^\gamma \,; \qquad \gamma \approx 3.3 \,.
		\label{eq:RRG-cross-precrit-crit}
	\end{equation}
	
	\begin{figure} 
		\includegraphics[width=0.5\textwidth]{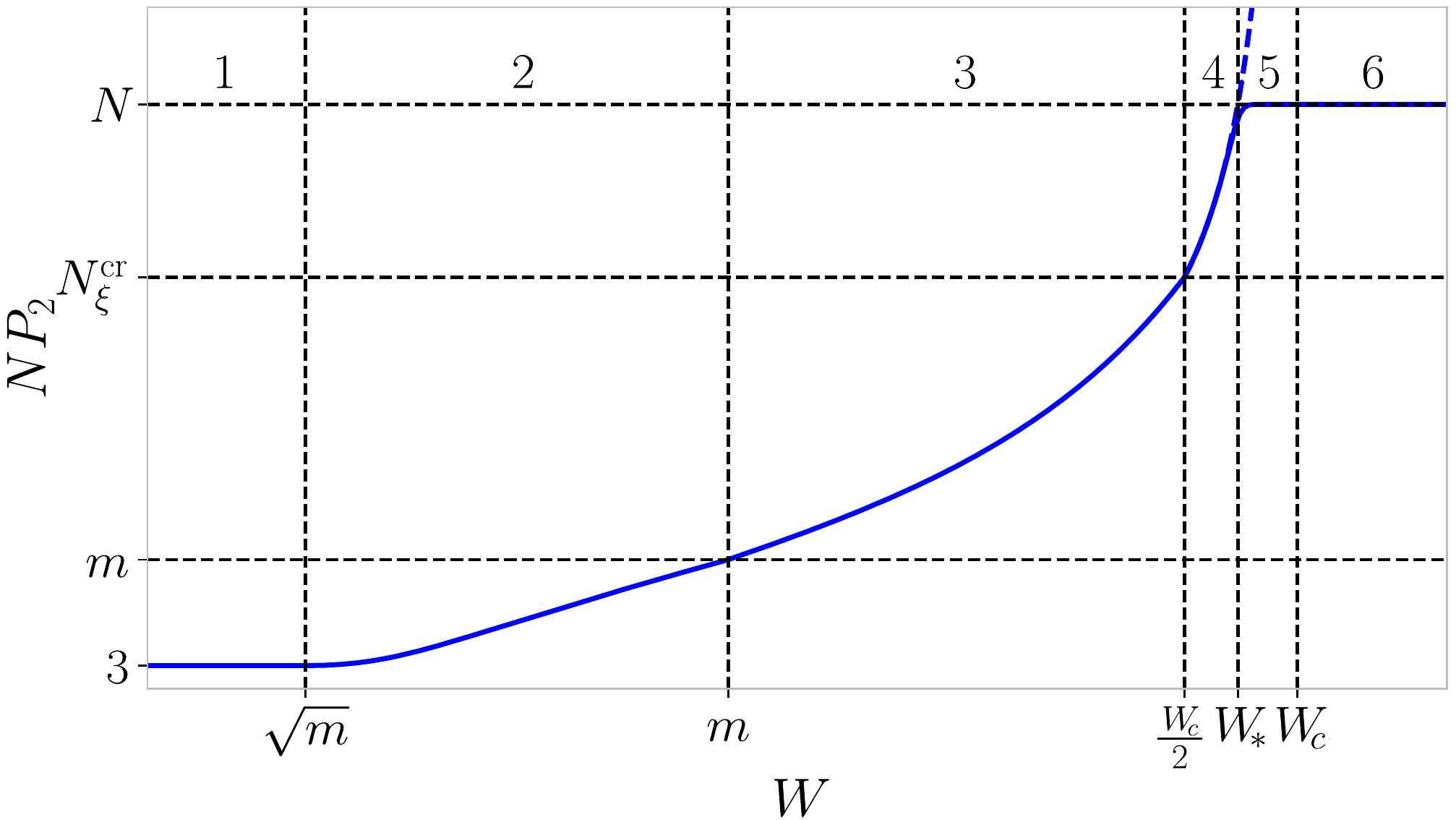}
		\caption{$NP_2$ as a function of disorder $W$ for  RRG with large connectivity m. Schematic presentation of the regimes on log-log scale: 1) RMT, 2) golden rule, 3) pre-critical, 4) ``critical metal'', 5) critical point, 6) localized. It is assumed in this figure that the system size $N$ is much larger than $N_\xi^{\rm cr}$ marking the crossover between the pre-critical and critical regimes,  Eq.~\eqref{eq:RRG-cross-precrit-crit}. In the opposite case, $N < N_\xi^{\rm cr}$, the ``critical metal'' regime will not be probed. Dashed line shows the behavior of $NP_2$ in the ``thermodynamic limit'' $N\to \infty$. In this limit, $W_* \to W_c$ and the regime 5 shrinks to a single point $W_c$. 
		}
		\label{fig:RRGschematic}
	\end{figure}
	
	It is worth recalling at this point that we consider the IPR in an RRG model with a finite (although large) number of sites $N$.  When the disorder $W$ increases towards $W_c$, the correlation volume $N_\xi$ at some point reaches $N$. Let us call this disorder $W_*(N)$; it is defined by the condition $N_\xi(W_*) = N$.  Clearly, $\lim_{N \to \infty} W_*(N) = W_c$, so that $W_*(N)$ can be viewed as a finite-size approximation to the critical disorder $W_c$. When $W$ increases beyond $W_*(N)$, the IPR $P_2$ exhibits saturation to its maximal value $NP_2 = N$. Depending on the value of $N$ with respect to $N_\xi^{\rm cr}$, the point $W_*(N)$ can be located either in the critical or in the pre-critical regime (or in the crossover between them).   In Fig.~\ref{fig:RRGschematic}, a sketch of the dependence of $NP_2$ on disorder $W$ is shown, with all regimes indicated. In this sketch, it is assumed that $N$ is very large (much larger than $N_\xi^{\rm cr}$), so that $W_*(N)$ is deeply in the critical regime.  We note that, for $m > 10$, the crossover value $N_\xi^{\rm cr}$, Eq.~\eqref{eq:RRG-cross-precrit-crit}, becomes larger than system sizes $N$ accessible by exact diagonalization. Thus, for large $m$, the point $W_*(N)$ will be within the pre-critical regime if exact diagonalization is used, as we are going to see in Sec.~\ref{sec:rrg-ED}. In that case, the curve $NP_2 (W)$ will not probe the  ``critical metal'' regime.

	\subsection{Exact-diagonalization numerical simulations}
	\label{sec:rrg-ED}  
	
\begin{figure}
		\includegraphics[width=0.5\textwidth]{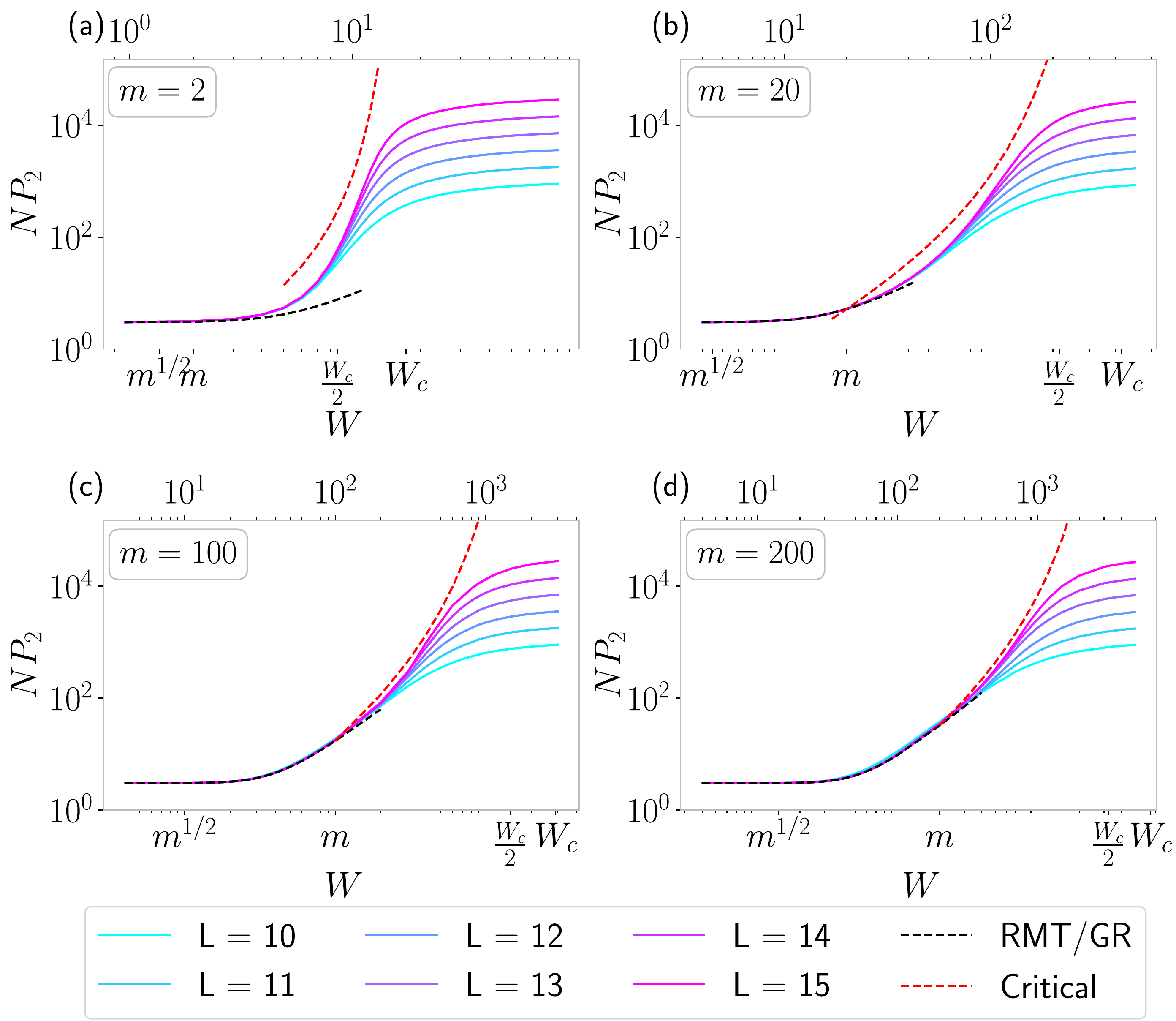}
		\caption{$NP_2$ of RRG as a function of W for various system sizes:  (a) $m = 2$, (b) $m = 20$, (c) $m = 100$, (d) $m = 200$.  Full lines are results of exact diagonalization for various systems sizes $N=2^L$ (Sec.~\ref{sec:rrg-ED}), while dashed lines represent $N\to \infty$ analytical prediction for the RMT and Golden-rule regimes (Sec.~\ref{sec:RRG-RMT-GR}) and for the pre-critical and critical regimes (Sec.~\ref{sec:RRG-critical}). 
		}
		\label{fig:RRG_NP2_largem}
	\end{figure}

	We have calculated the averaged IPR on RRG with large coordination number, $m = 20$, 100, and 200 by numerical diagonalization of systems of sizes $N$ from $2^{10}$ to $2^{15}$. Note that for large $m$ the Hamiltonian matrix is much less sparse (in comparison with, say, $m=2$), which affects the upper boundary for system sizes than can be efficiently treated by exact diagonalization. For each realization of disorder, $N/16$ eigenstates in the middle of the spectrum (i.e., around $E=0$) were determined by the shift-invert techniques. In addition, the IPR was averaged over 100 disorder realizations. The results are shown in Fig.~\ref{fig:RRG_NP2_largem}. To better illustrate the evolution with increasing $m$, we include also the data for $m=2$ for the same range of $N$. The black and red dashed lines in this figure represent large-$m$ analytical results for the RMT and Golden-rule regimes (Sec.~\ref{sec:RRG-RMT-GR}) and for the pre-critical and critical regimes (Sec.~\ref{sec:RRG-critical}), respectively. By inspecting the theoretical curves, we see that the Golden-rule regime, where the theoretical curves agree with each other, becomes well developed for $m=200$. Furthermore, we observe that the agreement between exact-diagonalization data and analytical predictions improves with increasing $m$, becoming essentially excellent for $m=200$. This agreement with the numerics serves as an additional confirmation of the validity of our large-$m$ analytical treatment. 
	
	We recall that the analytical curves shown in Fig.~\ref{fig:RRG_NP2_largem} are derived under the assumption $N\gg N_\xi$. As expected, the numerical data follow the analytical curve up to the point $W_*$ where $NP_2 \sim  N_\xi$ approaches the system size $N$ and then saturate at the value $NP_2 = N$. With increasing system size $N$, the ``finite-size critical disorder'' $W_*(N)$ increases towards $W_c$ as predicted, i.e., the numerical data reproduce the analytical curve up to an increasingly large disorder. At the same time, even for our largest system size $N = 2^{15}$, the point $W_*(N)$ for $m=100$ and $m=200$ models is located well on the pre-localized side, $W_*(N) < W_c / 2$, of the pre-localized-to-localized crossover. Therefore, an exact-diagonalization numerical investigation of the asymptotic ``critical-metal'' regime ($W_c-W \ll W_c$ and $N\gg N_\xi$)  is essentially impossible for large-$m$ RRG models due to system-size limitations.
	
	\begin{figure}
		\includegraphics[width=0.5\textwidth]{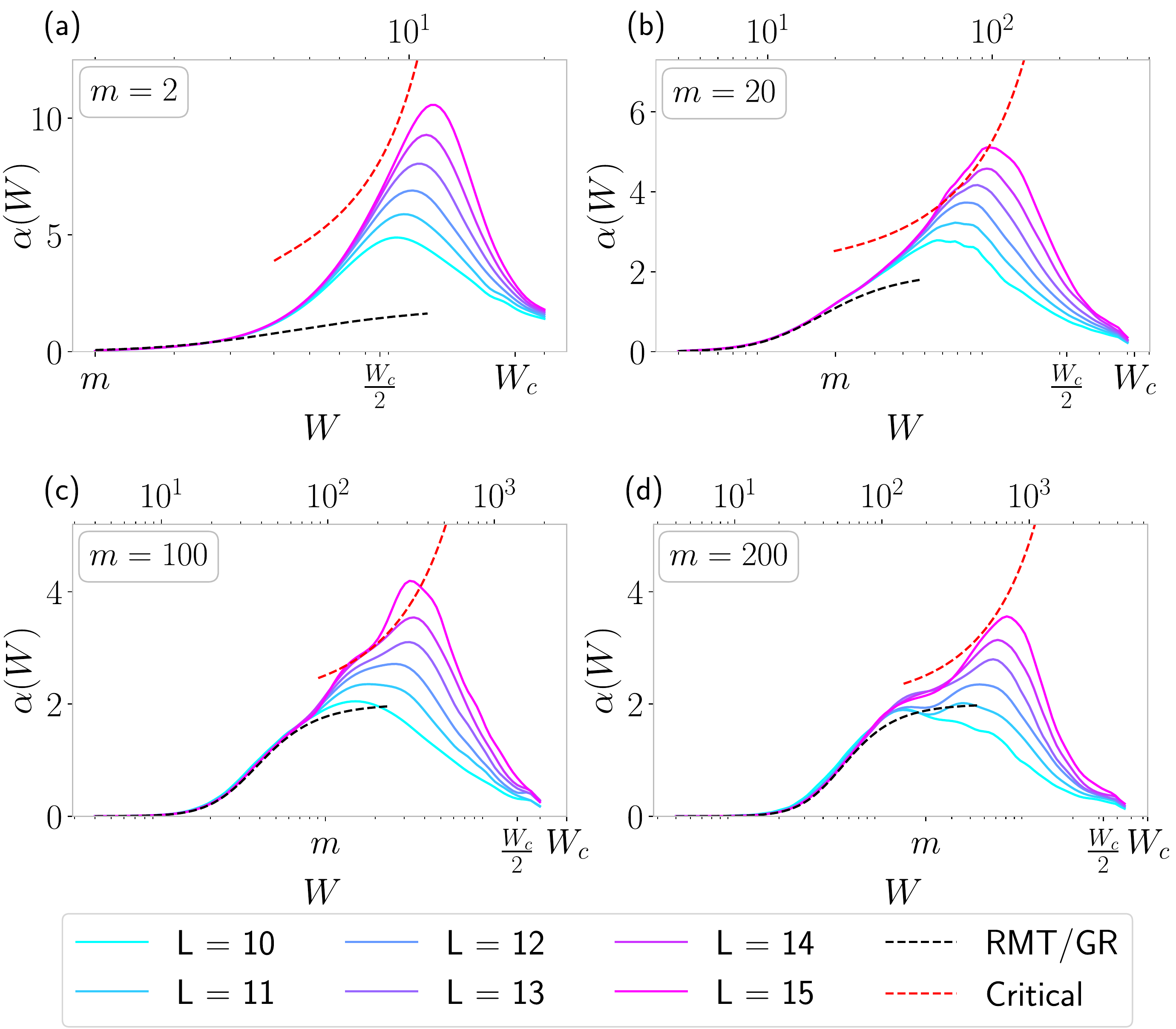}
		\caption{The same as Fig. \ref{fig:RRG_NP2_largem} but in the form of a logarithmic derivative $\alpha(W) = \d \ln (NP_2) / \d \ln W$.  At $m=200$, development of the Golden-rule plateau at $\alpha=2$ is clearly visible. The maxima of curves, which yield finite-size estimates $W_*(N)$ of the transition point, move towards $W_c$ with increasing $N$. For large $m=20$, 100, and 200, the maxima are in the pre-critical (rather than critical) regime for all accessible system sizes, $W_*(N) < W_c/2$. 
		}
		\label{fig:RRG_NP2_dlogdlog_largem}
	\end{figure}
	
	In Figure \ref{fig:RRG_NP2_dlogdlog_largem} we present the same numerical and analytical results as in Fig. \ref{fig:RRG_NP2_largem} but in the form of a logarithmic derivative,
	\begin{equation}
		\alpha(W) = \frac{ \d \ln (NP_2)}{\d \ln W} \,.
		\label{eq:RRG-alphaW}
	\end{equation}
	This presentation of the data is useful in two respects. 
	
	First, maxima of the curves nicely visualize the position of $W_*(N)$ (finite-size approximation  to the critical disorder) defined above. It is seen that $W_*(N)$ increases towards the true ($N\to \infty$) critical point $W_c$ as predicted analytically. At the same time, the maximum remains well below $W_c/2$ (i.e., in the pre-critical regime) for $m=20$, and especially for $m=100$ and 200. This shows that the critical regime (in the sense of ``critical metal'': $W_c-W \ll W_c$ and $N\gg N_\xi$) is totally inaccessible to exact diagonalization for large $m$. 
	
	Second, the logarithmic derivative $\alpha(W)$ is a sensitive indicator of the Golden-rule regime. Indeed, a power-law scaling $NP_2 \propto W^2$ in this regime implies $\alpha(W) = 2$. Therefore, for large $m$, when the Golden-rule regime is predicted, we expect to see a plateau at $\alpha(W) = 2$. Indeed, the exact-diagonalization data for $m=200$ and sufficiently large $N$ in Fig.~\ref{fig:RRG_NP2_dlogdlog_largem}d clearly exhibit a shoulder near $\alpha = 2$, demonstrating the Golden-rule regime. Furthermore, the fact that $\alpha(W)$ for our largest $N$ raises to values substantially larger than 2 around the maximum (i.e., to the right of the Golden-rule regime) demonstrates that the system probes well also the pre-critical regime, where $\alpha(W)$ grows as 
	\begin{equation} 
		\alpha(W) = 2 + \frac{W}{2m} 
	\end{equation}
	according to Eq.~\eqref{RRG-Nxi-precrit-final}.
	
	\subsection{Population dynamics}
	\label{sec:rrg-population-dyn}
	
	In the above, we used two distinct methods to calculate the disorder dependence of IPR in an RRG model with large $m$: an analytical approach (based on supersymmetric field theory)  in Sec.~\ref{sec:rrg-analytical} and numerical exact-diagonalization approach in Sec.~\ref{sec:rrg-ED}. Now, we are going to present results of one more approach that combines analytical (field theory) and numerical (population dynamics) tools. This approach is described in detail in Ref.~\cite{tikhonov19critical}, where it was implemented for the $m=2$ model. We first follow the analytical route (see Sec.~\ref{sec:RRG-analytical-general}), by using the field-theoretical formalism, leading, in the large-$N$ limit, to the saddle-point equation \eqref{pool_sc}   (equivalent to the self-consistency equation on an infinite Bethe lattice). This is a non-linear integral equation for the probability distribution $\mathcal{P}(\Re\,G,\Im\,G)$ of the Green function $G^{(m)} \equiv G$. We now solve this equation numerically by using the population dynamics (also known as pool method). Within this computational approach, a large pool of values of $G$ represents its distribution. We iterate the self-consitsency equation \eqref{pool_sc} starting with a randomly initialized pool until the convergence is reached. Once the distribution $\mathcal{P}(\Re\,G,\Im\,G)$ is determined in this way, we calculate the IPR (at $N \gg N_\xi$) by using Eq.~\eqref{eq:RRG-P2-general},
	\begin{equation}
		NP_2 = 3\frac{\langle \nu^2 \rangle}{\rho^2} \comma
		\label{eq:PD_IPR}
	\end{equation} 
	where $\nu = (-1/\pi) \Im\,G$ is the (fluctuating) local density of states and $\rho$ is its average, $\rho = \langle \nu \rangle$.
	
	A numerical limitation to the method is set by the pool size $M$. We have implemented the method for the $m=200$ model, by using the pool size $M = 2^{22}$. Remarkably, it was found in Ref.~\cite{tikhonov19critical} (in the analysis of the $m=2$ model) that, within this method, one can proceed controllably up to $NP_2 \sim N_\xi$ scaling as $N_\xi \sim M^{1/\kappa}$ with $\kappa \approx 0.46$, i.e.,  many orders of magnitude larger than the pool size $M$. The same happens in our simulations for $m=200$, allowing us to get population-dynamics results  up to $NP_2$ as large as $10^{13}$. 
	
	The imaginary part $\eta$ in Eq.\eqref{pool_sc} sets an upper bound $\sim \eta^{-1}$ on the correlation volume that can be calculated within this approach. Thus, $\eta$ should be chosen to be sufficiently small. We have taken $\ln \eta = -73.6$, so that $\eta$ is negligibly small in comparison with $N_\xi^{-1}$ for the largest $N_\xi$ that  can be reached. 
	
\begin{figure}
		\includegraphics[width=0.5\textwidth]{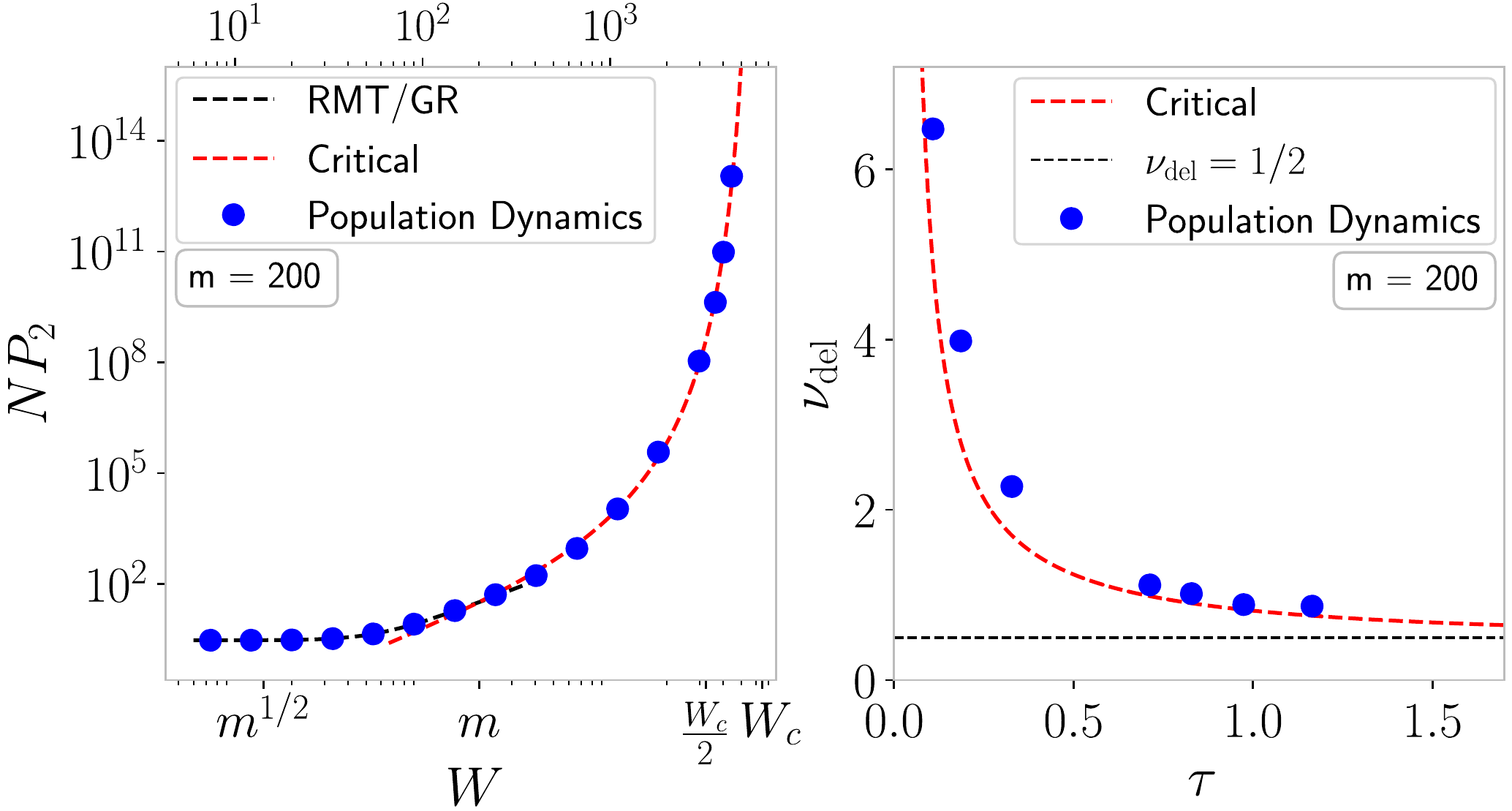}
		\caption{Population-dynamics results for RRG with $m=200$ (blue dots), in comparison with purely analytical results for the RMT and Golden-rule regime (black dashed line) and for the pre-critical and critical regimes (red dashed line). (a) $NP_2$ as a function of disorder $W$. (b) Flowing correlation-length exponent $\nu_{\mathrm{del}}(\tau)$, Eq.~\eqref{eq:RRG-nu-del-flowing}, as a function of $\tau = - \ln (1-W/W_c)$. The asymptotic ($\tau \to \infty$, i.e., $W \to W_c$) value $\nu_{\mathrm{del}}(\tau) = 1/2$ is shown by a horizontal dashed line.  
		}
		\label{fig:RRG_NP2_Population_Dynamics}
	\end{figure}

	Results of the population-dynamics calculation of $NP_2$ are presented in Fig.~\ref{fig:RRG_NP2_Population_Dynamics}a. For comparison, we also show the results obtained by purely analytical means (for RMT and Golden-rule regimes as well as for pre-critical and critical regimes). An excellent agreement is observed, providing an additional evidence of the validity and accuracy of all approaches used. It is seen that the population dynamics allows us to proceed up to disorder $W$ somewhat above $W_c/2$, i.e., to cover the whole pre-localized regime and also an initial part of the critical regime. 
	
	In Fig.~\ref{fig:RRG_NP2_Population_Dynamics}b, we display results for the flowing correlation-length exponent
	\begin{equation}
		\nu_{\mathrm{del}} (\tau) = \frac{\d \ln \ln NP_2}{\d \tau} \comma
		\label{eq:RRG-nu-del-flowing}
	\end{equation}
	where $\tau$ is a parameter characterizing closeness to the criritcal point, 
	\begin{equation}
		\tau = - \ln \left(1 -\frac{W}{W_c} \right) \fullstop
	\end{equation}
	In the critical regime, $\tau \gg 1$, the IPR scales as $\ln (NP_2) \sim \tau^{1/2}$, see Eq.~\eqref{eq:AsymptoticCriticalBehavior}.  Thus, $\nu_{\mathrm{del}} (\tau)$ tends to the asymptotic value $\nu_{\mathrm{del}} = 1/2$ in the limit $\tau \to \infty$ (corresponding to $W \to W_c$). Figure \ref{fig:RRG_NP2_Population_Dynamics}b illustrates how this asymptotic value is approached. 
	
	It is worth noting that the position $W_*$ of the maximum in Fig.~\ref{fig:RRG_NP2_dlogdlog_largem}d for the largest $N$ in exact-diagonalization simulations is nearly one order of magnitude smaller than $W_c$. Thus, exact diagonalization allows us to proceed (within the regime $N \gg N_\xi$) only up to $\tau \approx 0.1$ for the $m=200$ model. It is seen from Fig.~\ref{fig:RRG_NP2_Population_Dynamics}b how far the corresponding value $\nu_{\mathrm{del}} (\tau)$ is from its $\tau \to \infty$ asymptotics. This demonstrates once more that it is impossible to extract controllably the asymptotic critical behavior (i.e., the exponent $\nu_{\mathrm{del}} (\infty) = 1/2$)  using solely the exact-diagonalization data.  
	
	\section{Fermionic and spin quantum-dot models: Numerical study}
	\label{sec:quantum-dots}
	
	In this section, we perform a numerical study of the IPR and the level statistics in fermionic and spin quantum dot models by means of exact diagonalization. Our main focus in this section is one the scaling of the critical disorder $W_c(n)$ with the number of fermions (respectively, spins) $n$ and on the sharpness of the transition. 
	We begin by defining the models to be studied. 
	
	\begin{figure}
		\includegraphics[width=0.5\textwidth]{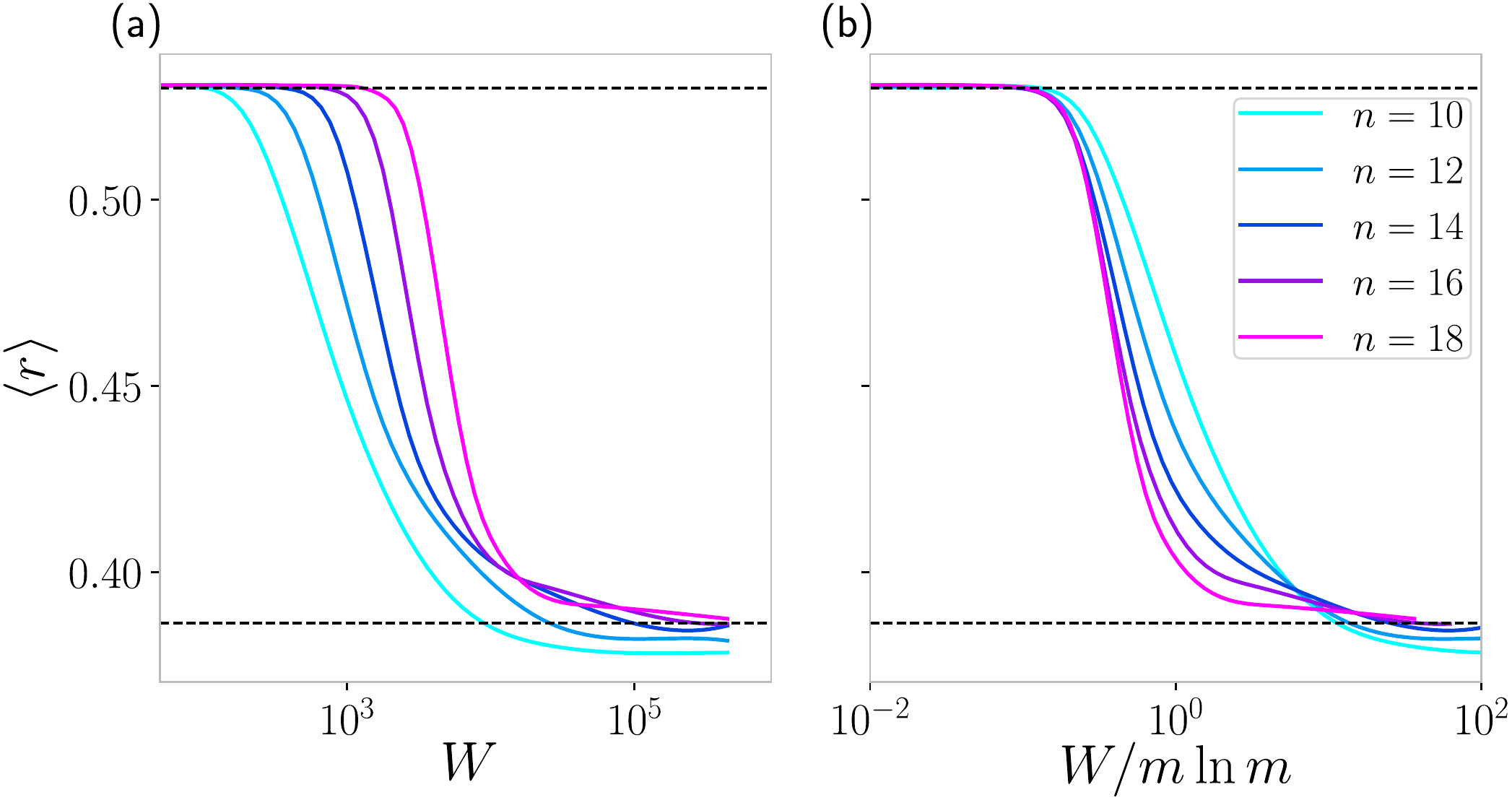}
		\caption{Level statistics from exact diagonalization of the fermionic quantum dot with $n=10$, 12, 14, 16, and 18.  {\it Left:} Mean adjacent gap ratio $r(W)$  for various system sizes; {\it Right:} same data plotted as $r(W / (m \ln m))$, where $m$ is the coordination number \eqref{eq:m-FQD}.}
		\label{fig:spectral_fermi}
	\end{figure}

	\begin{figure*}
		\includegraphics[width=\textwidth]{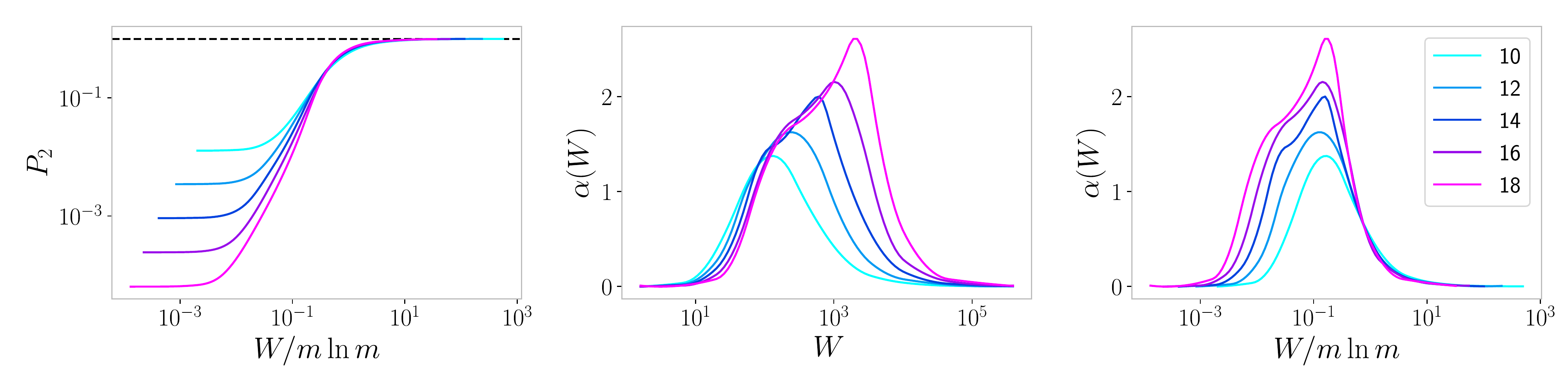}
		\caption{Average IPR $P_2$ of eigenstates in the fermionic quantum dot model with $n=10$, 12, 14, 16, and 18 as obtained by exact diagonalization.  {\it Left:} $NP_2$ as function of disorder $W$ for various system sizes. {\it Middle:} same data in the form of a logarithmic derivative $\alpha(W)$, Eq.~\eqref{eq:RRG-alphaW}.  {\it Right:} same data with rescaled disorder axis:  $\alpha$ as a function of $W / (m\ln m)$, where $m$ is the coordination number. The position of the maximum is nearly independent on $n$  for $n \ge 12$. For the largest system size, a shoulder near $\alpha = 2$ is observed, which corresponds to the Golden-rule regime.}
		\label{fig:ipr_fermi}
	\end{figure*}
	
	The fermionic quantum dot can be described by the following Hamiltonian written in the basis of exact eigenstates of the non-interacting problem:
	\be
	\label{hamiltonian_qdot}
	\hat H = \sum_{i} \varepsilon_i \hat c_i^\dag \hat c_i + \sum_{ijkl} V_{ijkl}\left(\hat c_i^\dag \hat c_j^\dag \hat c_k \hat c_l+h.c.\right).
	\ee
	Here single-particle orbital energies $\varepsilon_i$ are random, we choose them to be uniformly distributed on the interval $[-W,W]$. Further, the interaction matrix elements $V_{ijkl}$ are random as well; they are chosen as Gaussian random variables with zero mean and variance unity. (We can always rescale $\hat H$ to make the variance equal to unity, so that this assumption does not restrict the generality.) The number of particles is conserved, and we consider $n/2$ fermions occupying $n$ orbitals. 
	
Models of the type \eqref{hamiltonian_qdot}, with some variations, were proposed for description of complex nuclei and atoms under the names ``two-body random ensemble'', ``embedded two-body ensemble'', and ``two-body random interaction model'' \cite{brody1981random-matrix,flambaum1996towards}. More recently, they attracted attention in the context of Fock-space MBL physics \cite{altshuler1997quasiparticle,jacquod1997emergence,mirlin1997localization,silvestrov1997decay,silvestrov1998chaos,gornyi2016many,gornyi2017spectral, jacquod1997emergence, georgeot1997breit, leyronas2000scaling, shepelyansky2001quantum,  PhysRevE.62.R7575, jacquod2001duality, rivas2002numerical,bulchandani2022onset} as ``quantum dot'' models; we use this terminology in the present paper. 
(We refer the reader to Refs.~\cite{blanter1996electron,altshuler1997quasiparticle} for a derivation of the Hamiltonian \eqref{hamiltonian_qdot} for a model of electrons in a disordered quantum dot with Coulomb interaction.) In the last few years, a very similar model (usually defined in terms of Majorana fermions) was studied under the name of $\text{SYK}_2 + \text{SYK}_4$ model (where SYK is an abbreviation for Sachdev-Ye-Kitaev) in several works \cite{garcia-garcia2018chaotic,micklitz2019nonergodic,monteiro2020minimal,monteiro2020quantum,nandy2022delayed, larzul2022quenches}. We note that in the $\text{SYK}_2 + \text{SYK}_4$ formulation, the quadratic part of the Hamiltonian is defined as random Gaussian matrix. Transforming to a basis in which the quadratic part is diagonal, one gets the form \eqref{hamiltonian_qdot}. There is a small difference with respect to our model, as single-particle energies $\varepsilon_i$ will then exhibit level repulsion, while we assume them to be uncorrelated for simplicity. This difference is, however, immaterial since interaction-induced transitions corresponding to moving two (in the lowest order) or more (in higher orders) particles, and the corresponding many-body energies do not exhibit level repulsion.

	The model of spin quantum dot that we explore is defined by the Hamiltonian
	\be
	\hat H = \sum_{i=1}^n  \varepsilon_i \hat S_i^z + \sum_{i,j=1}^n \sum_{\alpha, \beta \in \{x,y,z\}}V_{ij}^{\alpha\beta}\left(\hat S_i^\alpha \hat S_j^\beta+h.c. \right).
	\label{H-spin-quantum-dot}
	\ee
	Here $\hat S_i^\alpha = \frac12 \sigma_i^{\alpha}$ with $i=1, \ldots, n$ and $\alpha = x,y,z$ are spin-$\frac{1}{2}$ operators ($\sigma_i^{\alpha}$ are Pauli matrices), $\varepsilon_i$ are random fields uniformly distributed on $[-W,W]$, and interaction matrix elements $V_{ij}^{\alpha\beta}$ are gaussian random variables with zero mean and variance unity.
	
	For each disorder realization, we average over  $N/10$ states in the middle of the many-body band (where $N$ is the dimension of the Hilbert space). In addition, the average over disorder realizations is performed; their number ranges from $10^4$ for the smallest systems to $10^2$ for the largest systems. 
	
	In Fig.~\ref{fig:spectral_fermi}, we show the evolution of the level statistics in the fermionic quantum dot model \eqref{hamiltonian_qdot} for number of orbitals $n$ growing from $n=10$ to $n=18$. Specifically, we plot the mean adjacent gap ratio $r(W)$, which is known to be a very convenient spectral observables for detecting a transition from ergodicity to localization, as a function of disorder $W$.  The value $r = 0.536$ corresponds to an ergodic system, with level statistics as in the Gaussian Orthogonal Ensemble (GOE), whereas $r=0.386$ corresponds to Poisson statistics, i.e., localization.  As is clearly seen in the left panel, the upper border of the ``ergodicity region'' (i.e., of the range of $W$ in which $r(W)$ is close to its GOE value) quickly moves to the right with increasing $n$. In other words, for any fixed $W$ the system becomes ergodic when $n$ is sufficiently large. This is a manifestation of the fact that the critical disorder $W_c(n)$ grows fast with $n$.  To analyze this growth, we now test the RRG approximation for the scaling of critical disorder, $W_c^{\rm RRG} \sim m \ln m$, where $m$ is the coordination number for the fermionic quantum dot model,
	\begin{equation}
		m = \frac{\left[ \frac{n}{2} \left( \frac{n}{2} - 1\right) \right]^2 }{4} \,.
		\label{eq:m-FQD}
	\end{equation}
	For this purpose, we plot in the right panel of Fig.~\ref{fig:spectral_fermi} the same data for the gap ratio with the disorder axis rescaled as $W / (m \ln m)$. In these rescaled coordinates, the ergodicity region is nearly identical for four largest system sizes ($n$ from 12 to 18). Thus, the level statistics data support the RRG-like scaling of the critical disorder: 
	\begin{equation}
		W_c(n) \sim W_c^{\rm RRG} \sim m \ln m \,.
		\label{Wcn-QD-scaling}
	\end{equation}
	
	The following comment is in order here. As always, one should be cautious when trying to make a conclusion on the behavior in the limit of a large system size on the basis of numerical results for not too large systems. Obviously, these numerical data cannot rigorously prove anything concerning the asymptotic behavior. When analyzing the data, we make a plausible assumption that the trends that we observe in numerical simulations survive in the large-$n$ limit. 
	
	In Fig.~\ref{fig:ipr_fermi}, we show the evolution of the IPR  $P_2$ in the fermionic quantum-dot model. The left panel displays the dependence of $NP_2$ on disorder $W$ for different values of $n$. Here $N = n! / [(n/2)!]^2 $ is the dimension of the Hilbert space. For weak disorder we have $P_2 \simeq 3/N$ as in GOE; for strong disorder $P_2 \simeq 1$ as expected in the localized regime. The evolution between these two limits is qualitatively similar to that in the RRG model, see Fig.~\ref{fig:RRG_NP2_largem}; we relegate a quantitative comparison to Sec.~\ref{sec:quantum-dot-rrg}, focussing now on the position of the transition point $W_c(n)$. 
	For this purpose, we use the numerical data to calculate the logarithmic derivative $\alpha(W)$ defined by Eq.~\eqref{eq:RRG-alphaW}. The dependences $\alpha(W)$ for different system sizes are shown in the middle panel of Fig.~\ref{fig:ipr_fermi}. The maximum of $\alpha(W)$ provides an estimate for the finite-size position of the transition $W_c(n)$, cf.
	Fig.~\ref{fig:RRG_NP2_dlogdlog_largem} for the RRG model in Sec.~\ref{sec:rrg-ED}.  We observe that the position of the maximum (and thus $W_c(n)$) rapidly increases with $n$, in full agreement with the analogous conclusion made on the basis of level statistics, Fig.~\ref{fig:spectral_fermi}. In the right panel of Fig.~\ref{fig:ipr_fermi}, the date are plotted as a function of $W/(m\ln m)$. It is seen that, upon this rescaling, the position of the maximum is nearly independent on $n$ (in fact, even slightly drifting to the right) for $n$ from 12 to 18. This provides a further support to the RRG-like scaling of $W_c(n)$, Eq.~\eqref{Wcn-QD-scaling}.

	\begin{figure}
		\includegraphics[width=0.4\textwidth]{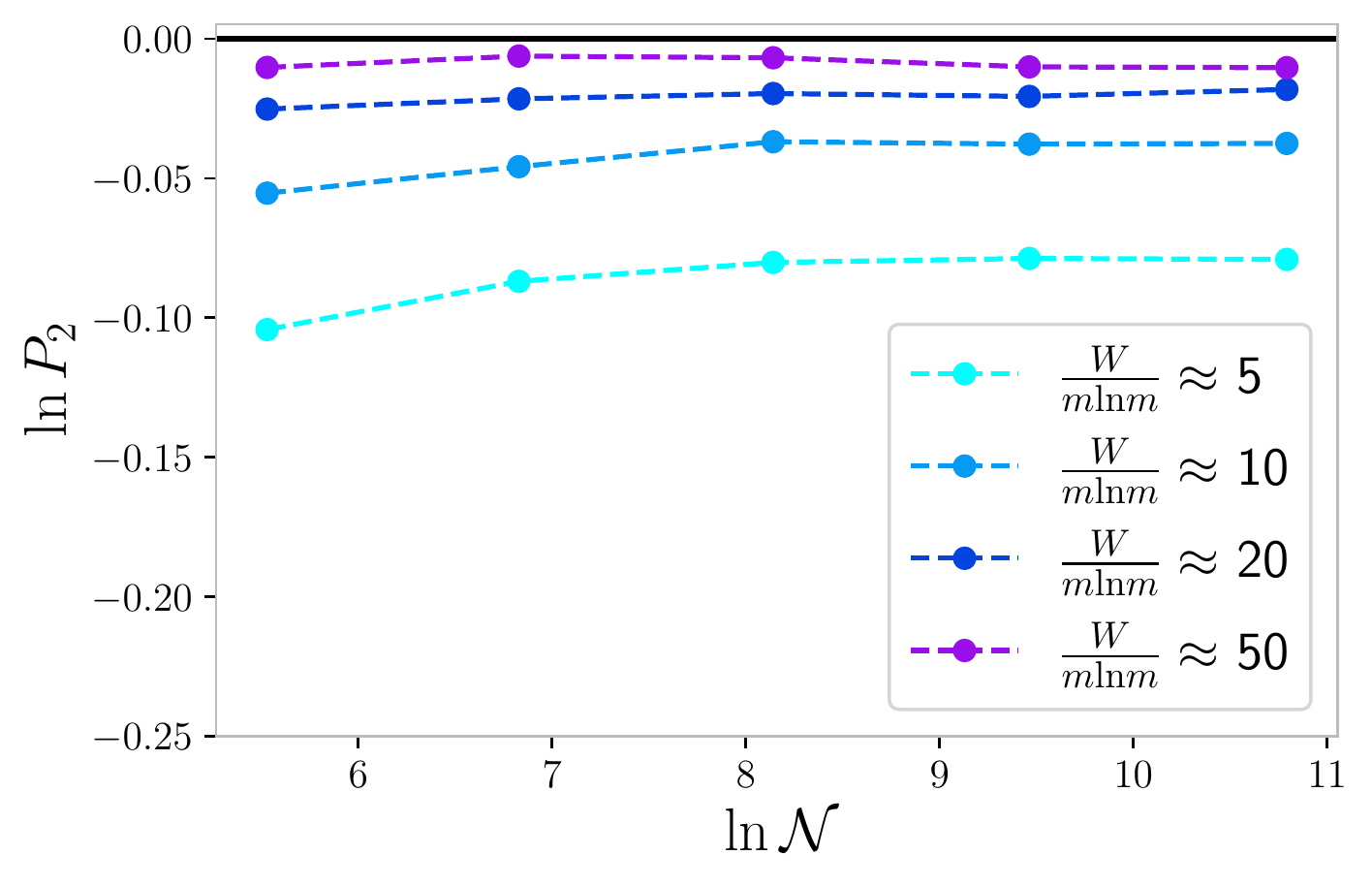}
		\caption{IPR in the localized phase of the fermionic quantum-dot model as a function of the Hilbert space dimension $N$ for several values of $W / m \ln m$.}
		\label{fig:ipr_localized_fermi}
	\end{figure}

A comment on the observed behavior of the IPR in the MBL phase is in order at this point.  In Fig.~\ref{fig:ipr_localized_fermi}, we plot $\ln P_2$ as a function of $\ln N$ for several values of $W / m \ln m$. (Fixing $W/ m \ln m$ corresponds, at least approximately, to fixing $W / W_c$ as discussed above.) We see that $P_2$ stays close to unity and is essentially independent on $N$.
This behavior, $P \sim 1$, in the MBL phase of a quantum dot model can be contrasted to the fractal scaling of IPR in the MBL phase of models with spatial structure, $P_2 \sim N^{-\tau}$, see Sec.~\ref{sec:observables} for more detail, including relevant references. This difference has the following reason. In the MBL phase of a large system with spatially localized one-particle states, there are many short-scale (adjacent-site) resonances, which lead to the above fractal scaling. On the other hand, in quantum-dot models, the critical disorder scales (at least approximately) as $W_c \sim m \ln m$, where $m$ is a coordination number on the Fock-space lattice. Thus, in the MBL phase ($W > W_c$), we also have $W > m$, so that typically there is no resonances at all. This explains the behavior $P_2 \sim 1$, which is analogous to that in the localized phase of the RRG model.

	\begin{figure}
		\includegraphics[width=0.5\textwidth]{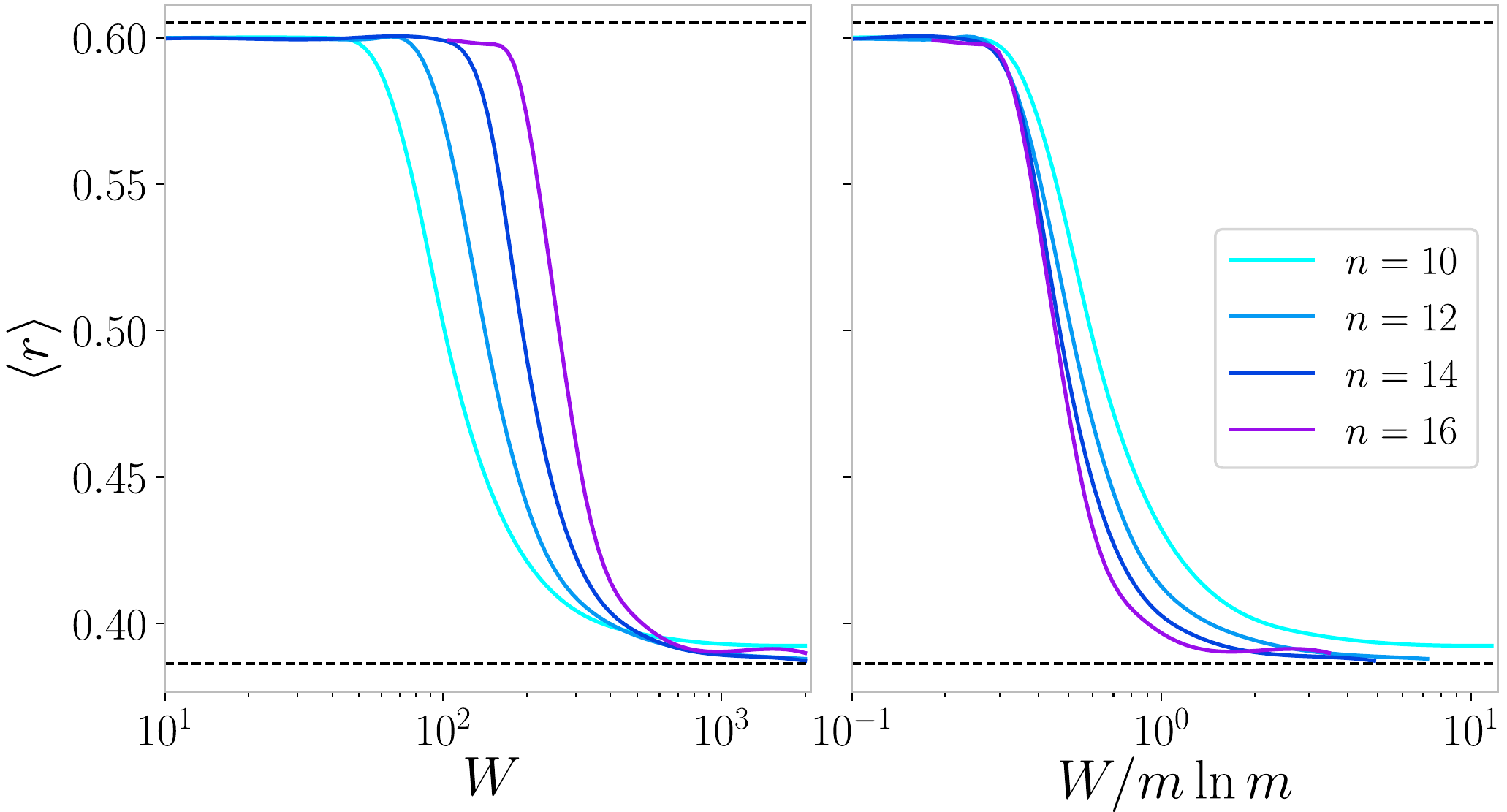}
		\caption{Level statistics from exact diagonalization of the spin quantum dot with $n=10$, 12, 14, and 16.  {\it Left:}   mean adjacent gap ratio $r(W)$  for various system sizes. {\it Right:} same data plotted as $r(W / (m \ln m))$, where $m$ is the coordination number \eqref{eq:m-SQD}.}
		\label{fig:spectral_spin}
	\end{figure}
	
\begin{figure*}
		\includegraphics[width=\textwidth]{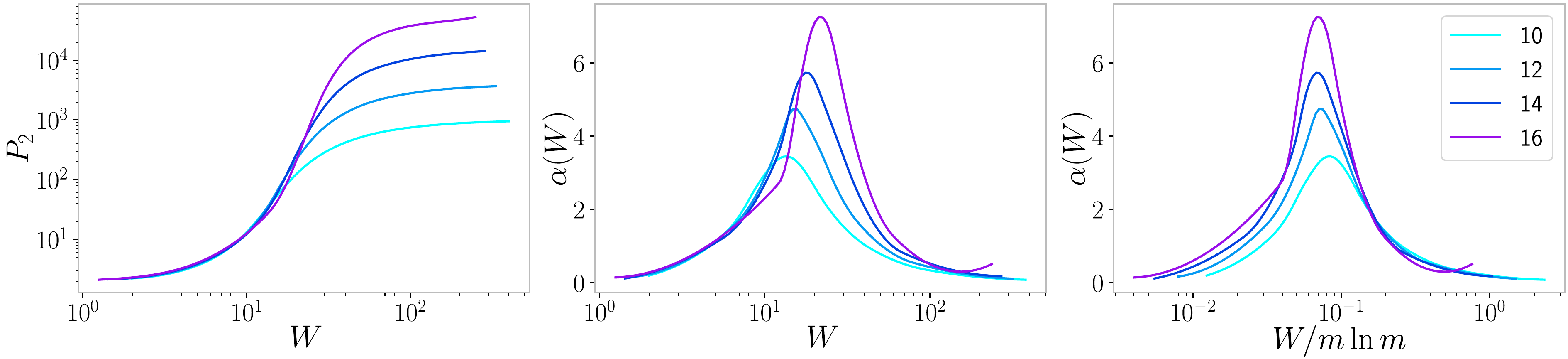}
		\caption{Average IPR $P_2$ of eigenstates in the spin quantum dot model with $n=10$, 12, 14, and 16, as obtained by exact diagonalization. {\it Left:} $NP_2$ as function of disorder $W$ for various system sizes. {\it Middle:} same data in the form of a logarithmic derivative $\alpha(W)$. {\it Right:} same data with rescaled disorder  axis:  $\alpha$ as a function of $W / (m\ln m)$, where $m$ is the coordination number.}
		\label{fig:ipr_spin}
	\end{figure*}

	In Figs.~\ref{fig:spectral_spin} and \ref{fig:ipr_spin}, numerical results for the level statistics (mean gap ratio) and the IPR of the spin quantum dot model \eqref{H-spin-quantum-dot} are shown. The presentation  of data in these two figures is fully analogous to that for fermionic quantum dot in Figs.~\ref{fig:spectral_fermi} and \ref{fig:ipr_fermi}, respectively. 
	The rescaling of disorder $W$ in the right panels is done in the same way, $W \to W/(m\ln m)$, where $m$ is now the coordination number of the spin quantum dot model,
	\begin{equation}
		m = \frac{n(n-1)}{2}\,.
		\label{eq:m-SQD}
	\end{equation}
	The Hilbert space dimension in this case is $N=2^n$. 
	The conclusion that can be made from these figures is essentially the same as for the case of a fermionic quantum dot. The critical disorder $W_c(n)$ rapidly increases with increasing $n$, as indicated both by level statistics and by IPR. Upon rescaling of $W$ by $m \ln m$, the upper border of the ergodic regime in the right panel of  Fig.~\ref{fig:spectral_spin} and the position of the maximum in the right panel of Fig.~\ref{fig:ipr_spin} become essentially $n$-independent for our larger values, $n=14$, 16. 
	This provides support to the RRG-like scaling \eqref{Wcn-QD-scaling} also for the spin quantum dot model.
	
\begin{figure}
		\includegraphics[width=.5\textwidth]{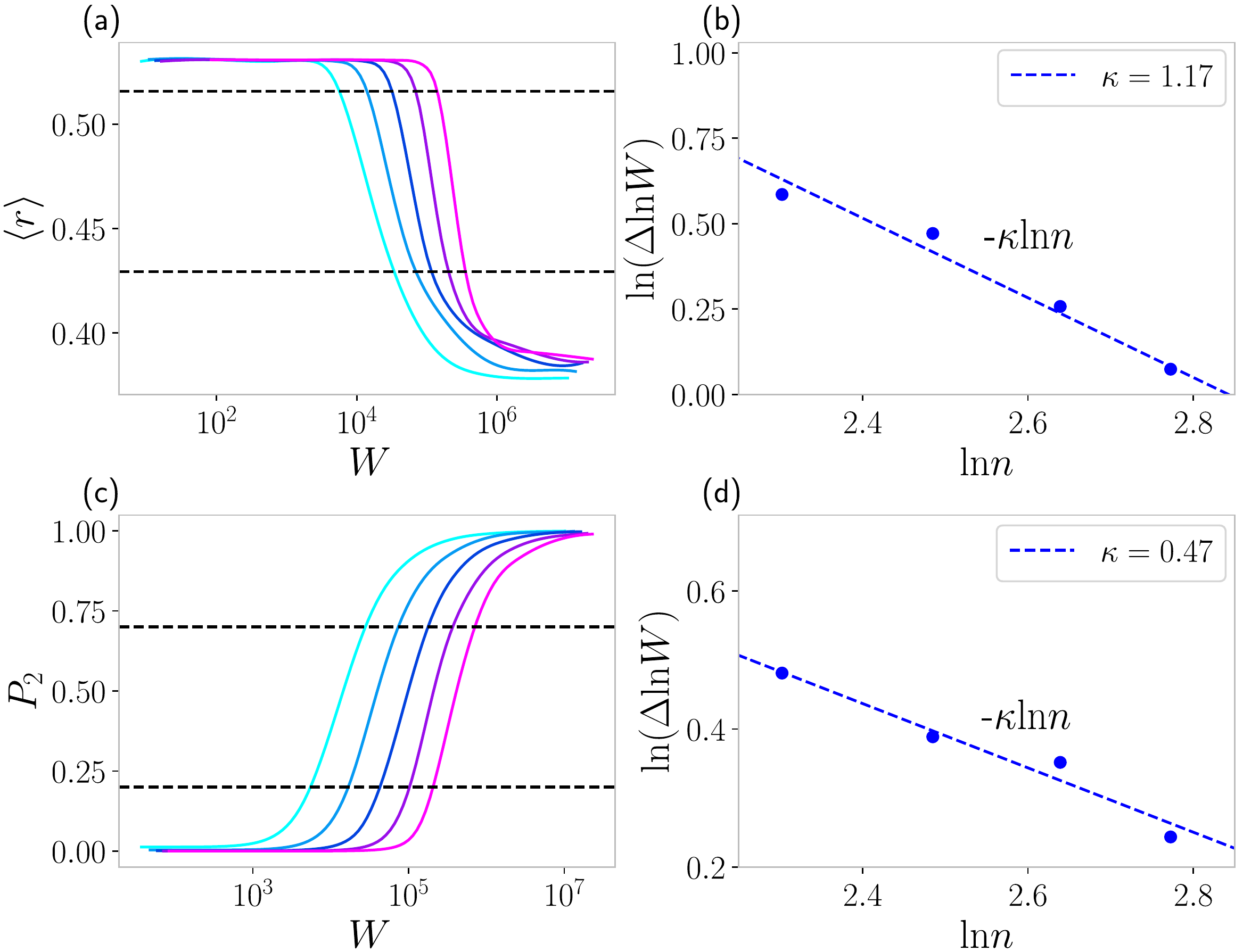}
		\caption{Width of the transition for fermionic quantum dot. (a) $r (W)$ for various system sizes with two horizontal lines showing a fixed interval $\Delta r$; (b) corresponding width $\Delta \ln W$ as a function of $n$ on log-log scale, with a power-law fit (straight line); (c) $P_2 (W)$  with two horizontal lines showing a fixed interval $\Delta P_2$; (d) corresponding width $\Delta \ln W$ as a function of $n$ on log-log scale, with a power-law fit.}
		\label{fig:fermi_transition_width}
	\end{figure}

\begin{figure}
		\includegraphics[width=.5\textwidth]{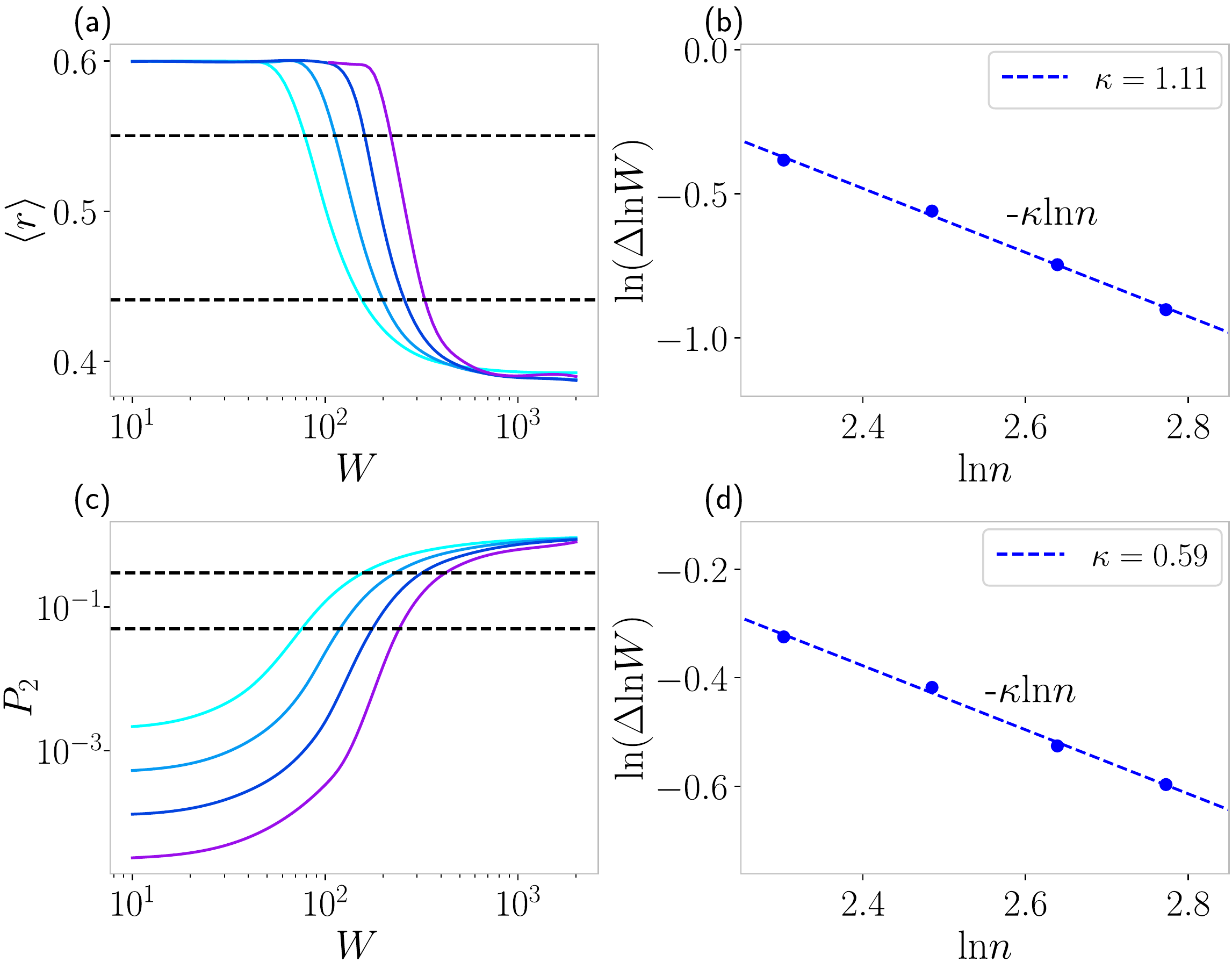}
		\caption{Same as Fig. \ref{fig:fermi_transition_width} but for spin quantum dot. }
		\label{fig:SQD_transition_width}
	\end{figure}

	We turn now to the question of sharpness of the localization transition; the corresponding criterion was discussed in the end of Sec.~\ref{sec:intro-Fock-MBL}. 
	For a finite $n$, we have a crossover from ergodicity to localization of a certain width $\Delta W_c(n)$. If the ratio $\Delta W_c(n) / W_c(n) \simeq \Delta \ln W_c(n)$ tends to zero at $n \to \infty$, we can speak about a sharp transition in the large-$n$ limit; otherwise, it remains a crossover also in this limit. 
	A quick look at Figs.~\ref{fig:spectral_fermi}  and  \ref{fig:spectral_spin} for the level statistics is sufficient to see that the crossover becomes sharper (on the logarithmic scale) with increasing $n$, suggesting  that 
	\begin{equation}
		\Delta \ln W_c(n) \to 0 \ \ \ {\rm at}  \ \ \ n\to \infty \,,
		\label{eq:sharp-transition}
	\end{equation} 
	i.e., a sharp transition.  This is made more quantitative in Figs.~\ref{fig:fermi_transition_width}a,b 
	and  \ref{fig:SQD_transition_width}a,b for fermionic quantum dot and spin quantum dot, respectively. In these figures, we analyze the width of the transition region from the Wigner-Dyson to Poisson statistics. The results are consistent with a power-law sharpening of the transition, 
	\begin{equation} 
		\Delta \ln W_c(n) \sim n^{-\kappa} \,,
		\label{eq:Delta-log-Wc}
	\end{equation} 
	with $\kappa \approx 1$. Notably, the behavior for the fermionic and spin quantum dots is very similar.
	In Figs.~\ref{fig:fermi_transition_width}c,d and \ref{fig:SQD_transition_width}c,d, a similar analysis is performed on the basis of IPR data. Specifically, we estimate the evolution of the width of the transition region between the ergodic $P_2 \ll 1$ and localized $P_2 \simeq 1$ regimes. We observe again the behavior of the type  \eqref{eq:Delta-log-Wc}, although with a smaller value of the exponent, $\kappa \simeq 0.5 - 0.6$.   
	
	As we showed in Sec.~\ref{sec:rrg-ED}, the exact diagonalization is not capable to probe the asymptotic critical behavior for the large-$m$ RRG model and can only reach the pre-critical regime, where the flowing exponent $\nu_{\rm del}$ is far from its asymptotic value. It is plausible that the situation is similar in the case of quantum dot models. In view of this,  the above values of the exponent $\kappa$ should be taken with a grain of salt, and one should not be too surprised by the difference between $\kappa$ obtained by using different observables. It is crucial that the data for both types of quantum dots and both observables consistently indicate that the crossover sharpens with increasing $n$, thus suggesting a true transition in the sense of  Eq.~\eqref{eq:sharp-transition}.

	\section{Spin quantum dot:  Quantitative comparison with RRG-like approximation}
	\label{sec:quantum-dot-rrg}
	
	In Sec.~\ref{sec:quantum-dots} we reported exact-diagonalization numerics for fermionic and spin quantum dots. We have found that the scaling of the critical disorder $W_c(n)$ supports the analytical predictions of Ref.~\cite{gornyi2017spectral}, Eqs.~\eqref{Wc-fermi-quantum-dot} and \eqref{Wc-spin-quantum-dot}. Moreover, the data are consistent with the exponent $\mu$ in these formulas being equal to the upper border, $\mu=1$, i.e., with the RRG scaling \eqref{Wcn-QD-scaling}. In the present section, we perform a detailed quantitative comparison of the quantum dot model with the corresponding RRG-like approximation. For this purpose, we choose the spin quantum dot since the corresponding coordination number \eqref{eq:m-SQD} is smaller  than for the fermionic quantum dot [Eq.~\eqref{eq:m-FQD}]. As a result, the exact-diagonalization numerics for the spin quantum dot model accesses substantially better the physics beyond the Golden rule (the counterpart of the pre-critical regime of the RRG model). This is clear from a comparison of Figs.~\ref{fig:ipr_fermi} and \ref{fig:ipr_spin}: whereas for the fermionic quantum dot the maximal numerically reached value of $\alpha(W)$ is 2.6, i.e., not so much above the Golden-rule value $\alpha=2$, for the spin quantum dot a much larger value $\alpha(W) = 8$ is reached. 
	
	We proceed now with calculating the IPR of the spin-quantum-dot model in the RRG-like approximation. In analogy with Sec.~\ref{sec:rrg}, we first perform estimates of the regimes
	(as in Sec.~\ref{sec:rrg-model}) and then carry out a careful calculation extending the analysis of Sec.~\ref{sec:rrg-analytical}. 
	
	\subsection{Overview of regimes}
	\label{sec:quantum-dot-rrg-regimes}
	
	The (approximate) correspondence between the spin-quantum-dot and the RRG models is established in the following way. 
	The nodes of the graph correspond to basis states in the many-body Hilbert space, i.e., spin configurations that are eigenstates of the operators $\sigma_i^z$ for all $i=1,2, \ldots, n$. These states are coupled by spin-flip terms that thus play a role of hopping terms, i.e., links in the random-graph model. Most important are terms that flip two spins, which yields the coordination number
	\begin{equation}
		m = \frac{n(n-1)}{2} \simeq \frac{n^2}{2} \,.
		\label{eq:SQD-m}
	\end{equation}
	(The Hamiltonian \eqref{H-spin-quantum-dot} contains also terms that flip only one spin. They give a much smaller contribution ($n$) to the coordination number and can safely be neglected for the estimate of the regimes. We will discuss such terms in more detail later and will show that they play only a minor role.) 
	All basis states coupled to a given one by two-spin-flip terms are in an interval of width $ \sim W$. The magnitude of spin-flip terms is $V \sim 1$ in our choice of the quantum-dot Hamiltonian. The (approximate) mapping to RRG now yields, according to Eq.~\eqref{eq:RRG-Gamma}, the broadening $\Gamma$:
	\begin{equation}
		\Gamma \sim \frac{m}{{\rm max} \{W, \Gamma\} }.
		\label{eq:SQD-Gamma}
	\end{equation}
	This implies a sequence of regimes that we are going to discuss, ordering them from weak to strong disorder, as in Sec.~\ref{sec:rrg-model}.
	
	If disorder $W$ is smaller than the level broadening $\Gamma$, the self-consistency in Eq.~\eqref{eq:SQD-Gamma} is important, $\Gamma \sim m / \Gamma$, i.e., 
	\begin{equation}
		\Gamma \sim \sqrt{m} \sim n \fullstop
		\label{eq:SQD-Gamma-self-cons}
	\end{equation}
	This regime can be further subdivided into two regimes:
	\begin{itemize}
		
		\item RMT regime,
		\begin{equation}
			W \ll n^{1/2} \,.
		\end{equation}
		In this regime, the broadening $\Gamma$ exceeds the characteristic energies $\sim n^{1/2}W$ of many body states. This implies that essentilally all many-body states are mixed and thus the IPR is
		\begin{equation}
			NP_2 \simeq 2 \,,
		\end{equation}
		as in the Gaussian unitary ensemble (GUE) of RMT. We note that GUE (rather than GOE) is applicable here since spin-flip terms in the Hamiltonian \eqref{H-spin-quantum-dot} break the time-reversal symmetry. 
		
		\item Self-consistent Golden-rule regime,
		\begin{equation}
			n^{1/2} \ll W \ll n \,,
		\end{equation}
		in which case $W < \Gamma < Wn^{1/2}$.  In this situation, the energy window $\Gamma$, within which the many-body states are strongly mixed, is smaller than the total many-body bandwidth $Wn^{1/2}$, implying that 
		\begin{equation}
			NP_2 \sim \frac{Wn^{1/2}}{\Gamma} \sim \frac{W}{n^{1/2}} \fullstop 
		\end{equation}
		
	\end{itemize}
	
	With further increase of disorder, Eq.~\eqref{eq:SQD-Gamma} takes the form of the conventional Golden-rule formula, 
	\begin{equation}
		\Gamma \sim m / W \sim n^2 /W \,.
	\end{equation} 
	
	\begin{itemize}
		
		\item Golden-rule regime,
		\begin{equation}
			n \ll  W \ll n^2 \,,
		\end{equation}
		with the IPR scaling
		\begin{equation}
			NP_2 \sim \frac{Wn^{1/2}}{\Gamma} \sim  \frac{W^2}{n^{3/2}} \fullstop
			\label{eq:SQD-IPR-GR}
		\end{equation} 
		The upper border of the Golden-rule regime is determined by the condition $\Gamma \sim \Delta$, where $\Delta \sim W / n^2$ is the level spacing of basis states (sites in the random-graph representation) directly connected to the given one. 
		
		\item  Pre-critical and critical regimes,
		\begin{equation}
			n^2 \ll W < W_c(n)\,.
		\end{equation}
		Within the mapping to RRG, the critical disorder is given by
		\begin{equation}
			W_c(n) = W_c^{\rm RRG}(n) \sim n^2 \ln n \fullstop
		\end{equation}
		
		\item Localized regime,
		\begin{equation}
			W  > W_c \,, 
		\end{equation}
		with $P_2 \approx 1$. 
		
	\end{itemize}
	
	We turn now to an accurate calculation (within the RRG approximation) of $P_2$ in these regimes by extending the analysis of Sec.~\ref{sec:rrg-analytical}.
	
	\subsection{Golden-rule regimes} 
	\label{Section: SQD Golden Rule}
	
	We proceed in analogy with Sec.~\ref{sec:RRG-RMT-GR}. Several modifications are, however, needed to take into account that distributions of energies and of transition matrix elements in the RRG-like approximation to the quantum dot model differ from those in the RRG model considered in Sec.~\ref{sec:rrg}.
	
	Consider a certain basis many-body state (i.e., a site in the random-graph representation) with an energy $\varepsilon^{(0)}$. We focus on states near the band center and can set $\varepsilon^{(0)} = 0$.  
	Consider now all sites (basis states) that are connected to this chosen state by a direct link (i.e., two-spin-flip process). The corresponding energies are 
	\begin{equation}
		\varepsilon^{(1)} = \varepsilon^{(0)} \pm \varepsilon_i\pm \varepsilon_j \,,
		\label{eq:SQD-energies}
	\end{equation}
	where $i$ and $j$ are indices of flipped spins. The energies $\varepsilon_i$ and $\varepsilon_j $ are taken independently from  the box distribution on $[-W,W]$ in our model. 
	Calculation the distribution of $\pm \varepsilon_i\pm \varepsilon_j$, we get a ``triangular'' distribution $\gamma_1(\varepsilon)$ for energies of states coupled to a given one, 
	\begin{equation}
		\gamma_1(\varepsilon) = \begin{cases}
			\displaystyle \frac{\varepsilon + 2W}{4W^2} \comma \;\;\;\; -2W < \varepsilon \leq 0 \comma \\[0.4cm]
			\displaystyle \frac{2W - \varepsilon}{4W^2} \comma \;\;\;\;\;\;\;\; 0 \leq \varepsilon < 2W \comma
		\end{cases} 
		\label{eq:Triangular_Distribution}
	\end{equation}
	and $\gamma_1(\varepsilon)=0$ everywhere else. The distribution $\gamma_1(\varepsilon)$ replaces the distribution $\gamma(\varepsilon)$ (box distribution on $[-W/2, W/2]$) that we assumed for site energies in the RRG model in Sec.~\ref{sec:RRG-RMT-GR}.
	
	The second modification is related to the distribution of transition matrix elements. In the RRG model of Sec.~\ref{sec:RRG-RMT-GR} they were simply constant (equal to $V$) and we subsequently set $V=1$.  Let us calculate the distribution of couplings $M$ between directly coupled states in the spin-quantum-dot model  \eqref{H-spin-quantum-dot}. 
	Consider two basis states that differ from each other by flips of spins on position $i$ and $j$, with $M$ given by
	\begin{equation}
		M = \langle...\underset{i}{\downarrow}...\underset{j}{\downarrow}...| \hat H |...\underset{i}{\uparrow}...\underset{j}{\uparrow}...\rangle \,.
		\label{eq:SQD-M}
	\end{equation} 
	We recall that  $V_{ij}^{\alpha\beta}$ are independent real Gaussian variables with zero mean and unit variance.  It is then easy to see that $M$ is a complex Gaussian variable with zero mean and
	\begin{equation}
		\langle |M|^2\rangle = \frac{1}{16} \cdot 2 \cdot 4 \cdot 4 = 2 \fullstop
		\label{eq:SQD-M-distribution}
	\end{equation}
	Here the factor 1/16 originates from the factors 1/2 relating spin operators to Pauli matrices, the factor 2 from accounting for $(i,j)$ and $(j,i)$ contributions in Eq.~\eqref{H-spin-quantum-dot}, one factor 4 from summing over $\alpha = x,y$ and $\beta = x, y$, and another factor 4 from the Hermitian conjugated contribution in Eq.~\eqref{H-spin-quantum-dot} that doubles all relevant  transition amplitudes $M$. Clearly, the constant $V^2$ is replaced by $\langle |M|^2 \rangle$ in the case of fluctuating matrix elements $M$ in the Golden-rule (and self-consistent Golden rule) regime, i.e., in Eq.\eqref{eq:generalg0sce}. 
	
	It is worth noting that the above modifications in distributions of diagonal energies and matrix elements do not affect the parametric estimates of the regimes performed in Sec.~\ref{sec:quantum-dot-rrg-regimes}.

	The self-consistency equation \eqref{eq:generalg0sce} thus becomes
	\begin{equation}
		\begin{split}
			g_0 &= \frac{\langle |M|^2 \rangle}{(4W)^2} \left[\int_{-2W}^{0} \dd \varepsilon \frac{\varepsilon + 2W}{mg_0 + \frac{i}{2}\varepsilon} + \int_{0}^{2W} \dd \varepsilon \frac{2W - \varepsilon}{mg_0 + \frac{i}{2}\varepsilon}\right].
		\end{split}\label{eq:g0ThreeIntegrals}
	\end{equation}
	Evaluating the integrals, we bring this equation to the form
	\begin{equation}
		g_0 = \frac{1}{W} \arctan \frac{W}{mg_0}  + \frac{mg_0}{2W^2} \ln \left(1 + \frac{W^2}{(mg_0)^2} \right)\fullstop \label{eq:SQD_g0_sce}
	\end{equation}
	This is a counterpart of Eq.~\eqref{eq:g0_arctan} for the RRG model in Sec.~\ref{sec:rrg-analytical}.
	
	Equation \eqref{eq:RRG-rho1-result} for the local density of states (at zero energy, $E=0$) $\rho_1 (0,\varepsilon)$ remains unchanged:
	\begin{equation}
		\rho_1 (0,\varepsilon) = \frac{1}{\pi} \frac{2mg_0}{(2mg_0)^2 + \varepsilon^2} \fullstop 
		\label{eq:SQD-rho1}
	\end{equation}
	To calculate the global density of states $\rho(0)$ we use a counterpart of Eq.~\eqref{eq:RRG-rho},
	\begin{equation}
		\rho(E) = \int \d \varepsilon \gamma_{\rm tot}(\varepsilon) \rho_1(E,\varepsilon) \,,
		\label{eq:SQD-rho}
	\end{equation}
	where $\gamma_{\rm tot}(\varepsilon)$ is now the distribution of energies $\varepsilon$ of all many-body basis states. Such an energy is a sum of random energies of individual spins, with random signs:
	\begin{equation}
		\varepsilon  = \frac{1}{2} \sum_i s_i \varepsilon_i  \,, \qquad s_i = \pm 1 \,.
	\end{equation}
	For large $n$, the central limit theorem applies, yielding a Gaussian probability distribution
	\begin{equation}
		\gamma_{\rm tot} (\varepsilon) = \frac{1}{\sqrt{2\pi}\sigma_{\rm tot}} \exp \left(- \frac{\varepsilon^2}{2 \sigma_{\rm tot}^2}\right), \ \ {\rm with}\ \  \sigma_{\rm tot}^2 = \frac{nW^2}{12}\,.
		\label{eq:SQD-gamma2}
	\end{equation}
	Substituting Eqs.~\eqref{eq:SQD-rho1} and \eqref{eq:SQD-gamma2} into Eq.~\eqref{eq:SQD-rho}, we get the following result for the global density of states at the band center ($E=0$):
	\begin{equation}
		\rho(0) = \int_{-\infty}^{\infty} \dd \varepsilon \gamma_{\rm tot}(\varepsilon) \rho_1(0,\varepsilon)  = \frac{1}{\sqrt{2\pi}\sigma_{\rm tot}}  e^{y^2} \mathrm{erfc} \left( y \right)\comma
	\end{equation}
	where $\mathrm{erfc}(y)$ denotes the complementary error function and
	\begin{equation}
		y = \frac{\sqrt{2} m g_0}{\sigma_{\rm tot}} = \sqrt{6} \: \frac{n^{1/2} (n-1) g_0}{W} \fullstop
		\label{eq:SQD-y}
	\end{equation}
	Finally, the IPR is given, in full analogy with Eq.~\eqref{eq:NP2_Integral}, by
	\begin{equation}
		\begin{split}
			NP_2 &= 2 \int_{-\infty}^{\infty} \dd \varepsilon \gamma_{\rm tot}(\varepsilon) \left[\frac{\rho_1(0,\varepsilon)}{\rho(0)}\right]^2 \\
			&= \frac{2}{\pi}e^{-y^2} \mathrm{erfc}^{-1}\left(y \right) \left[ e^{-y^2}\mathrm{erfc}^{-1}\left(y \right) - \frac{\sqrt{\pi}}{ y}   \left( y^2 - \frac{1}{2}\right)\right]  \fullstop \label{eq:SQD_GoldenRule_NP2}
		\end{split}
	\end{equation}
	
	Solving Eq.~\eqref{eq:SQD_g0_sce} numerically for a given disorder $W$, we obtain $g_0$. Using then Eq.~\eqref{eq:SQD-y}  for $y$ and plugging the result in Eq.~\eqref{eq:SQD_GoldenRule_NP2}, we find IPR $P_2$ as a function of $W$ in the whole Golden-rule range $W \lesssim n^2$, including the proper Golden-rule, self-consistent Golden-rule, and RMT regimes (and crossovers between them). We will show the corresponding curves in Sec.~\ref{sec:SQD-ED-RRG-comparison}, where we will confront them with the results of exact diagonalization. 
	
	We expand now on a comment below Eq.~\eqref{eq:SQD-m} concerning the terms with a single spin flip. These are terms of the type $\hat S_i^z \hat S_j^\alpha$ with $\alpha = x,y$. Considering only such terms, we would have the coordination number $\tilde{m} = n$, the zero-energy density $\tilde{\gamma}_1(0) = 1/2W$, and the averaged squared matrix element $\langle |\tilde{M}|^2 \rangle = n$. (We use a tilde to label  quantities associated with single-spin-flip processes.) The Golden-rule rate $\Gamma$ is controlled by the product $\gamma_1(0) m \langle |M|^2 \rangle$, which is equal to $n^2 / 2W$ for the two-spin-flip processes.  Interestingly, we obtain exactly the same value for single-spin-flip processes,  
	$\tilde{\gamma}_1(0)  \tilde{m} \langle |\tilde{M}|^2 \rangle = n^2/2W$. This implies that $\Gamma$ gets an additional factor 2 in the Golden-rule regime and factor $\sqrt{2}$ in the self-consistent Golden-rule regime, so that $NP_2$ acquires the factors $1/2$ and $1/\sqrt{2}$, respectively. Thus, including single-spin-flip terms would only lead to relatively minor modifications. Furthermore, the upper border of applicability of the Golden-rule formula for single-spin-flip processes is parametrically lower than for two-spin-flip processes. Indeed, comparing the characteristic matrix element $\tilde{M} \sim n^{1/2}$ with the level spacing $\tilde{\Delta} \sim W / \tilde{m} = W / n$, we get the upper border $W \sim n^{3/2}$, much lower than that  for the two-spin-flip processes ($W \sim n^2$). Thus, single-spin-flip processes do not play any role for $W \gtrsim n^{3/2}$, i.e., in the upper part of the Golden-rule regime and in the pre-critical and critical regimes.

	\subsection{Critical behavior in RRG-like approximation} 
	\label{Section: SQD Critical}  
	
	When the disorder $W$ increases above $m$, the RRG model enters the pre-critical regime where fluctuations get large and the IPR grows exponentially. With further increase of disorder, the system enters the critical regime, where $NP_2$ tends to diverge at $W \to W_c$.
	The corresponding calculations for the RRG model in the disorder range $m \lesssim W < W_c$ were  performed in Sec.~\ref{sec:RRG-critical}.   
	We derived there that the IPR is given by
	\begin{equation}
		P_2 \sim \frac{N_\xi}{N} \,, \qquad N_{\xi} \sim \frac{1}{m} \exp\left( \frac{\pi}{\sigma} \right) \comma 
		\label{eq:criticalNxi}
	\end{equation}
	where $\sigma$ is the solution of Eqs.~\eqref{eq:Sinc}, \eqref{eq:RRG-crit-x}.
	Here (in Sec.~\ref{Section: SQD Critical})   we make an assumption (without a priori justification) that the same mechanism is operative in the pre-critical and critical regimes of the quantum-dot model. In this way, we extend the formulas of Sec.~\ref{sec:RRG-critical} to the quantum-dot model. In Sec.~\ref{sec:SQD-ED-RRG-comparison}, we will compare the results with exact diagonalization, which will allow us to judge on the accuracy of the assumption.  
	
	In analogy with Sec.~\ref{Section: SQD Golden Rule}, we have to take into account that distributions of energies and transition matrix elements in the effective random-graph representation of the quantum-dot model differ from those in the RRG model considered in Sec.~\ref{sec:RRG-critical}. Most importantly, we need to find the correct numerical prefactor in the formula for $W_c$. In the standard RRG model, it is determined by the equation [this is Eq.~\eqref{eq:Critical_Disorder} with restored hopping $V$]:
	\begin{equation}
		1 = 4m \frac{V}{W} \ln \frac{W}{2V} \,,
		\label{eq:SQD-RRG-Wc}
	\end{equation}
	originally derived in Ref.~\cite{abou1973selfconsistent}.  Equation \eqref{eq:SQD-RRG-Wc} holds for a box distribution of energies on $[-W/2,W/2]$ and a constant hopping equal to $V$. 
	We have instead the distribution $\gamma_1(\varepsilon)$ of energies given by Eq.~\eqref{eq:Triangular_Distribution} (and parametrized by the disorder $W$) and a Gaussian-distributed complex hopping $M$ with zero mean and variance 2,   Eq.~\eqref{eq:SQD-M-distribution}. We focus on determining the correct numerical prefactor in the correspondingly modified Eq.~\eqref{eq:SQD-RRG-Wc}. (The numerical factor of order unity in the argument of logarithm is clearly mush less important; we do not attempt to determine it accurately for the case of generic distributions but rather replace it by  unity.)
	Inspecting the derivation of the criterion \eqref{eq:SQD-RRG-Wc} in Ref.~\cite{abou1973selfconsistent} and extending it to the case of generic distribution $\gamma_1(\varepsilon)$ and $P(M)$, we get a generalized criterion for the transition point $W_c$:  
	\begin{equation}
		1 = 4m \langle |M|\rangle \gamma_1(0) \ln \frac{1}{\gamma_1(0)\langle |M|\rangle} \fullstop 
		\label{eq:Generalized_AbouChacra}
	\end{equation} 
	The fact that $2/W$ in Eq.~\eqref{eq:SQD-RRG-Wc} is replaced by $\gamma_1(\varepsilon = 0) $ has a simple explanation. The logarithmic factor in Eq.~\eqref{eq:SQD-RRG-Wc}  originates from an integral $\int d\varepsilon / \varepsilon$, with the lower and upper limits being $V$ and $W/2$, respectively. For a generic distribution of energies, this becomes $\int_V^\infty (d\varepsilon / \varepsilon) \gamma_1(\varepsilon )$.  Since the large logarithm comes from energies $\varepsilon \ll W$, one can replace here, to the leading approximation, $\gamma_1(\varepsilon) \to \gamma_1(0) \equiv \langle \delta(\varepsilon) \rangle$. The same conclusion can be made by an inspection of the derivation of the condition for $W_c$ by inspection of the convergence of the perturbative expansion as performed in Refs.~\cite{altshuler1997quasiparticle} and \cite{gornyi2016many} (see, in particular, the Supplemental Material of the latter paper). To take into account also fluctuations of the transition matrix elements $M$ (at variance with the standard RRG model, where they are equal to a constant $V$), we note that the energies $\varepsilon$ enter in the perturbative expansion always in a form of a ratio $\varepsilon /M$. This implies a replacement $V \langle \delta(\varepsilon) \rangle \to \langle \delta(\varepsilon / M) \rangle = \langle \delta(\varepsilon)  \langle | M| \rangle = \gamma_1(0)  \langle | M| \rangle$, yielding Eq.~\eqref{eq:Generalized_AbouChacra}.
	
	For the triangular distribution \eqref{eq:Triangular_Distribution}, we have 
	\begin{equation} 
		\gamma_1(0) = \frac{1}{2W} \,.
		\label{eq:SQD-gamma10}
	\end{equation}  
	Further, the transition amplitude $M$, Eq.~\eqref{eq:SQD-M}, has 
	real ($M_1$) and imaginary ($M_2$) parts with equal Gaussian distributions: $M=M_1 + iM_2$ with
	\begin{equation}
		P(M_1,M_2) = \frac{1}{2\pi} e^{-\frac{1}{2}(M_1^2 + M_2^2)} \,,
	\end{equation} 
	so that
	\begin{equation}
		\langle |M|\rangle = \int \frac{\dd M_1 \dd M_2}{2\pi} |M| e^{-|M|^2/2}
		= \sqrt{\frac{\pi}{2}} \fullstop
		\label{eq:SQD-abs-M}
	\end{equation}
	Using Eqs.~\eqref{eq:SQD-gamma10} and \eqref{eq:SQD-abs-M}, we bring Eq.~\eqref{eq:Generalized_AbouChacra} for $W_c$ to an explicit form
	\begin{equation}
		W_c = \sqrt{\frac{\pi}{2}} n(n-1) \ln \left(\sqrt{\frac{8}{\pi}}W_c\right)\fullstop
		\label{eq:Generalized Wc}
	\end{equation}
	Asymptotically (at $n\to \infty$), the solution of Eq.~\eqref{eq:Generalized Wc} is
	\begin{equation}
		W_c \simeq \sqrt{2\pi} \: n^2 \ln n \approx 2.51 n^2 \ln n \fullstop \label{eq:Wc_asymptotic}
	\end{equation}
	Let us emphasize that this result corresponds to an RRG-like model inspired by spin quantum dot [in particular, with coordination number \eqref{eq:SQD-m} determined by $n$] in the ``thermodynamic limit'' $N \to \infty$. Of course, in a true quantum-dot model, the Hilbert-space volume $N$ is exponentially large but still finite and, furthermore, is also determined by $n$, namely, $N= 2^n$.  We will return to the discussion of the effect of finite $N$ below. 
	
	To get the critical behavior of the correlation volume $N_\xi$ and of the IPR $P_2$, we use Eqs.~\eqref{eq:Sinc}, \eqref{eq:RRG-crit-x} derived for the standard RRG model, with the modifications just discussed:
	\begin{equation}
		\begin{split}
			\frac{\sin x}{x} =& \frac{f(W)}{f(W_c)} \comma \;\;\; x= 2\sigma \ln \frac{1}{\gamma_1(0)\langle |M|\rangle}\comma \;\;\; \\
			&f(W) = \frac{W}{\ln \frac{1}{\gamma_1(0)\langle |M|\rangle}} \fullstop
		\end{split} 
		\label{eq:Generalized Sinc}
	\end{equation}
	Substitution of the solution $\sigma$ in Eq.~\eqref{eq:criticalNxi} yields $N_\xi$ and $P_2$. 
	
	\begin{figure}
		\includegraphics[width=0.5\textwidth]{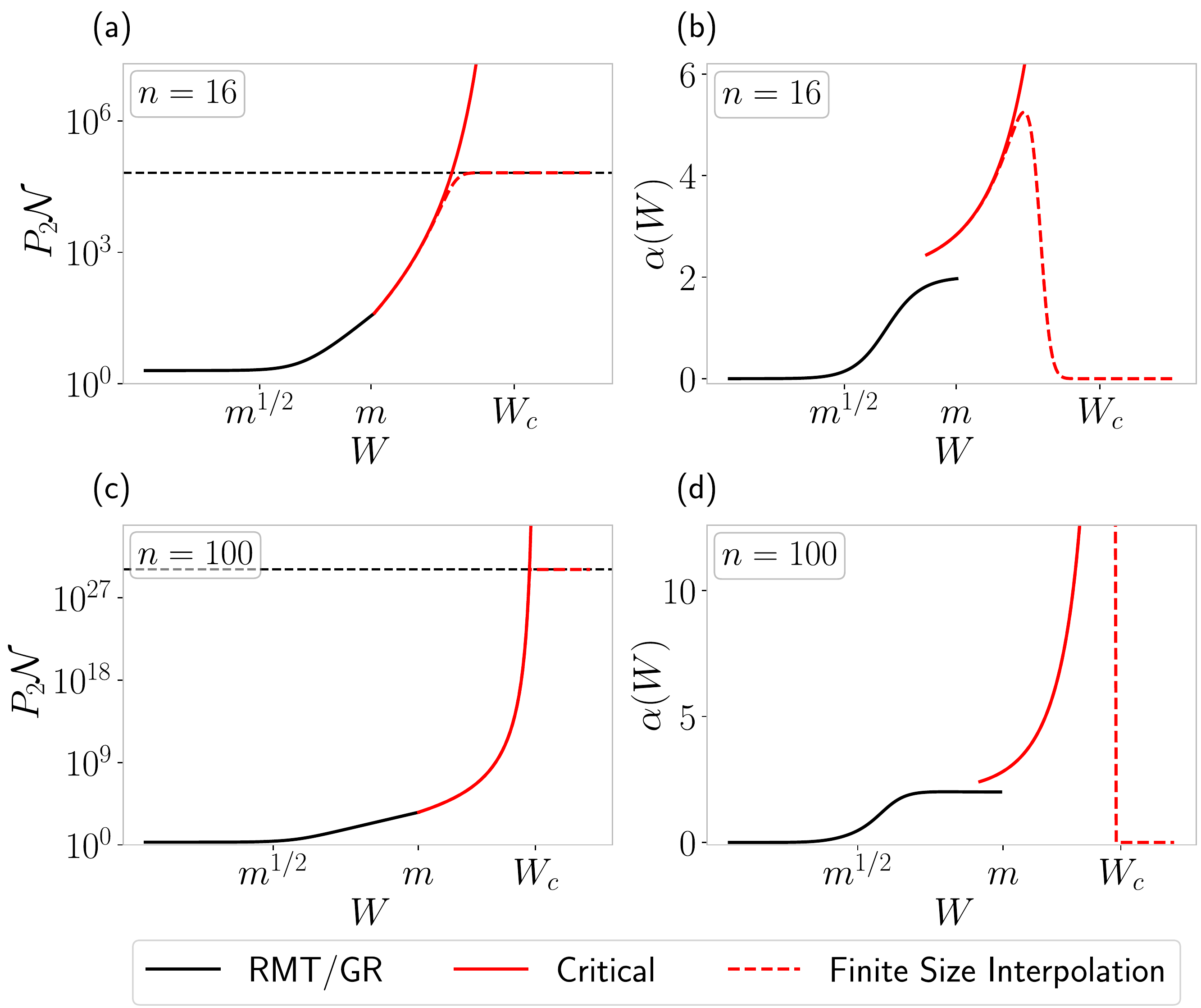}
		\caption{RRG-like approximation for IPR  $P_2$ in spin quantum dot. Black lines correspond to Golden-rule regimes (including RMT and self-consistent Golden rule), 
			Sec.~\ref{Section: SQD Golden Rule}, whereas red lines correspond to pre-critical and critical regimes, Sec.~\ref{Section: SQD Critical}.
			(a) $NP_2$ as function of disorder $W$ for a quantum dot with $n=16$ spins; (b) same data plotted as logarithmic derivative $\alpha(W)$, Eq.~\eqref{eq:RRG-alphaW}; (c), (d) same as (a), (b) but for $n= 100$.  Red dashed lines are finite size interpolation \eqref{eq:SQD-finite-size-interp} taking into account the finite size $N = 2^n$ of the Hilbert space.   
		}
		\label{fig:SQD_RRGlike_approximation}
	\end{figure}
	
	In Fig.~\ref{fig:SQD_RRGlike_approximation}, the dependence of $NP_2$ on disorder $W$ calculated in the RRG approximation is shown for $n=16$ [panel (a)] and $n=100$ [panel (c)].   Panel (b) and (d) display the corresponding logarithmic derivative $\alpha(W)$, Eq.~\eqref{eq:RRG-alphaW}. Curves in this figure are analogous to analytical curves
	in Figs.~\ref{fig:RRG_NP2_largem} and  \ref{fig:RRG_NP2_dlogdlog_largem}.  Black lines are the Golden-rule result given by Eq.~\eqref{eq:SQD_GoldenRule_NP2} 
	in combination with Eqs.~\eqref{eq:SQD_g0_sce} and \eqref{eq:SQD-y}. Red lines represent the pre-critical and critical regime as given by Eq.~\eqref{eq:criticalNxi} with $\sigma$  calculated from Eq.~\eqref{eq:Generalized Sinc}.  As in Sec.~\ref{sec:RRG-critical}, the results in the pre-critical and critical regime are applicable as long as $NP_2$ is smaller than its maximal value $N$.   Since $N=2^n$ grows exponentially with $n$, while $m \sim n^2$ only as a power-law, this restriction does not prevent the system to go deeply into the critical regime for large $n$, as is seen in the panel (c) of Fig.~\ref{fig:SQD_RRGlike_approximation}, where the results for $n=100$ are shown. In this case, the upper cutoff for $NP_2$ is as big as $2^{100} \approx 10^{30}$. On the other hand, for moderately large $n$ accessible for exact diagonalization, such as $n=16$,  the upper cutoff intervenes much earlier. To illustrate this, we show by red dashed lines an interpolation
	\begin{equation}
		NP_2 = \left( \frac{1}{2^n} + \frac{1}{N_{\xi}} \right)^{-1} \,,
		\label{eq:SQD-finite-size-interp}
	\end{equation}
	where $N_\xi$ is calculated in the $N\to \infty$ limit as above.  It is worth emphasizing that this interpolation is somewhat simplistic; if applied to a standard RRG model, it would yield a sharper crossover to the saturated value than the exact diagonalization (Fig.~\ref{fig:RRG_NP2_largem}). A rigorous derivation of the crossover around the ``finite-size transition point'' $W_*(N)$ in the RRG model remains a problem for future research.

	\subsection{Comparison of exact diagonalization and predictions of RRG-like approximations }
	\label{sec:SQD-ED-RRG-comparison}
	
\begin{figure}
		\includegraphics[width=0.5\textwidth]{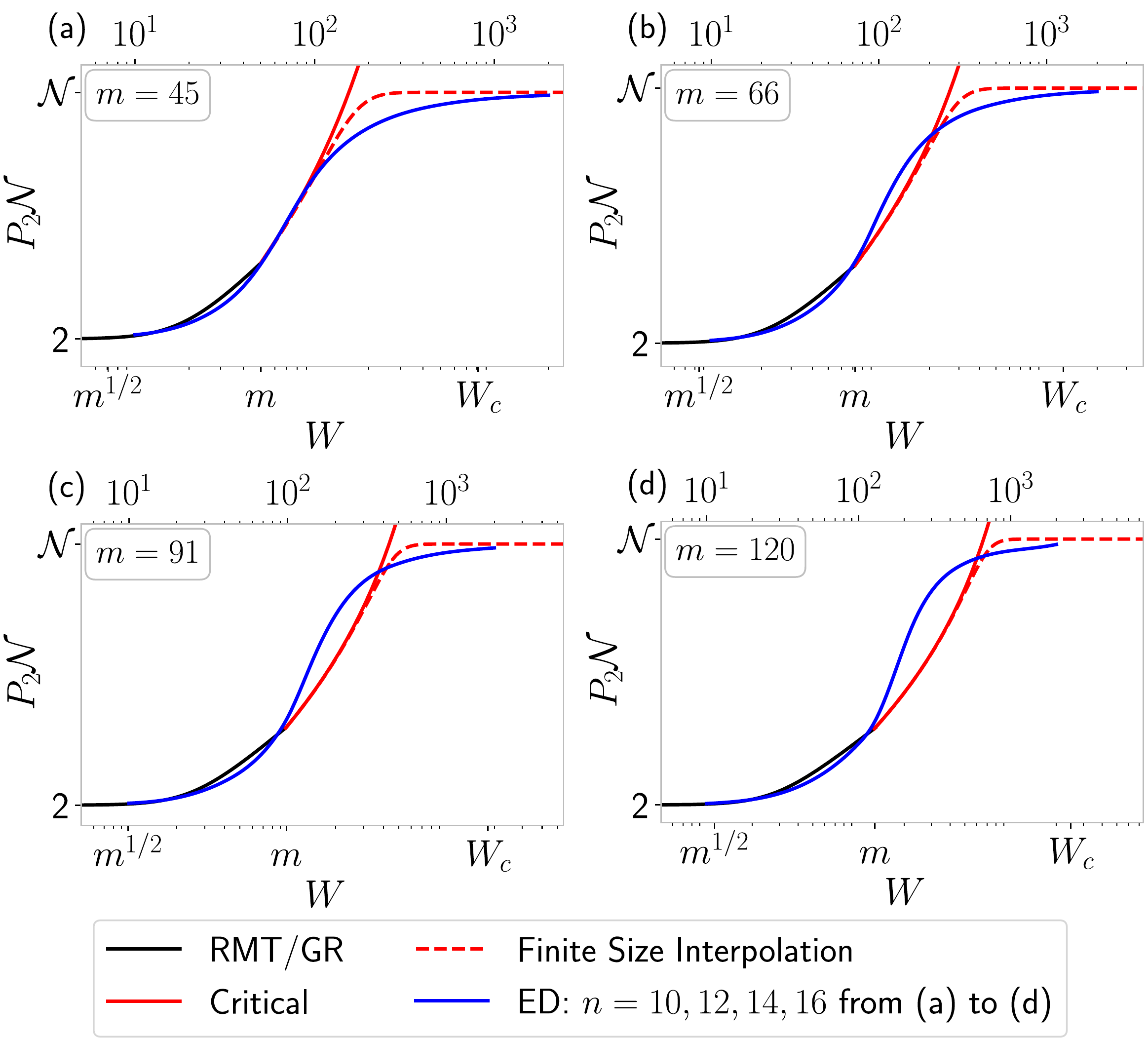}
		\caption{Spin quantum dot: comparison between exact-diagonalization numerics for $NP_2$ (Sec.~\eqref{sec:quantum-dots}, Fig.~\ref{fig:ipr_spin}) and RRG-like approximation  
			Sections~\ref{Section: SQD Golden Rule} and \ref{Section: SQD Critical}, Fig.~\ref{fig:SQD_RRGlike_approximation}).
			(a) $n=10$, (b) $n=12$, (c) $n=14$, (d) $n=16$.  The coordination number $m = n (n-1)/2$ is indicated on each panel.
		}
		\label{fig:SQD_RRGlike_ED_comparison}
	\end{figure}

	We are now in a position to compare the exact-diagonalization results for the IPR in spin-quantum dot model (Sec.~\eqref{sec:quantum-dots}) with the predictions of the RRG approximation, Sections~\ref{Section: SQD Golden Rule} and \ref{Section: SQD Critical}. This comparison of the corresponding curves $NP_2(W)$ is presented in  Fig.~\ref{fig:SQD_RRGlike_ED_comparison}. 
	
	First of all, we see that there is a very good agreement between the numerical (exact-diagonalization) and analytical (RRG-like approximation) results for $NP_2$ in the Golden-rule range, $W\le m$. There is only a small (within $\sim 20\%$) systematic downward deviation of numerically obtained $NP_2$ from the analytical curve in a part of this regime. This implies that the Golden-rule width $\Gamma$ is somewhat larger (again, within $\sim 20\%$) than its analytical value. We attribute this to the effect of single-spin-flip terms, which we have neglected in the analytical calculation; see a discussion in the end of Sec.~\ref{Section: SQD Golden Rule}.
	
\begin{figure}
		\includegraphics[width=0.5\textwidth]{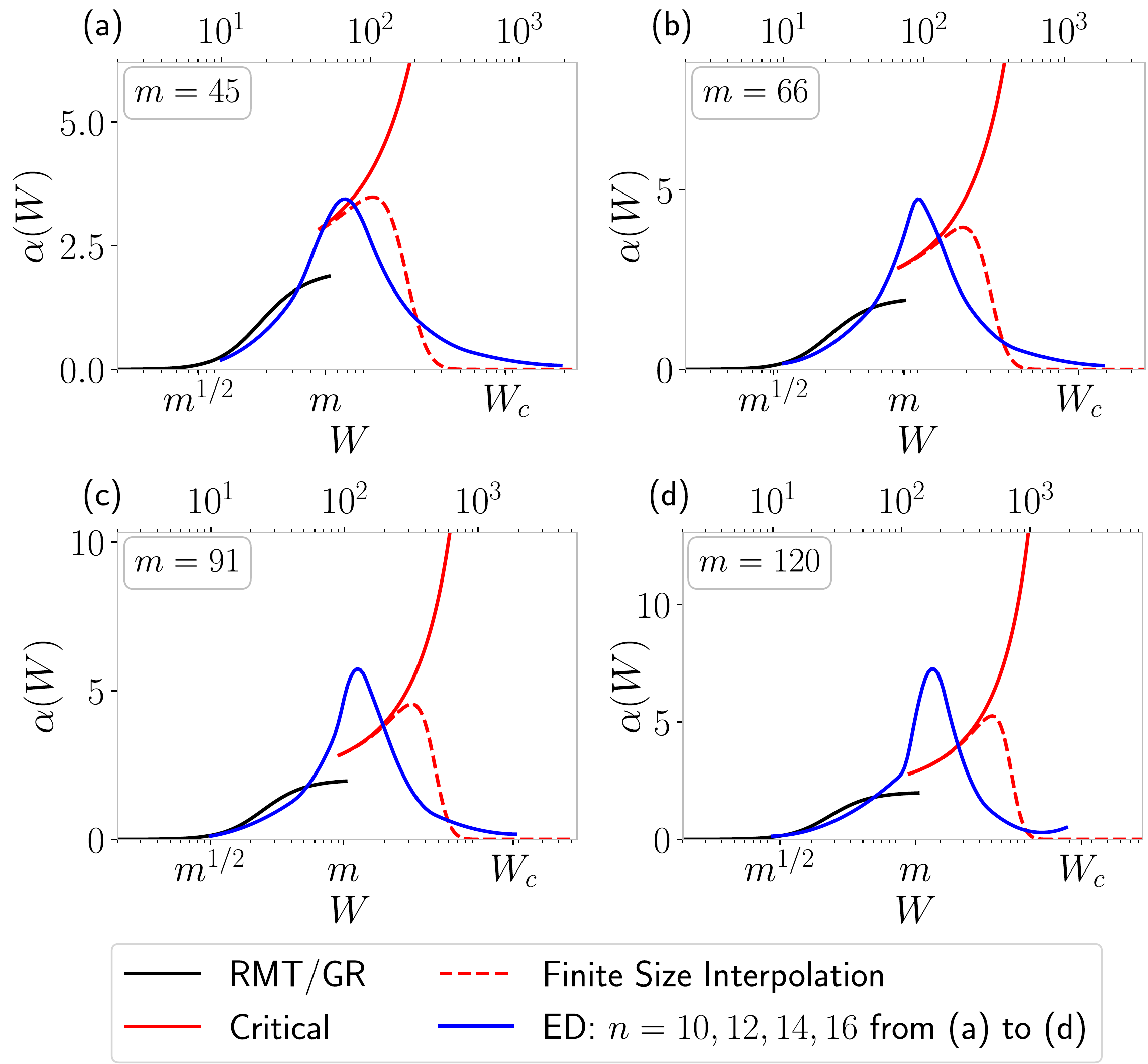}
		\caption{Same as Fig. \ref{fig:SQD_RRGlike_ED_comparison} in the form of a logarithmic derivative, $\alpha(W)$, Eq.~\eqref{eq:RRG-alphaW}.
		}
		\label{fig:SQD_RRGlike_ED_comparison_dlogdlog}
	\end{figure}

	With further increase of $W$, i.e., in the precritical and critical regimes, $m < W < W_c$, a substantial difference emerges between the numerical data and the RRG-like approximation.  Specifically, the numerical $NP_2$ increases considerably faster towards the  maximum (localized) value $NP_2 = N$ than predicted by the analytic approximation. 
	Importantly, this deviation becomes increasingly more pronounced with increasing $n$, suggesting that a large difference remains in the $n \to \infty$ limit.
	The same data are presented in the form of the logarithmic derivative $\alpha(W)$ in Fig.~\ref{fig:SQD_RRGlike_ED_comparison_dlogdlog}; this representation additionally emphasizes deviations.
	
	Our results thus provide an indication that, while the RRG-like approximation describes very well the physics of the quantum-dot model in the Golden-rule range, $W < m$, it becomes much less accurate in the critical domain, $m < W < W_c$.  Since the deviation of numerics in the range $W > m$ in the direction of stronger localization, our results indicate that the critical disorder of the quantum-dot model satisfies $m < W_c < W_c^{\rm RRG}$.  These conclusions are in full agreement with the results of Ref.~\cite{gornyi2017spectral}
	implying that [see Eq.~\eqref{Wc-spin-quantum-dot}]
	\be
	\label{Wc-spin-quantum-dot-repeated}
	W_c \sim m \ln^\mu m \,, \qquad \mu \le 1 \,.
	\ee
	Our numerical data suggest that $0 \le \mu \le 1$.  If the exponent $\mu$ has the same value $\mu = 1$ as in the RRG model
	(the scenario suggested by the right panel of  Fig.~\ref{fig:ipr_spin}; see a discussion  in Sec.~\ref{sec:quantum-dots}),
	a prefactor in Eq.~\eqref{Wc-spin-quantum-dot-repeated} for the quantum dot should be considerably smaller than for the associated RRG-like model, so that $W_c < W_c^{\rm RRG}$. 
	
	Substantial difference between the quantum-dot model and RRG approximation in the critical region can be attributed to the effect of correlations between transition matrix elements that are discarded in the RRG approximation \cite{gornyi2017spectral}, see a discussion in Sec.~\ref{sec:intro-Fock-MBL}. Whether the critical behavior in the quantum-dot model is the same as in the RRG model (and, if not, how essential are differences) remains an important open question. 
	
	A part of our analysis in Sec.~\ref{sec:quantum-dot-rrg} bears similarity with the investigation of a fermionic (Majorana) quantum dot in Ref.~\cite{monteiro2020minimal}, where the RRG-like approximation was derived as an effective-medium approximation within the supersymmetric field-theory formalism. A very good agreement between the results of this approximation and exact-diagonalization data in the Golden-rule regime  (``regime III'' in notations of Ref.~\cite{monteiro2020minimal}) was found there, in full consistency with our findings. The pre-critical and critical regimes were not analyzed in Ref.~\cite{monteiro2020minimal}. (In fact, these regimes are essentially not accessible by exact diagonalization for a fermionic quantum dot, as we discussed in the beginning of Sec.~\ref{sec:quantum-dot-rrg}.) It is also worth mentioning that the Golden-rule regime was classified as ``non-ergodic'' in Ref.~\cite{monteiro2020minimal}. This is a misleading terminology: there is a full hybridization of states within the energy window $\Gamma$ (Golden-rule width) containing a macroscopically large number of states (resulting in the Wigner-Dyson level statistics and in the IPR \eqref{eq:SQD-IPR-GR}), which is in full correspondence to the standard notion of ergodicity in statistical physics. This terminological issue was corrected by the authors of Ref.~\cite{monteiro2020minimal} in a subsequent publication, Ref.~\cite{monteiro2020quantum}.

	\section{Summary and outlook}
	\label{sec:summary}
	
	In this paper, we have carried out a detailed analytical and numerical study of the transition from ergodicity to localization in two closely related classes of models. First, we explored RRG models with large connectivity $m$. Second, we studied many-body quantum-dot models, in the fermionic and spin versions. (The fermionic version is essentially equivalent to a model termed $\text{SYK}_2 + \text{SYK}_4$ in recent literature.)  Arguably, the quantum-dot models are the simplest models of the (Fock-space) MBL transition (by analogy with all-to-all interacting spin models, which are the simplest models of second-order phase transition).
Understanding the evolution from ergodicity to MBL in these models is thus of fundamental importance. The large-$m$ RRG models serve as toy models for MBL transitions in the Fock  space of many-body quantum dots and are amenable to a controllable analytical investigation. We have studied what parametric regimes occur on a way from ergodicity to localization in large-$m$ RRG and in quantum-dot models and which of these regimes can be observed in systems of realistic size (accessible to numerical or experimental studies). For the quantum-dot models, we have used the numerical data to study the scaling of critical disorder $W_c$ with the system size and to analyze whether a sharp MBL transition emerges in the thermodynamic limit. Further, we have developed an RRG-like approximation to a quantum-dot model and compared it to the numerical simulations. We used the IPR $P_2$ to characterize eigenfunctions of the system. For quantum-dot models, we also used the mean adjacent gap ratio $r$ as a complementary observable characterizing spectral properties of the system. Below, we summarize the most salient results of the work.

We have shown that, for large $m$, the ergodicity-to-localization evolution of the RRG model  includes the RMT, Golden-rule, pre-critical, critical, and localized regimes, and calculated analytically the IPR $P_2$ in all of them.  A particularly interesting property of the large-$m$ model is the emergence of a parametrically broad pre-critical regime, in which the IPR decreases exponentially (or, equivalently, the correlation volume $N_\xi$ increases exponentially) with disorder $W$. For the $m=200$ RRG model, we have complemented the purely analytical treatment by a numerical solution of the (analytically derived) self-consistency equation by population dynamics. This has allowed us to study  numerically systems with correlation volume as large as $N \sim 10^{13}$, thus reaching  the critical regime, $W_c/2 < W < W_c$. The population-dynamics results are in excellent agreement with those of the purely analytical study. We have also carried out exact diagonalization of the RRG model with a large coordination number up to $m=200$, and evaluated the average IPR as a function of $W$,  in an excellent agreement with our analytical predictions. In view of limitations on the system size $N$, the exact diagonalization does not allow one to probe the ``critical metal'' regime for large $m$ (such as $m=100$ or $m=200$), since in the corresponding disorder range $W_c/2 < W< W_c$ the correlation volume $N_\xi$  becomes much larger than $N$. At the same time, the exact diagonalization permits us to proceed sufficiently far into the pre-critical regime, $m < W < W_c/2$, for these values of $m$. 
	
Exploring by exact diagonalization  the evolution of the level statistics and of the IPR in fermionic and spin quantum-dot models, we have found that the transition from the ergodicity to localization becomes sharper with increasing number $n$ of orbitals (respectively, spins). Thus, importantly, exact-diagonalization data provide an evidence of a true phase transition in the large-$n$ limit, i.e., $\Delta \ln W_c(n) \to 0$. Our results further support the analytically predicted scaling of the transition point $W_c(n)$ with $n$, Eqs.~\eqref{Wc-fermi-quantum-dot} and \eqref{Wc-spin-quantum-dot}. While a limited range of $n$ accessible to exact diagonalization makes it difficult to unambiguously determine the power $\mu$ of logarithm, the data are consistent with its upper limit, $\mu = 1$, as in the RRG model.  We have constructed an RRG-like approximation for the spin-quantum-dot model, which takes into account the Hilbert-space structure and the statistics of random energies but discards correlations between transition matrix elements. Comparing $NP_2$ calculated within this approximation with the results of exact diagonalization, we find a good agreement  in the Golden-rule range, $W < m$. At the same time, substantial deviations emerge closer to the transition point, i.e., in the pre-critical and critical regime: a stronger tendency to localization is found in the spin-quantum-dot model in comparison to the RRG approximation. This is consistent with the fact that the RRG-like approximation neglects correlations between matrix elements on links of the effective graph in the many-body space, thus overestimating delocalization effects. 
			
	This work paves a way to addressing a number of challenging open questions in the physics of the MBL transition. Before closing the paper, we briefly discuss some of  them. 
	
	\begin{itemize}
			
		\item  
		Reaching full understanding of the critical and  pre-critical regimes in quantum-dot models remains a challenge. 
It will be useful to perform a further study of RRG-like models inspired by quantum dots within a combined analytical, population-dynamics, and exact-diagonalization treatment. A comparison to exact diagonalization for genuine quantum-dot models will permit to understand better the role of correlations discarded in the RRG-like approximation.  Also, it will be instructive to extend the analysis to a modified spin-quantum-dot model, with only single-spin-flip processes allowed (so that $m=n$).  A model of this type was recently considered in Ref.~\cite{bulchandani2022onset}, with a focus on level statistics.  A not so large $m$ is advantageous for addressing numerically the critical regime. We expect that insights gained by these approaches will be helpful for developing a controllable analytical theory of the MBL transition in quantum-dot models.

	\item 
	
	While our analysis of many-body systems focussed on quantum dots, we expect that the ideas and results of this work will also permit to better understand the MBL transition in models with real-space structure.  Such models exhibit more complex MBL physics, in view of effects of rare anomalously ergodic or anomalously localized spatial regions. At the same time, a large body of recent work shows that the Fock-space view (properties of many-body eigenstates, matrix elements, and resonances) is very useful also for systems with spatial structure, with many analogies between them, quantum-dot models, and RRG models \cite{serbyn2017thouless,tikhonov18,mace19multifractal,tikhonov2021from,tikhonov2021eigenstate,nag2019many-body,tarzia2020many,roy2021fock,crowley2022constructive,bulchandani2022onset,creed2023probability}.
Importantly, there is also an impressive progress in experimental studies of associated properties (such as Fock-space dynamics and statistics of many-body energies) of systems of coupled qubits  across the MBL transition	\cite{smith2016many-body,roushan2017spectroscopic,xu2018emulating,lukin2019probing,yao2022observation}. It is also interesting to extend our analysis to MBL transitions in models with long-range interactions induced by a central spin, qudit, or cavity \cite{Sierant2019many,ng2019many,koshkaki2022inverted,hetterich2018detection}.
		
		\item When analyzing properties of eigenstates, we focussed in this work on the average IPR. While this is a primary eigenstate observable characterizing the evolution from ergodicity to localization, it is interesting to consider a broader class of observables describing eigenstate correlations  \cite{tikhonov2021from,tikhonov19statistics,tikhonov2021eigenstate}.
		
	\end{itemize}

	\section*{Acknowledgement}
	J.-N.H., J.F.K., and A.D.M. acknowledge support by the Deutsche Forschungsgemeinschaft (DFG) via the grant MI 658/14-1.

	\bibliography{rrg}

\begin{thebibliography}{90}%
\makeatletter
\providecommand \@ifxundefined [1]{%
 \@ifx{#1\undefined}
}%
\providecommand \@ifnum [1]{%
 \ifnum #1\expandafter \@firstoftwo
 \else \expandafter \@secondoftwo
 \fi
}%
\providecommand \@ifx [1]{%
 \ifx #1\expandafter \@firstoftwo
 \else \expandafter \@secondoftwo
 \fi
}%
\providecommand \natexlab [1]{#1}%
\providecommand \enquote  [1]{``#1''}%
\providecommand \bibnamefont  [1]{#1}%
\providecommand \bibfnamefont [1]{#1}%
\providecommand \citenamefont [1]{#1}%
\providecommand \href@noop [0]{\@secondoftwo}%
\providecommand \href [0]{\begingroup \@sanitize@url \@href}%
\providecommand \@href[1]{\@@startlink{#1}\@@href}%
\providecommand \@@href[1]{\endgroup#1\@@endlink}%
\providecommand \@sanitize@url [0]{\catcode `\\12\catcode `\$12\catcode
  `\&12\catcode `\#12\catcode `\^12\catcode `\_12\catcode `\%12\relax}%
\providecommand \@@startlink[1]{}%
\providecommand \@@endlink[0]{}%
\providecommand \url  [0]{\begingroup\@sanitize@url \@url }%
\providecommand \@url [1]{\endgroup\@href {#1}{\urlprefix }}%
\providecommand \urlprefix  [0]{URL }%
\providecommand \Eprint [0]{\href }%
\providecommand \doibase [0]{https://doi.org/}%
\providecommand \selectlanguage [0]{\@gobble}%
\providecommand \bibinfo  [0]{\@secondoftwo}%
\providecommand \bibfield  [0]{\@secondoftwo}%
\providecommand \translation [1]{[#1]}%
\providecommand \BibitemOpen [0]{}%
\providecommand \bibitemStop [0]{}%
\providecommand \bibitemNoStop [0]{.\EOS\space}%
\providecommand \EOS [0]{\spacefactor3000\relax}%
\providecommand \BibitemShut  [1]{\csname bibitem#1\endcsname}%
\let\auto@bib@innerbib\@empty
\bibitem [{\citenamefont {Anderson}(1958)}]{anderson58}%
  \BibitemOpen
  \bibfield  {author} {\bibinfo {author} {\bibfnamefont {P.~W.}\ \bibnamefont
  {Anderson}},\ }\bibfield  {title} {\bibinfo {title} {Absence of diffusion in
  certain random lattices},\ }\href@noop {} {\bibfield  {journal} {\bibinfo
  {journal} {Phys. Rev.}\ }\textbf {\bibinfo {volume} {109}},\ \bibinfo {pages}
  {1492} (\bibinfo {year} {1958})}\BibitemShut {NoStop}%
\bibitem [{\citenamefont {Evers}\ and\ \citenamefont {Mirlin}(2008)}]{evers08}%
  \BibitemOpen
  \bibfield  {author} {\bibinfo {author} {\bibfnamefont {F.}~\bibnamefont
  {Evers}}\ and\ \bibinfo {author} {\bibfnamefont {A.~D.}\ \bibnamefont
  {Mirlin}},\ }\bibfield  {title} {\bibinfo {title} {Anderson transitions},\
  }\href@noop {} {\bibfield  {journal} {\bibinfo  {journal} {Reviews of Modern
  Physics}\ }\textbf {\bibinfo {volume} {80}},\ \bibinfo {pages} {1355}
  (\bibinfo {year} {2008})}\BibitemShut {NoStop}%
\bibitem [{\citenamefont {Gornyi}\ \emph {et~al.}(2005)\citenamefont {Gornyi},
  \citenamefont {Mirlin},\ and\ \citenamefont
  {Polyakov}}]{gornyi2005interacting}%
  \BibitemOpen
  \bibfield  {author} {\bibinfo {author} {\bibfnamefont {I.}~\bibnamefont
  {Gornyi}}, \bibinfo {author} {\bibfnamefont {A.}~\bibnamefont {Mirlin}},\
  and\ \bibinfo {author} {\bibfnamefont {D.}~\bibnamefont {Polyakov}},\
  }\bibfield  {title} {\bibinfo {title} {Interacting electrons in disordered
  wires: Anderson localization and low-{T} transport},\ }\href@noop {}
  {\bibfield  {journal} {\bibinfo  {journal} {Physical Review Letters}\
  }\textbf {\bibinfo {volume} {95}},\ \bibinfo {pages} {206603} (\bibinfo
  {year} {2005})}\BibitemShut {NoStop}%
\bibitem [{\citenamefont {Basko}\ \emph {et~al.}(2006)\citenamefont {Basko},
  \citenamefont {Aleiner},\ and\ \citenamefont {Altshuler}}]{basko2006metal}%
  \BibitemOpen
  \bibfield  {author} {\bibinfo {author} {\bibfnamefont {D.}~\bibnamefont
  {Basko}}, \bibinfo {author} {\bibfnamefont {I.}~\bibnamefont {Aleiner}},\
  and\ \bibinfo {author} {\bibfnamefont {B.}~\bibnamefont {Altshuler}},\
  }\bibfield  {title} {\bibinfo {title} {Metal--insulator transition in a
  weakly interacting many-electron system with localized single-particle
  states},\ }\href@noop {} {\bibfield  {journal} {\bibinfo  {journal} {Annals
  Of Physics}\ }\textbf {\bibinfo {volume} {321}},\ \bibinfo {pages} {1126}
  (\bibinfo {year} {2006})}\BibitemShut {NoStop}%
\bibitem [{\citenamefont {Nandkishore}\ and\ \citenamefont
  {Huse}(2015)}]{nandkishore15}%
  \BibitemOpen
  \bibfield  {author} {\bibinfo {author} {\bibfnamefont {R.}~\bibnamefont
  {Nandkishore}}\ and\ \bibinfo {author} {\bibfnamefont {D.~A.}\ \bibnamefont
  {Huse}},\ }\bibfield  {title} {\bibinfo {title} {Many-body localization and
  thermalization in quantum statistical mechanics},\ }\href@noop {} {\bibfield
  {journal} {\bibinfo  {journal} {Annu. Rev. Condens. Matter Phys.}\ }\textbf
  {\bibinfo {volume} {6}},\ \bibinfo {pages} {15} (\bibinfo {year}
  {2015})}\BibitemShut {NoStop}%
\bibitem [{\citenamefont {Alet}\ and\ \citenamefont
  {Laflorencie}(2018)}]{Alet2018a}%
  \BibitemOpen
  \bibfield  {author} {\bibinfo {author} {\bibfnamefont {F.}~\bibnamefont
  {Alet}}\ and\ \bibinfo {author} {\bibfnamefont {N.}~\bibnamefont
  {Laflorencie}},\ }\bibfield  {title} {\bibinfo {title} {Many-body
  localization: An introduction and selected topics},\ }\href@noop {}
  {\bibfield  {journal} {\bibinfo  {journal} {Comptes Rendus Physique}\
  }\textbf {\bibinfo {volume} {19}},\ \bibinfo {pages} {498} (\bibinfo {year}
  {2018})}\BibitemShut {NoStop}%
\bibitem [{\citenamefont {Abanin}\ \emph {et~al.}(2019)\citenamefont {Abanin},
  \citenamefont {Altman}, \citenamefont {Bloch},\ and\ \citenamefont
  {Serbyn}}]{abanin2019colloquium}%
  \BibitemOpen
  \bibfield  {author} {\bibinfo {author} {\bibfnamefont {D.~A.}\ \bibnamefont
  {Abanin}}, \bibinfo {author} {\bibfnamefont {E.}~\bibnamefont {Altman}},
  \bibinfo {author} {\bibfnamefont {I.}~\bibnamefont {Bloch}},\ and\ \bibinfo
  {author} {\bibfnamefont {M.}~\bibnamefont {Serbyn}},\ }\bibfield  {title}
  {\bibinfo {title} {Colloquium: Many-body localization, thermalization, and
  entanglement},\ }\href@noop {} {\bibfield  {journal} {\bibinfo  {journal}
  {Reviews of Modern Physics}\ }\textbf {\bibinfo {volume} {91}},\ \bibinfo
  {pages} {021001} (\bibinfo {year} {2019})}\BibitemShut {NoStop}%
\bibitem [{\citenamefont {Gopalakrishnan}\ and\ \citenamefont
  {Parameswaran}(2020)}]{gopalakrishnan2020dynamics}%
  \BibitemOpen
  \bibfield  {author} {\bibinfo {author} {\bibfnamefont {S.}~\bibnamefont
  {Gopalakrishnan}}\ and\ \bibinfo {author} {\bibfnamefont {S.}~\bibnamefont
  {Parameswaran}},\ }\bibfield  {title} {\bibinfo {title} {Dynamics and
  transport at the threshold of many-body localization},\ }\href@noop {}
  {\bibfield  {journal} {\bibinfo  {journal} {Physics Reports}\ }\textbf
  {\bibinfo {volume} {862}},\ \bibinfo {pages} {1} (\bibinfo {year}
  {2020})}\BibitemShut {NoStop}%
\bibitem [{\citenamefont {Tikhonov}\ and\ \citenamefont
  {Mirlin}(2021{\natexlab{a}})}]{tikhonov2021from}%
  \BibitemOpen
  \bibfield  {author} {\bibinfo {author} {\bibfnamefont {K.~S.}\ \bibnamefont
  {Tikhonov}}\ and\ \bibinfo {author} {\bibfnamefont {A.~D.}\ \bibnamefont
  {Mirlin}},\ }\bibfield  {title} {\bibinfo {title} {From anderson localization
  on random regular graphs to many-body localization},\ }\href@noop {}
  {\bibfield  {journal} {\bibinfo  {journal} {Annals of Physics}\ }\textbf
  {\bibinfo {volume} {435}},\ \bibinfo {pages} {168525} (\bibinfo {year}
  {2021}{\natexlab{a}})}\BibitemShut {NoStop}%
\bibitem [{\citenamefont {Ros}\ \emph {et~al.}(2015)\citenamefont {Ros},
  \citenamefont {M{\"u}ller},\ and\ \citenamefont {Scardicchio}}]{Ros2015420}%
  \BibitemOpen
  \bibfield  {author} {\bibinfo {author} {\bibfnamefont {V.}~\bibnamefont
  {Ros}}, \bibinfo {author} {\bibfnamefont {M.}~\bibnamefont {M{\"u}ller}},\
  and\ \bibinfo {author} {\bibfnamefont {A.}~\bibnamefont {Scardicchio}},\
  }\bibfield  {title} {\bibinfo {title} {Integrals of motion in the many-body
  localized phase},\ }\href@noop {} {\bibfield  {journal} {\bibinfo  {journal}
  {Nuclear Physics B}\ }\textbf {\bibinfo {volume} {891}},\ \bibinfo {pages}
  {420 } (\bibinfo {year} {2015})}\BibitemShut {NoStop}%
\bibitem [{\citenamefont {Gornyi}\ \emph {et~al.}(2017)\citenamefont {Gornyi},
  \citenamefont {Mirlin}, \citenamefont {Polyakov},\ and\ \citenamefont
  {Burin}}]{gornyi2017spectral}%
  \BibitemOpen
  \bibfield  {author} {\bibinfo {author} {\bibfnamefont {I.}~\bibnamefont
  {Gornyi}}, \bibinfo {author} {\bibfnamefont {A.}~\bibnamefont {Mirlin}},
  \bibinfo {author} {\bibfnamefont {D.}~\bibnamefont {Polyakov}},\ and\
  \bibinfo {author} {\bibfnamefont {A.}~\bibnamefont {Burin}},\ }\bibfield
  {title} {\bibinfo {title} {Spectral diffusion and scaling of many-body
  delocalization transitions},\ }\href@noop {} {\bibfield  {journal} {\bibinfo
  {journal} {Annalen der Physik}\ }\textbf {\bibinfo {volume} {529}},\ \bibinfo
  {pages} {1600360} (\bibinfo {year} {2017})}\BibitemShut {NoStop}%
\bibitem [{\citenamefont {Agarwal}\ \emph {et~al.}(2016)\citenamefont
  {Agarwal}, \citenamefont {Altman}, \citenamefont {Demler}, \citenamefont
  {Gopalakrishnan}, \citenamefont {Huse},\ and\ \citenamefont
  {Knap}}]{Agarwal2016a}%
  \BibitemOpen
  \bibfield  {author} {\bibinfo {author} {\bibfnamefont {K.}~\bibnamefont
  {Agarwal}}, \bibinfo {author} {\bibfnamefont {E.}~\bibnamefont {Altman}},
  \bibinfo {author} {\bibfnamefont {E.}~\bibnamefont {Demler}}, \bibinfo
  {author} {\bibfnamefont {S.}~\bibnamefont {Gopalakrishnan}}, \bibinfo
  {author} {\bibfnamefont {D.~A.}\ \bibnamefont {Huse}},\ and\ \bibinfo
  {author} {\bibfnamefont {M.}~\bibnamefont {Knap}},\ }\bibfield  {title}
  {\bibinfo {title} {Rare-region effects and dynamics near the many-body
  localization transition},\ }\href@noop {} {\bibfield  {journal} {\bibinfo
  {journal} {Ann. Phys. (Berl.)}\ }\textbf {\bibinfo {volume} {529}},\ \bibinfo
  {pages} {1600326} (\bibinfo {year} {2016})}\BibitemShut {NoStop}%
\bibitem [{\citenamefont {Thiery}\ \emph {et~al.}(2018)\citenamefont {Thiery},
  \citenamefont {Huveneers}, \citenamefont {M\"uller},\ and\ \citenamefont
  {De~Roeck}}]{Thiery2017a}%
  \BibitemOpen
  \bibfield  {author} {\bibinfo {author} {\bibfnamefont {T.}~\bibnamefont
  {Thiery}}, \bibinfo {author} {\bibfnamefont {F.}~\bibnamefont {Huveneers}},
  \bibinfo {author} {\bibfnamefont {M.}~\bibnamefont {M\"uller}},\ and\
  \bibinfo {author} {\bibfnamefont {W.}~\bibnamefont {De~Roeck}},\ }\bibfield
  {title} {\bibinfo {title} {Many-body delocalization as a quantum avalanche},\
  }\href@noop {} {\bibfield  {journal} {\bibinfo  {journal} {Phys. Rev. Lett.}\
  }\textbf {\bibinfo {volume} {121}},\ \bibinfo {pages} {140601} (\bibinfo
  {year} {2018})}\BibitemShut {NoStop}%
\bibitem [{\citenamefont {Dumitrescu}\ \emph {et~al.}(2019)\citenamefont
  {Dumitrescu}, \citenamefont {Goremykina}, \citenamefont {Parameswaran},
  \citenamefont {Serbyn},\ and\ \citenamefont {Vasseur}}]{Dumitrescu2019a}%
  \BibitemOpen
  \bibfield  {author} {\bibinfo {author} {\bibfnamefont {P.~T.}\ \bibnamefont
  {Dumitrescu}}, \bibinfo {author} {\bibfnamefont {A.}~\bibnamefont
  {Goremykina}}, \bibinfo {author} {\bibfnamefont {S.~A.}\ \bibnamefont
  {Parameswaran}}, \bibinfo {author} {\bibfnamefont {M.}~\bibnamefont
  {Serbyn}},\ and\ \bibinfo {author} {\bibfnamefont {R.}~\bibnamefont
  {Vasseur}},\ }\bibfield  {title} {\bibinfo {title} {Kosterlitz-{T}houless
  scaling at many-body localization phase transitions},\ }\href@noop {}
  {\bibfield  {journal} {\bibinfo  {journal} {Phys. Rev. B}\ }\textbf {\bibinfo
  {volume} {99}},\ \bibinfo {pages} {094205} (\bibinfo {year}
  {2019})}\BibitemShut {NoStop}%
\bibitem [{\citenamefont {Morningstar}\ \emph {et~al.}(2020)\citenamefont
  {Morningstar}, \citenamefont {Huse},\ and\ \citenamefont
  {Imbrie}}]{morningstar2020a}%
  \BibitemOpen
  \bibfield  {author} {\bibinfo {author} {\bibfnamefont {A.}~\bibnamefont
  {Morningstar}}, \bibinfo {author} {\bibfnamefont {D.~A.}\ \bibnamefont
  {Huse}},\ and\ \bibinfo {author} {\bibfnamefont {J.~Z.}\ \bibnamefont
  {Imbrie}},\ }\bibfield  {title} {\bibinfo {title} {Many-body localization
  near the critical point},\ }\href@noop {} {\bibfield  {journal} {\bibinfo
  {journal} {Phys. Rev. B}\ }\textbf {\bibinfo {volume} {102}},\ \bibinfo
  {pages} {125134} (\bibinfo {year} {2020})}\BibitemShut {NoStop}%
\bibitem [{\citenamefont {Altshuler}\ \emph {et~al.}(1997)\citenamefont
  {Altshuler}, \citenamefont {Gefen}, \citenamefont {Kamenev},\ and\
  \citenamefont {Levitov}}]{altshuler1997quasiparticle}%
  \BibitemOpen
  \bibfield  {author} {\bibinfo {author} {\bibfnamefont {B.~L.}\ \bibnamefont
  {Altshuler}}, \bibinfo {author} {\bibfnamefont {Y.}~\bibnamefont {Gefen}},
  \bibinfo {author} {\bibfnamefont {A.}~\bibnamefont {Kamenev}},\ and\ \bibinfo
  {author} {\bibfnamefont {L.~S.}\ \bibnamefont {Levitov}},\ }\bibfield
  {title} {\bibinfo {title} {Quasiparticle lifetime in a finite system: A
  nonperturbative approach},\ }\href@noop {} {\bibfield  {journal} {\bibinfo
  {journal} {Physical Review Letters}\ }\textbf {\bibinfo {volume} {78}},\
  \bibinfo {pages} {2803} (\bibinfo {year} {1997})}\BibitemShut {NoStop}%
\bibitem [{\citenamefont {Tikhonov}\ and\ \citenamefont
  {Mirlin}(2016)}]{tikhonov2016fractality}%
  \BibitemOpen
  \bibfield  {author} {\bibinfo {author} {\bibfnamefont {K.}~\bibnamefont
  {Tikhonov}}\ and\ \bibinfo {author} {\bibfnamefont {A.}~\bibnamefont
  {Mirlin}},\ }\bibfield  {title} {\bibinfo {title} {Fractality of wave
  functions on a cayley tree: Difference between tree and locally treelike
  graph without boundary},\ }\href@noop {} {\bibfield  {journal} {\bibinfo
  {journal} {Phys. Rev. B}\ }\textbf {\bibinfo {volume} {94}},\ \bibinfo
  {pages} {184203} (\bibinfo {year} {2016})}\BibitemShut {NoStop}%
\bibitem [{\citenamefont {Sonner}\ \emph {et~al.}(2017)\citenamefont {Sonner},
  \citenamefont {Tikhonov},\ and\ \citenamefont {Mirlin}}]{sonner17}%
  \BibitemOpen
  \bibfield  {author} {\bibinfo {author} {\bibfnamefont {M.}~\bibnamefont
  {Sonner}}, \bibinfo {author} {\bibfnamefont {K.~S.}\ \bibnamefont
  {Tikhonov}},\ and\ \bibinfo {author} {\bibfnamefont {A.~D.}\ \bibnamefont
  {Mirlin}},\ }\bibfield  {title} {\bibinfo {title} {Multifractality of wave
  functions on a {Cayley} tree: From root to leaves},\ }\href@noop {}
  {\bibfield  {journal} {\bibinfo  {journal} {Phys. Rev. B}\ }\textbf {\bibinfo
  {volume} {96}},\ \bibinfo {pages} {214204} (\bibinfo {year}
  {2017})}\BibitemShut {NoStop}%
\bibitem [{\citenamefont {Biroli}\ \emph {et~al.}(2012)\citenamefont {Biroli},
  \citenamefont {Ribeiro-Teixeira},\ and\ \citenamefont
  {Tarzia}}]{biroli2012difference}%
  \BibitemOpen
  \bibfield  {author} {\bibinfo {author} {\bibfnamefont {G.}~\bibnamefont
  {Biroli}}, \bibinfo {author} {\bibfnamefont {A.}~\bibnamefont
  {Ribeiro-Teixeira}},\ and\ \bibinfo {author} {\bibfnamefont {M.}~\bibnamefont
  {Tarzia}},\ }\bibfield  {title} {\bibinfo {title} {Difference between level
  statistics, ergodicity and localization transitions on the bethe lattice},\
  }\href@noop {} {\bibfield  {journal} {\bibinfo  {journal} {arXiv:1211.7334}\
  } (\bibinfo {year} {2012})}\BibitemShut {NoStop}%
\bibitem [{\citenamefont {De~Luca}\ \emph {et~al.}(2014)\citenamefont
  {De~Luca}, \citenamefont {Altshuler}, \citenamefont {Kravtsov},\ and\
  \citenamefont {Scardicchio}}]{de2014anderson}%
  \BibitemOpen
  \bibfield  {author} {\bibinfo {author} {\bibfnamefont {A.}~\bibnamefont
  {De~Luca}}, \bibinfo {author} {\bibfnamefont {B.}~\bibnamefont {Altshuler}},
  \bibinfo {author} {\bibfnamefont {V.}~\bibnamefont {Kravtsov}},\ and\
  \bibinfo {author} {\bibfnamefont {A.}~\bibnamefont {Scardicchio}},\
  }\bibfield  {title} {\bibinfo {title} {Anderson localization on the {B}ethe
  lattice: nonergodicity of extended states},\ }\href@noop {} {\bibfield
  {journal} {\bibinfo  {journal} {Physical Review Letters}\ }\textbf {\bibinfo
  {volume} {113}},\ \bibinfo {pages} {046806} (\bibinfo {year}
  {2014})}\BibitemShut {NoStop}%
\bibitem [{\citenamefont {Tikhonov}\ \emph {et~al.}(2016)\citenamefont
  {Tikhonov}, \citenamefont {Mirlin},\ and\ \citenamefont
  {Skvortsov}}]{tikhonov2016anderson}%
  \BibitemOpen
  \bibfield  {author} {\bibinfo {author} {\bibfnamefont {K.}~\bibnamefont
  {Tikhonov}}, \bibinfo {author} {\bibfnamefont {A.}~\bibnamefont {Mirlin}},\
  and\ \bibinfo {author} {\bibfnamefont {M.}~\bibnamefont {Skvortsov}},\
  }\bibfield  {title} {\bibinfo {title} {Anderson localization and ergodicity
  on random regular graphs},\ }\href@noop {} {\bibfield  {journal} {\bibinfo
  {journal} {Phys. Rev. B}\ }\textbf {\bibinfo {volume} {94}},\ \bibinfo
  {pages} {220203} (\bibinfo {year} {2016})}\BibitemShut {NoStop}%
\bibitem [{\citenamefont {Garc{\'\i}a-Mata}\ \emph {et~al.}(2017)\citenamefont
  {Garc{\'\i}a-Mata}, \citenamefont {Giraud}, \citenamefont {Georgeot},
  \citenamefont {Martin}, \citenamefont {Dubertrand},\ and\ \citenamefont
  {Lemari{\'e}}}]{garcia-mata17}%
  \BibitemOpen
  \bibfield  {author} {\bibinfo {author} {\bibfnamefont {I.}~\bibnamefont
  {Garc{\'\i}a-Mata}}, \bibinfo {author} {\bibfnamefont {O.}~\bibnamefont
  {Giraud}}, \bibinfo {author} {\bibfnamefont {B.}~\bibnamefont {Georgeot}},
  \bibinfo {author} {\bibfnamefont {J.}~\bibnamefont {Martin}}, \bibinfo
  {author} {\bibfnamefont {R.}~\bibnamefont {Dubertrand}},\ and\ \bibinfo
  {author} {\bibfnamefont {G.}~\bibnamefont {Lemari{\'e}}},\ }\bibfield
  {title} {\bibinfo {title} {Scaling theory of the {A}nderson transition in
  random graphs: ergodicity and universality},\ }\href@noop {} {\bibfield
  {journal} {\bibinfo  {journal} {Physical Review Letters}\ }\textbf {\bibinfo
  {volume} {118}},\ \bibinfo {pages} {166801} (\bibinfo {year}
  {2017})}\BibitemShut {NoStop}%
\bibitem [{\citenamefont {Metz}\ and\ \citenamefont
  {Castillo}(2017)}]{metz2017level}%
  \BibitemOpen
  \bibfield  {author} {\bibinfo {author} {\bibfnamefont {F.~L.}\ \bibnamefont
  {Metz}}\ and\ \bibinfo {author} {\bibfnamefont {I.~P.}\ \bibnamefont
  {Castillo}},\ }\bibfield  {title} {\bibinfo {title} {Level compressibility
  for the {A}nderson model on regular random graphs and the eigenvalue
  statistics in the extended phase},\ }\href@noop {} {\bibfield  {journal}
  {\bibinfo  {journal} {Phys. Rev. B}\ }\textbf {\bibinfo {volume} {96}},\
  \bibinfo {pages} {064202} (\bibinfo {year} {2017})}\BibitemShut {NoStop}%
\bibitem [{\citenamefont {Biroli}\ and\ \citenamefont
  {Tarzia}(2017)}]{Biroli2017}%
  \BibitemOpen
  \bibfield  {author} {\bibinfo {author} {\bibfnamefont {G.}~\bibnamefont
  {Biroli}}\ and\ \bibinfo {author} {\bibfnamefont {M.}~\bibnamefont
  {Tarzia}},\ }\bibfield  {title} {\bibinfo {title} {Delocalized glassy
  dynamics and many-body localization},\ }\href@noop {} {\bibfield  {journal}
  {\bibinfo  {journal} {Phys. Rev. B}\ }\textbf {\bibinfo {volume} {96}},\
  \bibinfo {pages} {201114} (\bibinfo {year} {2017})}\BibitemShut {NoStop}%
\bibitem [{\citenamefont {Kravtsov}\ \emph {et~al.}(2018)\citenamefont
  {Kravtsov}, \citenamefont {Altshuler},\ and\ \citenamefont
  {Ioffe}}]{kravtsov2018non}%
  \BibitemOpen
  \bibfield  {author} {\bibinfo {author} {\bibfnamefont {V.}~\bibnamefont
  {Kravtsov}}, \bibinfo {author} {\bibfnamefont {B.}~\bibnamefont
  {Altshuler}},\ and\ \bibinfo {author} {\bibfnamefont {L.}~\bibnamefont
  {Ioffe}},\ }\bibfield  {title} {\bibinfo {title} {Non-ergodic delocalized
  phase in {A}nderson model on {B}ethe lattice and regular graph},\ }\href@noop
  {} {\bibfield  {journal} {\bibinfo  {journal} {Annals of Physics}\ }\textbf
  {\bibinfo {volume} {389}},\ \bibinfo {pages} {148} (\bibinfo {year}
  {2018})}\BibitemShut {NoStop}%
\bibitem [{\citenamefont {Biroli}\ and\ \citenamefont
  {Tarzia}(2018)}]{biroli2018}%
  \BibitemOpen
  \bibfield  {author} {\bibinfo {author} {\bibfnamefont {G.}~\bibnamefont
  {Biroli}}\ and\ \bibinfo {author} {\bibfnamefont {M.}~\bibnamefont
  {Tarzia}},\ }\bibfield  {title} {\bibinfo {title} {Delocalization and
  ergodicity of the {A}nderson model on {B}ethe lattices},\ }\href@noop {}
  {\bibfield  {journal} {\bibinfo  {journal} {arXiv:1810.07545}\ } (\bibinfo
  {year} {2018})}\BibitemShut {NoStop}%
\bibitem [{\citenamefont {Bera}\ \emph {et~al.}(2018)\citenamefont {Bera},
  \citenamefont {De~Tomasi}, \citenamefont {Khaymovich},\ and\ \citenamefont
  {Scardicchio}}]{PhysRevB.98.134205}%
  \BibitemOpen
  \bibfield  {author} {\bibinfo {author} {\bibfnamefont {S.}~\bibnamefont
  {Bera}}, \bibinfo {author} {\bibfnamefont {G.}~\bibnamefont {De~Tomasi}},
  \bibinfo {author} {\bibfnamefont {I.~M.}\ \bibnamefont {Khaymovich}},\ and\
  \bibinfo {author} {\bibfnamefont {A.}~\bibnamefont {Scardicchio}},\
  }\bibfield  {title} {\bibinfo {title} {Return probability for the {A}nderson
  model on the random regular graph},\ }\href@noop {} {\bibfield  {journal}
  {\bibinfo  {journal} {Phys. Rev. B}\ }\textbf {\bibinfo {volume} {98}},\
  \bibinfo {pages} {134205} (\bibinfo {year} {2018})}\BibitemShut {NoStop}%
\bibitem [{\citenamefont {Tikhonov}\ and\ \citenamefont
  {Mirlin}(2019{\natexlab{a}})}]{tikhonov19statistics}%
  \BibitemOpen
  \bibfield  {author} {\bibinfo {author} {\bibfnamefont {K.~S.}\ \bibnamefont
  {Tikhonov}}\ and\ \bibinfo {author} {\bibfnamefont {A.~D.}\ \bibnamefont
  {Mirlin}},\ }\bibfield  {title} {\bibinfo {title} {Statistics of eigenstates
  near the localization transition on random regular graphs},\ }\href@noop {}
  {\bibfield  {journal} {\bibinfo  {journal} {Phys. Rev. B}\ }\textbf {\bibinfo
  {volume} {99}},\ \bibinfo {pages} {024202} (\bibinfo {year}
  {2019}{\natexlab{a}})}\BibitemShut {NoStop}%
\bibitem [{\citenamefont {Tikhonov}\ and\ \citenamefont
  {Mirlin}(2019{\natexlab{b}})}]{tikhonov19critical}%
  \BibitemOpen
  \bibfield  {author} {\bibinfo {author} {\bibfnamefont {K.~S.}\ \bibnamefont
  {Tikhonov}}\ and\ \bibinfo {author} {\bibfnamefont {A.~D.}\ \bibnamefont
  {Mirlin}},\ }\bibfield  {title} {\bibinfo {title} {Critical behavior at the
  localization transition on random regular graphs},\ }\href@noop {} {\bibfield
   {journal} {\bibinfo  {journal} {Phys. Rev. B}\ }\textbf {\bibinfo {volume}
  {99}},\ \bibinfo {pages} {214202} (\bibinfo {year}
  {2019}{\natexlab{b}})}\BibitemShut {NoStop}%
\bibitem [{\citenamefont {Garc\'{\i}a-Mata}\ \emph {et~al.}(2020)\citenamefont
  {Garc\'{\i}a-Mata}, \citenamefont {Martin}, \citenamefont {Dubertrand},
  \citenamefont {Giraud}, \citenamefont {Georgeot},\ and\ \citenamefont
  {Lemari\'e}}]{PhysRevResearch.2.012020}%
  \BibitemOpen
  \bibfield  {author} {\bibinfo {author} {\bibfnamefont {I.}~\bibnamefont
  {Garc\'{\i}a-Mata}}, \bibinfo {author} {\bibfnamefont {J.}~\bibnamefont
  {Martin}}, \bibinfo {author} {\bibfnamefont {R.}~\bibnamefont {Dubertrand}},
  \bibinfo {author} {\bibfnamefont {O.}~\bibnamefont {Giraud}}, \bibinfo
  {author} {\bibfnamefont {B.}~\bibnamefont {Georgeot}},\ and\ \bibinfo
  {author} {\bibfnamefont {G.}~\bibnamefont {Lemari\'e}},\ }\bibfield  {title}
  {\bibinfo {title} {Two critical localization lengths in the {A}nderson
  transition on random graphs},\ }\href@noop {} {\bibfield  {journal} {\bibinfo
   {journal} {Phys. Rev. Research}\ }\textbf {\bibinfo {volume} {2}},\ \bibinfo
  {pages} {012020} (\bibinfo {year} {2020})}\BibitemShut {NoStop}%
\bibitem [{\citenamefont {Roy}\ and\ \citenamefont
  {Logan}(2020{\natexlab{a}})}]{PhysRevLett.125.250402}%
  \BibitemOpen
  \bibfield  {author} {\bibinfo {author} {\bibfnamefont {S.}~\bibnamefont
  {Roy}}\ and\ \bibinfo {author} {\bibfnamefont {D.~E.}\ \bibnamefont
  {Logan}},\ }\bibfield  {title} {\bibinfo {title} {Localization on certain
  graphs with strongly correlated disorder},\ }\href@noop {} {\bibfield
  {journal} {\bibinfo  {journal} {Phys. Rev. Lett.}\ }\textbf {\bibinfo
  {volume} {125}},\ \bibinfo {pages} {250402} (\bibinfo {year}
  {2020}{\natexlab{a}})}\BibitemShut {NoStop}%
\bibitem [{\citenamefont {Tikhonov}\ and\ \citenamefont
  {Mirlin}(2021{\natexlab{b}})}]{tikhonov2021eigenstate}%
  \BibitemOpen
  \bibfield  {author} {\bibinfo {author} {\bibfnamefont {K.~S.}\ \bibnamefont
  {Tikhonov}}\ and\ \bibinfo {author} {\bibfnamefont {A.~D.}\ \bibnamefont
  {Mirlin}},\ }\bibfield  {title} {\bibinfo {title} {Eigenstate correlations
  around the many-body localization transition},\ }\href
  {https://doi.org/10.1103/PhysRevB.103.064204} {\bibfield  {journal} {\bibinfo
   {journal} {Phys. Rev. B}\ }\textbf {\bibinfo {volume} {103}},\ \bibinfo
  {pages} {064204} (\bibinfo {year} {2021}{\natexlab{b}})}\BibitemShut
  {NoStop}%
\bibitem [{\citenamefont {Garc\'{\i}a-Mata}\ \emph {et~al.}(2022)\citenamefont
  {Garc\'{\i}a-Mata}, \citenamefont {Martin}, \citenamefont {Giraud},
  \citenamefont {Georgeot}, \citenamefont {Dubertrand},\ and\ \citenamefont
  {Lemari\'e}}]{garcia-mata2022critical}%
  \BibitemOpen
  \bibfield  {author} {\bibinfo {author} {\bibfnamefont {I.}~\bibnamefont
  {Garc\'{\i}a-Mata}}, \bibinfo {author} {\bibfnamefont {J.}~\bibnamefont
  {Martin}}, \bibinfo {author} {\bibfnamefont {O.}~\bibnamefont {Giraud}},
  \bibinfo {author} {\bibfnamefont {B.}~\bibnamefont {Georgeot}}, \bibinfo
  {author} {\bibfnamefont {R.}~\bibnamefont {Dubertrand}},\ and\ \bibinfo
  {author} {\bibfnamefont {G.}~\bibnamefont {Lemari\'e}},\ }\bibfield  {title}
  {\bibinfo {title} {Critical properties of the anderson transition on random
  graphs: Two-parameter scaling theory, kosterlitz-thouless type flow, and
  many-body localization},\ }\href
  {https://doi.org/10.1103/PhysRevB.106.214202} {\bibfield  {journal} {\bibinfo
   {journal} {Phys. Rev. B}\ }\textbf {\bibinfo {volume} {106}},\ \bibinfo
  {pages} {214202} (\bibinfo {year} {2022})}\BibitemShut {NoStop}%
\bibitem [{\citenamefont {Sierant}\ \emph {et~al.}(2022)\citenamefont
  {Sierant}, \citenamefont {Lewenstein},\ and\ \citenamefont
  {Scardicchio}}]{sierant2023universality}%
  \BibitemOpen
  \bibfield  {author} {\bibinfo {author} {\bibfnamefont {P.}~\bibnamefont
  {Sierant}}, \bibinfo {author} {\bibfnamefont {M.}~\bibnamefont
  {Lewenstein}},\ and\ \bibinfo {author} {\bibfnamefont {A.}~\bibnamefont
  {Scardicchio}},\ }\bibfield  {title} {\bibinfo {title} {Universality in
  anderson localization on random graphs with varying connectivity},\
  }\href@noop {} {\bibfield  {journal} {\bibinfo  {journal} {arXiv:2205.14614}\
  } (\bibinfo {year} {2022})}\BibitemShut {NoStop}%
\bibitem [{\citenamefont {Valba}\ and\ \citenamefont
  {Gorsky}(2022)}]{valba2022mobility}%
  \BibitemOpen
  \bibfield  {author} {\bibinfo {author} {\bibfnamefont {O.}~\bibnamefont
  {Valba}}\ and\ \bibinfo {author} {\bibfnamefont {A.}~\bibnamefont {Gorsky}},\
  }\bibfield  {title} {\bibinfo {title} {Mobility edge in the anderson model on
  partially disordered random regular graphs},\ }\href@noop {} {\bibfield
  {journal} {\bibinfo  {journal} {JETP LETTERS}\ }\textbf {\bibinfo {volume}
  {116}},\ \bibinfo {pages} {398} (\bibinfo {year} {2022})}\BibitemShut
  {NoStop}%
\bibitem [{\citenamefont {Mirlin}\ and\ \citenamefont
  {Fyodorov}(1991{\natexlab{a}})}]{mirlin1991universality}%
  \BibitemOpen
  \bibfield  {author} {\bibinfo {author} {\bibfnamefont {A.}~\bibnamefont
  {Mirlin}}\ and\ \bibinfo {author} {\bibfnamefont {Y.~V.}\ \bibnamefont
  {Fyodorov}},\ }\bibfield  {title} {\bibinfo {title} {Universality of level
  correlation function of sparse random matrices},\ }\href@noop {} {\bibfield
  {journal} {\bibinfo  {journal} {Journal of Physics A: Mathematical and
  General}\ }\textbf {\bibinfo {volume} {24}},\ \bibinfo {pages} {2273}
  (\bibinfo {year} {1991}{\natexlab{a}})}\BibitemShut {NoStop}%
\bibitem [{\citenamefont {Fyodorov}\ and\ \citenamefont
  {Mirlin}(1991)}]{fyodorov1991localization}%
  \BibitemOpen
  \bibfield  {author} {\bibinfo {author} {\bibfnamefont {Y.~V.}\ \bibnamefont
  {Fyodorov}}\ and\ \bibinfo {author} {\bibfnamefont {A.~D.}\ \bibnamefont
  {Mirlin}},\ }\bibfield  {title} {\bibinfo {title} {Localization in ensemble
  of sparse random matrices},\ }\href@noop {} {\bibfield  {journal} {\bibinfo
  {journal} {Physical Review Letters}\ }\textbf {\bibinfo {volume} {67}},\
  \bibinfo {pages} {2049} (\bibinfo {year} {1991})}\BibitemShut {NoStop}%
\bibitem [{\citenamefont {Fyodorov}\ \emph {et~al.}(1992)\citenamefont
  {Fyodorov}, \citenamefont {Mirlin},\ and\ \citenamefont
  {Sommers}}]{fyodorov1992novel}%
  \BibitemOpen
  \bibfield  {author} {\bibinfo {author} {\bibfnamefont {Y.~V.}\ \bibnamefont
  {Fyodorov}}, \bibinfo {author} {\bibfnamefont {A.~D.}\ \bibnamefont
  {Mirlin}},\ and\ \bibinfo {author} {\bibfnamefont {H.-J.}\ \bibnamefont
  {Sommers}},\ }\bibfield  {title} {\bibinfo {title} {A novel field theoretical
  approach to the {A}nderson localization: sparse random hopping model},\
  }\href@noop {} {\bibfield  {journal} {\bibinfo  {journal} {Journal de
  Physique I}\ }\textbf {\bibinfo {volume} {2}},\ \bibinfo {pages} {1571}
  (\bibinfo {year} {1992})}\BibitemShut {NoStop}%
\bibitem [{\citenamefont {Roy}\ and\ \citenamefont
  {Logan}(2020{\natexlab{b}})}]{roy2020fock-space}%
  \BibitemOpen
  \bibfield  {author} {\bibinfo {author} {\bibfnamefont {S.}~\bibnamefont
  {Roy}}\ and\ \bibinfo {author} {\bibfnamefont {D.~E.}\ \bibnamefont
  {Logan}},\ }\bibfield  {title} {\bibinfo {title} {Fock-space correlations and
  the origins of many-body localization},\ }\href@noop {} {\bibfield  {journal}
  {\bibinfo  {journal} {Phys. Rev. B}\ }\textbf {\bibinfo {volume} {101}},\
  \bibinfo {pages} {134202} (\bibinfo {year} {2020}{\natexlab{b}})}\BibitemShut
  {NoStop}%
\bibitem [{\citenamefont {Burin}(2006)}]{burin2006energy}%
  \BibitemOpen
  \bibfield  {author} {\bibinfo {author} {\bibfnamefont {A.~L.}\ \bibnamefont
  {Burin}},\ }\bibfield  {title} {\bibinfo {title} {Energy delocalization in
  strongly disordered systems induced by the long-range many-body
  interaction},\ }\href@noop {} {\bibfield  {journal} {\bibinfo  {journal}
  {arXiv:cond-mat/0611387}\ } (\bibinfo {year} {2006})}\BibitemShut {NoStop}%
\bibitem [{\citenamefont {Yao}\ \emph {et~al.}(2014)\citenamefont {Yao},
  \citenamefont {Laumann}, \citenamefont {Gopalakrishnan}, \citenamefont
  {Knap}, \citenamefont {Mueller}, \citenamefont {Demler},\ and\ \citenamefont
  {Lukin}}]{Demler14}%
  \BibitemOpen
  \bibfield  {author} {\bibinfo {author} {\bibfnamefont {N.~Y.}\ \bibnamefont
  {Yao}}, \bibinfo {author} {\bibfnamefont {C.~R.}\ \bibnamefont {Laumann}},
  \bibinfo {author} {\bibfnamefont {S.}~\bibnamefont {Gopalakrishnan}},
  \bibinfo {author} {\bibfnamefont {M.}~\bibnamefont {Knap}}, \bibinfo {author}
  {\bibfnamefont {M.}~\bibnamefont {Mueller}}, \bibinfo {author} {\bibfnamefont
  {E.~A.}\ \bibnamefont {Demler}},\ and\ \bibinfo {author} {\bibfnamefont
  {M.~D.}\ \bibnamefont {Lukin}},\ }\bibfield  {title} {\bibinfo {title}
  {Many-body localization in dipolar systems},\ }\href@noop {} {\bibfield
  {journal} {\bibinfo  {journal} {Physical Review Letters}\ }\textbf {\bibinfo
  {volume} {113}},\ \bibinfo {pages} {243002} (\bibinfo {year}
  {2014})}\BibitemShut {NoStop}%
\bibitem [{\citenamefont {Gutman}\ \emph {et~al.}(2016)\citenamefont {Gutman},
  \citenamefont {Protopopov}, \citenamefont {Burin}, \citenamefont {Gornyi},
  \citenamefont {Santos},\ and\ \citenamefont {Mirlin}}]{gutman2015energy}%
  \BibitemOpen
  \bibfield  {author} {\bibinfo {author} {\bibfnamefont {D.~B.}\ \bibnamefont
  {Gutman}}, \bibinfo {author} {\bibfnamefont {I.~V.}\ \bibnamefont
  {Protopopov}}, \bibinfo {author} {\bibfnamefont {A.~L.}\ \bibnamefont
  {Burin}}, \bibinfo {author} {\bibfnamefont {I.~V.}\ \bibnamefont {Gornyi}},
  \bibinfo {author} {\bibfnamefont {R.~A.}\ \bibnamefont {Santos}},\ and\
  \bibinfo {author} {\bibfnamefont {A.~D.}\ \bibnamefont {Mirlin}},\ }\bibfield
   {title} {\bibinfo {title} {Energy transport in the {A}nderson insulator},\
  }\href@noop {} {\bibfield  {journal} {\bibinfo  {journal} {Phys. Rev. B}\
  }\textbf {\bibinfo {volume} {93}},\ \bibinfo {pages} {245427} (\bibinfo
  {year} {2016})}\BibitemShut {NoStop}%
\bibitem [{\citenamefont {Burin}(2015)}]{burin2015many}%
  \BibitemOpen
  \bibfield  {author} {\bibinfo {author} {\bibfnamefont {A.~L.}\ \bibnamefont
  {Burin}},\ }\bibfield  {title} {\bibinfo {title} {Many-body delocalization in
  a strongly disordered system with long-range interactions: Finite-size
  scaling},\ }\href@noop {} {\bibfield  {journal} {\bibinfo  {journal} {Phys.
  Rev. B}\ }\textbf {\bibinfo {volume} {91}},\ \bibinfo {pages} {094202}
  (\bibinfo {year} {2015})}\BibitemShut {NoStop}%
\bibitem [{\citenamefont {Tikhonov}\ and\ \citenamefont
  {Mirlin}(2018)}]{tikhonov18}%
  \BibitemOpen
  \bibfield  {author} {\bibinfo {author} {\bibfnamefont {K.~S.}\ \bibnamefont
  {Tikhonov}}\ and\ \bibinfo {author} {\bibfnamefont {A.~D.}\ \bibnamefont
  {Mirlin}},\ }\bibfield  {title} {\bibinfo {title} {Many-body localization
  transition with power-law interactions: Statistics of eigenstates},\
  }\href@noop {} {\bibfield  {journal} {\bibinfo  {journal} {Phys. Rev. B}\
  }\textbf {\bibinfo {volume} {97}},\ \bibinfo {pages} {214205} (\bibinfo
  {year} {2018})}\BibitemShut {NoStop}%
\bibitem [{\citenamefont {Gopalakrishnan}\ and\ \citenamefont
  {Huse}(2019)}]{gopalakrishnan2019instability}%
  \BibitemOpen
  \bibfield  {author} {\bibinfo {author} {\bibfnamefont {S.}~\bibnamefont
  {Gopalakrishnan}}\ and\ \bibinfo {author} {\bibfnamefont {D.~A.}\
  \bibnamefont {Huse}},\ }\bibfield  {title} {\bibinfo {title} {Instability of
  many-body localized systems as a phase transition in a nonstandard
  thermodynamic limit},\ }\href@noop {} {\bibfield  {journal} {\bibinfo
  {journal} {Physical Review B}\ }\textbf {\bibinfo {volume} {99}},\ \bibinfo
  {pages} {134305} (\bibinfo {year} {2019})}\BibitemShut {NoStop}%
\bibitem [{\citenamefont {Jacquod}\ and\ \citenamefont
  {Shepelyansky}(1997)}]{jacquod1997emergence}%
  \BibitemOpen
  \bibfield  {author} {\bibinfo {author} {\bibfnamefont {P.}~\bibnamefont
  {Jacquod}}\ and\ \bibinfo {author} {\bibfnamefont {D.}~\bibnamefont
  {Shepelyansky}},\ }\bibfield  {title} {\bibinfo {title} {Emergence of quantum
  chaos in finite interacting {F}ermi systems},\ }\href@noop {} {\bibfield
  {journal} {\bibinfo  {journal} {Physical Review Letters}\ }\textbf {\bibinfo
  {volume} {79}},\ \bibinfo {pages} {1837} (\bibinfo {year}
  {1997})}\BibitemShut {NoStop}%
\bibitem [{\citenamefont {Mirlin}\ and\ \citenamefont
  {Fyodorov}(1997)}]{mirlin1997localization}%
  \BibitemOpen
  \bibfield  {author} {\bibinfo {author} {\bibfnamefont {A.~D.}\ \bibnamefont
  {Mirlin}}\ and\ \bibinfo {author} {\bibfnamefont {Y.~V.}\ \bibnamefont
  {Fyodorov}},\ }\bibfield  {title} {\bibinfo {title} {Localization and
  fluctuations of local spectral density on treelike structures with large
  connectivity: Application to the quasiparticle line shape in quantum dots},\
  }\href@noop {} {\bibfield  {journal} {\bibinfo  {journal} {Phys. Rev. B}\
  }\textbf {\bibinfo {volume} {56}},\ \bibinfo {pages} {13393} (\bibinfo {year}
  {1997})}\BibitemShut {NoStop}%
\bibitem [{\citenamefont {Silvestrov}(1997)}]{silvestrov1997decay}%
  \BibitemOpen
  \bibfield  {author} {\bibinfo {author} {\bibfnamefont {P.}~\bibnamefont
  {Silvestrov}},\ }\bibfield  {title} {\bibinfo {title} {Decay of a
  quasiparticle in a quantum dot: The role of energy resolution},\ }\href@noop
  {} {\bibfield  {journal} {\bibinfo  {journal} {Physical Review Letters}\
  }\textbf {\bibinfo {volume} {79}},\ \bibinfo {pages} {3994} (\bibinfo {year}
  {1997})}\BibitemShut {NoStop}%
\bibitem [{\citenamefont {Silvestrov}(1998)}]{silvestrov1998chaos}%
  \BibitemOpen
  \bibfield  {author} {\bibinfo {author} {\bibfnamefont {P.}~\bibnamefont
  {Silvestrov}},\ }\bibfield  {title} {\bibinfo {title} {Chaos thresholds in
  finite {F}ermi systems},\ }\href@noop {} {\bibfield  {journal} {\bibinfo
  {journal} {Physical Review E}\ }\textbf {\bibinfo {volume} {58}},\ \bibinfo
  {pages} {5629} (\bibinfo {year} {1998})}\BibitemShut {NoStop}%
\bibitem [{\citenamefont {Gornyi}\ \emph {et~al.}(2016)\citenamefont {Gornyi},
  \citenamefont {Mirlin},\ and\ \citenamefont {Polyakov}}]{gornyi2016many}%
  \BibitemOpen
  \bibfield  {author} {\bibinfo {author} {\bibfnamefont {I.}~\bibnamefont
  {Gornyi}}, \bibinfo {author} {\bibfnamefont {A.}~\bibnamefont {Mirlin}},\
  and\ \bibinfo {author} {\bibfnamefont {D.}~\bibnamefont {Polyakov}},\
  }\bibfield  {title} {\bibinfo {title} {Many-body delocalization transition
  and relaxation in a quantum dot},\ }\href@noop {} {\bibfield  {journal}
  {\bibinfo  {journal} {Physical Review B}\ }\textbf {\bibinfo {volume} {93}},\
  \bibinfo {pages} {125419} (\bibinfo {year} {2016})}\BibitemShut {NoStop}%
\bibitem [{\citenamefont {Georgeot}\ and\ \citenamefont
  {Shepelyansky}(1997)}]{georgeot1997breit}%
  \BibitemOpen
  \bibfield  {author} {\bibinfo {author} {\bibfnamefont {B.}~\bibnamefont
  {Georgeot}}\ and\ \bibinfo {author} {\bibfnamefont {D.~L.}\ \bibnamefont
  {Shepelyansky}},\ }\bibfield  {title} {\bibinfo {title} {Breit-{W}igner width
  and inverse participation ratio in finite interacting {F}ermi systems},\
  }\href@noop {} {\bibfield  {journal} {\bibinfo  {journal} {Physical Review
  Letters}\ }\textbf {\bibinfo {volume} {79}},\ \bibinfo {pages} {4365}
  (\bibinfo {year} {1997})}\BibitemShut {NoStop}%
\bibitem [{\citenamefont {Leyronas}\ \emph {et~al.}(2000)\citenamefont
  {Leyronas}, \citenamefont {Silvestrov},\ and\ \citenamefont
  {Beenakker}}]{leyronas2000scaling}%
  \BibitemOpen
  \bibfield  {author} {\bibinfo {author} {\bibfnamefont {X.}~\bibnamefont
  {Leyronas}}, \bibinfo {author} {\bibfnamefont {P.}~\bibnamefont
  {Silvestrov}},\ and\ \bibinfo {author} {\bibfnamefont {C.}~\bibnamefont
  {Beenakker}},\ }\bibfield  {title} {\bibinfo {title} {Scaling at the chaos
  threshold for interacting electrons in a quantum dot},\ }\href@noop {}
  {\bibfield  {journal} {\bibinfo  {journal} {Physical Review Letters}\
  }\textbf {\bibinfo {volume} {84}},\ \bibinfo {pages} {3414} (\bibinfo {year}
  {2000})}\BibitemShut {NoStop}%
\bibitem [{\citenamefont {Shepelyansky}(2001)}]{shepelyansky2001quantum}%
  \BibitemOpen
  \bibfield  {author} {\bibinfo {author} {\bibfnamefont {D.}~\bibnamefont
  {Shepelyansky}},\ }\bibfield  {title} {\bibinfo {title} {Quantum chaos and
  quantum computers},\ }\href@noop {} {\bibfield  {journal} {\bibinfo
  {journal} {Physica Scripta}\ }\textbf {\bibinfo {volume} {2001}},\ \bibinfo
  {pages} {112} (\bibinfo {year} {2001})}\BibitemShut {NoStop}%
\bibitem [{\citenamefont {Song}(2000)}]{PhysRevE.62.R7575}%
  \BibitemOpen
  \bibfield  {author} {\bibinfo {author} {\bibfnamefont {P.~H.}\ \bibnamefont
  {Song}},\ }\bibfield  {title} {\bibinfo {title} {Scaling near the quantum
  chaos border in interacting {F}ermi systems},\ }\href@noop {} {\bibfield
  {journal} {\bibinfo  {journal} {Phys. Rev. E}\ }\textbf {\bibinfo {volume}
  {62}},\ \bibinfo {pages} {R7575} (\bibinfo {year} {2000})}\BibitemShut
  {NoStop}%
\bibitem [{\citenamefont {Jacquod}\ and\ \citenamefont
  {Varga}(2001)}]{jacquod2001duality}%
  \BibitemOpen
  \bibfield  {author} {\bibinfo {author} {\bibfnamefont {P.}~\bibnamefont
  {Jacquod}}\ and\ \bibinfo {author} {\bibfnamefont {I.}~\bibnamefont
  {Varga}},\ }\bibfield  {title} {\bibinfo {title} {Duality between the weak
  and strong interaction limits of deformed fermionic two-body random
  ensembles},\ }\href@noop {} {\bibfield  {journal} {\bibinfo  {journal} {Phys.
  Rev. Lett.}\ }\textbf {\bibinfo {volume} {89}},\ \bibinfo {pages} {134101}
  (\bibinfo {year} {2001})}\BibitemShut {NoStop}%
\bibitem [{\citenamefont {Rivas}\ \emph {et~al.}(2002)\citenamefont {Rivas},
  \citenamefont {Mucciolo},\ and\ \citenamefont
  {Kamenev}}]{rivas2002numerical}%
  \BibitemOpen
  \bibfield  {author} {\bibinfo {author} {\bibfnamefont {A.~M.}\ \bibnamefont
  {Rivas}}, \bibinfo {author} {\bibfnamefont {E.~R.}\ \bibnamefont
  {Mucciolo}},\ and\ \bibinfo {author} {\bibfnamefont {A.}~\bibnamefont
  {Kamenev}},\ }\bibfield  {title} {\bibinfo {title} {Numerical study of
  quasiparticle lifetime in quantum dots},\ }\href@noop {} {\bibfield
  {journal} {\bibinfo  {journal} {Physical Review B}\ }\textbf {\bibinfo
  {volume} {65}},\ \bibinfo {pages} {155309} (\bibinfo {year}
  {2002})}\BibitemShut {NoStop}%
\bibitem [{\citenamefont {Bulchandani}\ \emph {et~al.}(2022)\citenamefont
  {Bulchandani}, \citenamefont {Huse},\ and\ \citenamefont
  {Gopalakrishnan}}]{bulchandani2022onset}%
  \BibitemOpen
  \bibfield  {author} {\bibinfo {author} {\bibfnamefont {V.~B.}\ \bibnamefont
  {Bulchandani}}, \bibinfo {author} {\bibfnamefont {D.~A.}\ \bibnamefont
  {Huse}},\ and\ \bibinfo {author} {\bibfnamefont {S.}~\bibnamefont
  {Gopalakrishnan}},\ }\bibfield  {title} {\bibinfo {title} {Onset of many-body
  quantum chaos due to breaking integrability},\ }\href@noop {} {\bibfield
  {journal} {\bibinfo  {journal} {Phys. Rev. B}\ }\textbf {\bibinfo {volume}
  {105}},\ \bibinfo {pages} {214308} (\bibinfo {year} {2022})}\BibitemShut
  {NoStop}%
\bibitem [{\citenamefont {Garc\'{\i}a-Garc\'{\i}a}\ \emph
  {et~al.}(2018)\citenamefont {Garc\'{\i}a-Garc\'{\i}a}, \citenamefont
  {Loureiro}, \citenamefont {Romero-Berm\'udez},\ and\ \citenamefont
  {Tezuka}}]{garcia-garcia2018chaotic}%
  \BibitemOpen
  \bibfield  {author} {\bibinfo {author} {\bibfnamefont {A.~M.}\ \bibnamefont
  {Garc\'{\i}a-Garc\'{\i}a}}, \bibinfo {author} {\bibfnamefont
  {B.}~\bibnamefont {Loureiro}}, \bibinfo {author} {\bibfnamefont
  {A.}~\bibnamefont {Romero-Berm\'udez}},\ and\ \bibinfo {author}
  {\bibfnamefont {M.}~\bibnamefont {Tezuka}},\ }\bibfield  {title} {\bibinfo
  {title} {Chaotic-integrable transition in the sachdev-ye-kitaev model},\
  }\href {https://doi.org/10.1103/PhysRevLett.120.241603} {\bibfield  {journal}
  {\bibinfo  {journal} {Phys. Rev. Lett.}\ }\textbf {\bibinfo {volume} {120}},\
  \bibinfo {pages} {241603} (\bibinfo {year} {2018})}\BibitemShut {NoStop}%
\bibitem [{\citenamefont {Micklitz}\ \emph {et~al.}(2019)\citenamefont
  {Micklitz}, \citenamefont {Monteiro},\ and\ \citenamefont
  {Altland}}]{micklitz2019nonergodic}%
  \BibitemOpen
  \bibfield  {author} {\bibinfo {author} {\bibfnamefont {T.}~\bibnamefont
  {Micklitz}}, \bibinfo {author} {\bibfnamefont {F.}~\bibnamefont {Monteiro}},\
  and\ \bibinfo {author} {\bibfnamefont {A.}~\bibnamefont {Altland}},\
  }\bibfield  {title} {\bibinfo {title} {Nonergodic extended states in the
  {S}achdev-{Y}e-{K}itaev model},\ }\href@noop {} {\bibfield  {journal}
  {\bibinfo  {journal} {Physical Review Letters}\ }\textbf {\bibinfo {volume}
  {123}},\ \bibinfo {pages} {125701} (\bibinfo {year} {2019})}\BibitemShut
  {NoStop}%
\bibitem [{\citenamefont {Monteiro}\ \emph {et~al.}(2020)\citenamefont
  {Monteiro}, \citenamefont {Micklitz}, \citenamefont {Tezuka},\ and\
  \citenamefont {Altland}}]{monteiro2020minimal}%
  \BibitemOpen
  \bibfield  {author} {\bibinfo {author} {\bibfnamefont {F.}~\bibnamefont
  {Monteiro}}, \bibinfo {author} {\bibfnamefont {T.}~\bibnamefont {Micklitz}},
  \bibinfo {author} {\bibfnamefont {M.}~\bibnamefont {Tezuka}},\ and\ \bibinfo
  {author} {\bibfnamefont {A.}~\bibnamefont {Altland}},\ }\bibfield  {title}
  {\bibinfo {title} {Minimal model of many-body localization},\ }\href@noop {}
  {\bibfield  {journal} {\bibinfo  {journal} {Physical Review Research}\
  }\textbf {\bibinfo {volume} {3}},\ \bibinfo {pages} {013023} (\bibinfo {year}
  {2020})}\BibitemShut {NoStop}%
\bibitem [{\citenamefont {Monteiro}\ \emph {et~al.}(2021)\citenamefont
  {Monteiro}, \citenamefont {Tezuka}, \citenamefont {Altland}, \citenamefont
  {Huse},\ and\ \citenamefont {Micklitz}}]{monteiro2020quantum}%
  \BibitemOpen
  \bibfield  {author} {\bibinfo {author} {\bibfnamefont {F.}~\bibnamefont
  {Monteiro}}, \bibinfo {author} {\bibfnamefont {M.}~\bibnamefont {Tezuka}},
  \bibinfo {author} {\bibfnamefont {A.}~\bibnamefont {Altland}}, \bibinfo
  {author} {\bibfnamefont {D.~A.}\ \bibnamefont {Huse}},\ and\ \bibinfo
  {author} {\bibfnamefont {T.}~\bibnamefont {Micklitz}},\ }\bibfield  {title}
  {\bibinfo {title} {Quantum ergodicity in the many-body localization
  problem},\ }\href@noop {} {\bibfield  {journal} {\bibinfo  {journal} {Phys.
  Rev. Lett.}\ }\textbf {\bibinfo {volume} {127}},\ \bibinfo {pages} {030601}
  (\bibinfo {year} {2021})}\BibitemShut {NoStop}%
\bibitem [{\citenamefont {Nandy}\ \emph {et~al.}(2022)\citenamefont {Nandy},
  \citenamefont {\ifmmode \check{C}\else \v{C}\fi{}ade\ifmmode~\check{z}\else
  \v{z}\fi{}}, \citenamefont {Dietz}, \citenamefont {Andreanov},\ and\
  \citenamefont {Rosa}}]{nandy2022delayed}%
  \BibitemOpen
  \bibfield  {author} {\bibinfo {author} {\bibfnamefont {D.~K.}\ \bibnamefont
  {Nandy}}, \bibinfo {author} {\bibfnamefont {T.}~\bibnamefont {\ifmmode
  \check{C}\else \v{C}\fi{}ade\ifmmode~\check{z}\else \v{z}\fi{}}}, \bibinfo
  {author} {\bibfnamefont {B.}~\bibnamefont {Dietz}}, \bibinfo {author}
  {\bibfnamefont {A.}~\bibnamefont {Andreanov}},\ and\ \bibinfo {author}
  {\bibfnamefont {D.}~\bibnamefont {Rosa}},\ }\bibfield  {title} {\bibinfo
  {title} {Delayed thermalization in the mass-deformed sachdev-ye-kitaev
  model},\ }\href {https://doi.org/10.1103/PhysRevB.106.245147} {\bibfield
  {journal} {\bibinfo  {journal} {Phys. Rev. B}\ }\textbf {\bibinfo {volume}
  {106}},\ \bibinfo {pages} {245147} (\bibinfo {year} {2022})}\BibitemShut
  {NoStop}%
\bibitem [{\citenamefont {Larzul}\ and\ \citenamefont
  {Schir\'o}(2022)}]{larzul2022quenches}%
  \BibitemOpen
  \bibfield  {author} {\bibinfo {author} {\bibfnamefont {A.}~\bibnamefont
  {Larzul}}\ and\ \bibinfo {author} {\bibfnamefont {M.}~\bibnamefont
  {Schir\'o}},\ }\bibfield  {title} {\bibinfo {title} {Quenches and
  (pre)thermalization in a mixed sachdev-ye-kitaev model},\ }\href
  {https://doi.org/10.1103/PhysRevB.105.045105} {\bibfield  {journal} {\bibinfo
   {journal} {Phys. Rev. B}\ }\textbf {\bibinfo {volume} {105}},\ \bibinfo
  {pages} {045105} (\bibinfo {year} {2022})}\BibitemShut {NoStop}%
\bibitem [{\citenamefont {{\AA}berg}(1990)}]{aaberg1990onset}%
  \BibitemOpen
  \bibfield  {author} {\bibinfo {author} {\bibfnamefont {S.}~\bibnamefont
  {{\AA}berg}},\ }\bibfield  {title} {\bibinfo {title} {Onset of chaos in
  rapidly rotating nuclei},\ }\href@noop {} {\bibfield  {journal} {\bibinfo
  {journal} {Physical Review Letters}\ }\textbf {\bibinfo {volume} {64}},\
  \bibinfo {pages} {3119} (\bibinfo {year} {1990})}\BibitemShut {NoStop}%
\bibitem [{\citenamefont {{\AA}berg}(1992)}]{aaberg1992quantum}%
  \BibitemOpen
  \bibfield  {author} {\bibinfo {author} {\bibfnamefont {S.}~\bibnamefont
  {{\AA}berg}},\ }\bibfield  {title} {\bibinfo {title} {Quantum choas and
  rotational damping},\ }\href@noop {} {\bibfield  {journal} {\bibinfo
  {journal} {Progress in Particle and Nuclear Physics}\ }\textbf {\bibinfo
  {volume} {28}},\ \bibinfo {pages} {11} (\bibinfo {year} {1992})}\BibitemShut
  {NoStop}%
\bibitem [{\citenamefont {Luitz}\ \emph {et~al.}(2015)\citenamefont {Luitz},
  \citenamefont {Laflorencie},\ and\ \citenamefont {Alet}}]{luitz2015many}%
  \BibitemOpen
  \bibfield  {author} {\bibinfo {author} {\bibfnamefont {D.~J.}\ \bibnamefont
  {Luitz}}, \bibinfo {author} {\bibfnamefont {N.}~\bibnamefont {Laflorencie}},\
  and\ \bibinfo {author} {\bibfnamefont {F.}~\bibnamefont {Alet}},\ }\bibfield
  {title} {\bibinfo {title} {Many-body localization edge in the random-field
  {H}eisenberg chain},\ }\href@noop {} {\bibfield  {journal} {\bibinfo
  {journal} {Phys. Rev. B}\ }\textbf {\bibinfo {volume} {91}},\ \bibinfo
  {pages} {081103} (\bibinfo {year} {2015})}\BibitemShut {NoStop}%
\bibitem [{\citenamefont {Mac\'e}\ \emph {et~al.}(2019)\citenamefont {Mac\'e},
  \citenamefont {Alet},\ and\ \citenamefont
  {Laflorencie}}]{mace19multifractal}%
  \BibitemOpen
  \bibfield  {author} {\bibinfo {author} {\bibfnamefont {N.}~\bibnamefont
  {Mac\'e}}, \bibinfo {author} {\bibfnamefont {F.}~\bibnamefont {Alet}},\ and\
  \bibinfo {author} {\bibfnamefont {N.}~\bibnamefont {Laflorencie}},\
  }\bibfield  {title} {\bibinfo {title} {Multifractal scalings across the
  many-body localization transition},\ }\href@noop {} {\bibfield  {journal}
  {\bibinfo  {journal} {Phys. Rev. Lett.}\ }\textbf {\bibinfo {volume} {123}},\
  \bibinfo {pages} {180601} (\bibinfo {year} {2019})}\BibitemShut {NoStop}%
\bibitem [{\citenamefont {Oganesyan}\ and\ \citenamefont
  {Huse}(2007)}]{oganesyan2007localization}%
  \BibitemOpen
  \bibfield  {author} {\bibinfo {author} {\bibfnamefont {V.}~\bibnamefont
  {Oganesyan}}\ and\ \bibinfo {author} {\bibfnamefont {D.~A.}\ \bibnamefont
  {Huse}},\ }\bibfield  {title} {\bibinfo {title} {Localization of interacting
  fermions at high temperature},\ }\href@noop {} {\bibfield  {journal}
  {\bibinfo  {journal} {Phys. Rev. B}\ }\textbf {\bibinfo {volume} {75}},\
  \bibinfo {pages} {155111} (\bibinfo {year} {2007})}\BibitemShut {NoStop}%
\bibitem [{\citenamefont {Atas}\ \emph {et~al.}(2013)\citenamefont {Atas},
  \citenamefont {Bogomolny}, \citenamefont {Giraud},\ and\ \citenamefont
  {Roux}}]{atas2013distribution}%
  \BibitemOpen
  \bibfield  {author} {\bibinfo {author} {\bibfnamefont {Y.}~\bibnamefont
  {Atas}}, \bibinfo {author} {\bibfnamefont {E.}~\bibnamefont {Bogomolny}},
  \bibinfo {author} {\bibfnamefont {O.}~\bibnamefont {Giraud}},\ and\ \bibinfo
  {author} {\bibfnamefont {G.}~\bibnamefont {Roux}},\ }\bibfield  {title}
  {\bibinfo {title} {Distribution of the ratio of consecutive level spacings in
  random matrix ensembles},\ }\href@noop {} {\bibfield  {journal} {\bibinfo
  {journal} {Physical review letters}\ }\textbf {\bibinfo {volume} {110}},\
  \bibinfo {pages} {084101} (\bibinfo {year} {2013})}\BibitemShut {NoStop}%
\bibitem [{\citenamefont {Giraud}\ \emph {et~al.}(2022)\citenamefont {Giraud},
  \citenamefont {Mac\'e}, \citenamefont {Vernier},\ and\ \citenamefont
  {Alet}}]{giraud2022probing}%
  \BibitemOpen
  \bibfield  {author} {\bibinfo {author} {\bibfnamefont {O.}~\bibnamefont
  {Giraud}}, \bibinfo {author} {\bibfnamefont {N.}~\bibnamefont {Mac\'e}},
  \bibinfo {author} {\bibfnamefont {E.}~\bibnamefont {Vernier}},\ and\ \bibinfo
  {author} {\bibfnamefont {F.}~\bibnamefont {Alet}},\ }\bibfield  {title}
  {\bibinfo {title} {Probing symmetries of quantum many-body systems through
  gap ratio statistics},\ }\href {https://doi.org/10.1103/PhysRevX.12.011006}
  {\bibfield  {journal} {\bibinfo  {journal} {Phys. Rev. X}\ }\textbf {\bibinfo
  {volume} {12}},\ \bibinfo {pages} {011006} (\bibinfo {year}
  {2022})}\BibitemShut {NoStop}%
\bibitem [{\citenamefont {Mirlin}\ and\ \citenamefont
  {Fyodorov}(1991{\natexlab{b}})}]{mirlin1991localization}%
  \BibitemOpen
  \bibfield  {author} {\bibinfo {author} {\bibfnamefont {A.~D.}\ \bibnamefont
  {Mirlin}}\ and\ \bibinfo {author} {\bibfnamefont {Y.~V.}\ \bibnamefont
  {Fyodorov}},\ }\bibfield  {title} {\bibinfo {title} {Localization transition
  in the {A}nderson model on the {B}ethe lattice: spontaneous symmetry breaking
  and correlation functions},\ }\href@noop {} {\bibfield  {journal} {\bibinfo
  {journal} {Nuclear Physics B}\ }\textbf {\bibinfo {volume} {366}},\ \bibinfo
  {pages} {507} (\bibinfo {year} {1991}{\natexlab{b}})}\BibitemShut {NoStop}%
\bibitem [{\citenamefont {Abou-Chacra}\ \emph {et~al.}(1973)\citenamefont
  {Abou-Chacra}, \citenamefont {Thouless},\ and\ \citenamefont
  {Anderson}}]{abou1973selfconsistent}%
  \BibitemOpen
  \bibfield  {author} {\bibinfo {author} {\bibfnamefont {R.}~\bibnamefont
  {Abou-Chacra}}, \bibinfo {author} {\bibfnamefont {D.}~\bibnamefont
  {Thouless}},\ and\ \bibinfo {author} {\bibfnamefont {P.}~\bibnamefont
  {Anderson}},\ }\bibfield  {title} {\bibinfo {title} {A selfconsistent theory
  of localization},\ }\href@noop {} {\bibfield  {journal} {\bibinfo  {journal}
  {Journal of Physics C: Solid State Physics}\ }\textbf {\bibinfo {volume}
  {6}},\ \bibinfo {pages} {1734} (\bibinfo {year} {1973})}\BibitemShut
  {NoStop}%
\bibitem [{\citenamefont {Brody}\ \emph {et~al.}(1981)\citenamefont {Brody},
  \citenamefont {Flores}, \citenamefont {French}, \citenamefont {Mello},
  \citenamefont {Pandey},\ and\ \citenamefont {Wong}}]{brody1981random-matrix}%
  \BibitemOpen
  \bibfield  {author} {\bibinfo {author} {\bibfnamefont {T.~A.}\ \bibnamefont
  {Brody}}, \bibinfo {author} {\bibfnamefont {J.}~\bibnamefont {Flores}},
  \bibinfo {author} {\bibfnamefont {J.~B.}\ \bibnamefont {French}}, \bibinfo
  {author} {\bibfnamefont {P.~A.}\ \bibnamefont {Mello}}, \bibinfo {author}
  {\bibfnamefont {A.}~\bibnamefont {Pandey}},\ and\ \bibinfo {author}
  {\bibfnamefont {S.~S.~M.}\ \bibnamefont {Wong}},\ }\bibfield  {title}
  {\bibinfo {title} {Random-matrix physics: spectrum and strength
  fluctuations},\ }\href {https://doi.org/10.1103/RevModPhys.53.385} {\bibfield
   {journal} {\bibinfo  {journal} {Rev. Mod. Phys.}\ }\textbf {\bibinfo
  {volume} {53}},\ \bibinfo {pages} {385} (\bibinfo {year} {1981})}\BibitemShut
  {NoStop}%
\bibitem [{\citenamefont {Flambaum}\ \emph {et~al.}(1996)\citenamefont
  {Flambaum}, \citenamefont {Izrailev},\ and\ \citenamefont
  {Casati}}]{flambaum1996towards}%
  \BibitemOpen
  \bibfield  {author} {\bibinfo {author} {\bibfnamefont {V.~V.}\ \bibnamefont
  {Flambaum}}, \bibinfo {author} {\bibfnamefont {F.~M.}\ \bibnamefont
  {Izrailev}},\ and\ \bibinfo {author} {\bibfnamefont {G.}~\bibnamefont
  {Casati}},\ }\bibfield  {title} {\bibinfo {title} {Towards a statistical
  theory of finite fermi systems and compound states: Random two-body
  interaction approach},\ }\href {https://doi.org/10.1103/PhysRevE.54.2136}
  {\bibfield  {journal} {\bibinfo  {journal} {Phys. Rev. E}\ }\textbf {\bibinfo
  {volume} {54}},\ \bibinfo {pages} {2136} (\bibinfo {year}
  {1996})}\BibitemShut {NoStop}%
\bibitem [{\citenamefont {Blanter}(1996)}]{blanter1996electron}%
  \BibitemOpen
  \bibfield  {author} {\bibinfo {author} {\bibfnamefont {Y.~M.}\ \bibnamefont
  {Blanter}},\ }\bibfield  {title} {\bibinfo {title} {Electron-electron
  scattering rate in disordered mesoscopic systems},\ }\href
  {https://doi.org/10.1103/PhysRevB.54.12807} {\bibfield  {journal} {\bibinfo
  {journal} {Phys. Rev. B}\ }\textbf {\bibinfo {volume} {54}},\ \bibinfo
  {pages} {12807} (\bibinfo {year} {1996})}\BibitemShut {NoStop}%
\bibitem [{\citenamefont {Serbyn}\ \emph {et~al.}(2017)\citenamefont {Serbyn},
  \citenamefont {Papi{\'c}},\ and\ \citenamefont
  {Abanin}}]{serbyn2017thouless}%
  \BibitemOpen
  \bibfield  {author} {\bibinfo {author} {\bibfnamefont {M.}~\bibnamefont
  {Serbyn}}, \bibinfo {author} {\bibfnamefont {Z.}~\bibnamefont {Papi{\'c}}},\
  and\ \bibinfo {author} {\bibfnamefont {D.~A.}\ \bibnamefont {Abanin}},\
  }\bibfield  {title} {\bibinfo {title} {Thouless energy and multifractality
  across the many-body localization transition},\ }\href@noop {} {\bibfield
  {journal} {\bibinfo  {journal} {Physical Review B}\ }\textbf {\bibinfo
  {volume} {96}},\ \bibinfo {pages} {104201} (\bibinfo {year}
  {2017})}\BibitemShut {NoStop}%
\bibitem [{\citenamefont {Nag}\ and\ \citenamefont
  {Garg}(2019)}]{nag2019many-body}%
  \BibitemOpen
  \bibfield  {author} {\bibinfo {author} {\bibfnamefont {S.}~\bibnamefont
  {Nag}}\ and\ \bibinfo {author} {\bibfnamefont {A.}~\bibnamefont {Garg}},\
  }\bibfield  {title} {\bibinfo {title} {Many-body localization in the presence
  of long-range interactions and long-range hopping},\ }\href
  {https://doi.org/10.1103/PhysRevB.99.224203} {\bibfield  {journal} {\bibinfo
  {journal} {Phys. Rev. B}\ }\textbf {\bibinfo {volume} {99}},\ \bibinfo
  {pages} {224203} (\bibinfo {year} {2019})}\BibitemShut {NoStop}%
\bibitem [{\citenamefont {Tarzia}(2020)}]{tarzia2020many}%
  \BibitemOpen
  \bibfield  {author} {\bibinfo {author} {\bibfnamefont {M.}~\bibnamefont
  {Tarzia}},\ }\bibfield  {title} {\bibinfo {title} {Many-body localization
  transition in hilbert space},\ }\href@noop {} {\bibfield  {journal} {\bibinfo
   {journal} {Physical Review B}\ }\textbf {\bibinfo {volume} {102}},\ \bibinfo
  {pages} {014208} (\bibinfo {year} {2020})}\BibitemShut {NoStop}%
\bibitem [{\citenamefont {Roy}\ and\ \citenamefont
  {Logan}(2021)}]{roy2021fock}%
  \BibitemOpen
  \bibfield  {author} {\bibinfo {author} {\bibfnamefont {S.}~\bibnamefont
  {Roy}}\ and\ \bibinfo {author} {\bibfnamefont {D.~E.}\ \bibnamefont
  {Logan}},\ }\bibfield  {title} {\bibinfo {title} {Fock-space anatomy of
  eigenstates across the many-body localization transition},\ }\href
  {https://doi.org/10.1103/PhysRevB.104.174201} {\bibfield  {journal} {\bibinfo
   {journal} {Phys. Rev. B}\ }\textbf {\bibinfo {volume} {104}},\ \bibinfo
  {pages} {174201} (\bibinfo {year} {2021})}\BibitemShut {NoStop}%
\bibitem [{\citenamefont {Crowley}\ and\ \citenamefont
  {Chandran}(2022)}]{crowley2022constructive}%
  \BibitemOpen
  \bibfield  {author} {\bibinfo {author} {\bibfnamefont {P.~J.~D.}\
  \bibnamefont {Crowley}}\ and\ \bibinfo {author} {\bibfnamefont
  {A.}~\bibnamefont {Chandran}},\ }\bibfield  {title} {\bibinfo {title} {{A
  constructive theory of the numerically accessible many-body localized to
  thermal crossover}},\ }\href {https://doi.org/10.21468/SciPostPhys.12.6.201}
  {\bibfield  {journal} {\bibinfo  {journal} {SciPost Phys.}\ }\textbf
  {\bibinfo {volume} {12}},\ \bibinfo {pages} {201} (\bibinfo {year}
  {2022})}\BibitemShut {NoStop}%
\bibitem [{\citenamefont {Creed}\ \emph {et~al.}(2023)\citenamefont {Creed},
  \citenamefont {Logan},\ and\ \citenamefont {Roy}}]{creed2023probability}%
  \BibitemOpen
  \bibfield  {author} {\bibinfo {author} {\bibfnamefont {I.}~\bibnamefont
  {Creed}}, \bibinfo {author} {\bibfnamefont {D.~E.}\ \bibnamefont {Logan}},\
  and\ \bibinfo {author} {\bibfnamefont {S.}~\bibnamefont {Roy}},\ }\bibfield
  {title} {\bibinfo {title} {Probability transport on the fock space of a
  disordered quantum spin chain},\ }\href
  {https://doi.org/10.1103/PhysRevB.107.094206} {\bibfield  {journal} {\bibinfo
   {journal} {Phys. Rev. B}\ }\textbf {\bibinfo {volume} {107}},\ \bibinfo
  {pages} {094206} (\bibinfo {year} {2023})}\BibitemShut {NoStop}%
\bibitem [{\citenamefont {Smith}\ \emph {et~al.}(2016)\citenamefont {Smith},
  \citenamefont {Lee}, \citenamefont {Richerme}, \citenamefont {Neyenhuis},
  \citenamefont {Hess}, \citenamefont {Hauke}, \citenamefont {Heyl},
  \citenamefont {Huse},\ and\ \citenamefont {Monroe}}]{smith2016many-body}%
  \BibitemOpen
  \bibfield  {author} {\bibinfo {author} {\bibfnamefont {J.}~\bibnamefont
  {Smith}}, \bibinfo {author} {\bibfnamefont {A.}~\bibnamefont {Lee}}, \bibinfo
  {author} {\bibfnamefont {P.}~\bibnamefont {Richerme}}, \bibinfo {author}
  {\bibfnamefont {B.}~\bibnamefont {Neyenhuis}}, \bibinfo {author}
  {\bibfnamefont {P.~W.}\ \bibnamefont {Hess}}, \bibinfo {author}
  {\bibfnamefont {P.}~\bibnamefont {Hauke}}, \bibinfo {author} {\bibfnamefont
  {M.}~\bibnamefont {Heyl}}, \bibinfo {author} {\bibfnamefont {D.~A.}\
  \bibnamefont {Huse}},\ and\ \bibinfo {author} {\bibfnamefont
  {C.}~\bibnamefont {Monroe}},\ }\bibfield  {title} {\bibinfo {title}
  {Many-body localization in a quantum simulator with programmable random
  disorder},\ }\href {https://doi.org/10.1038/nphys3783} {\bibfield  {journal}
  {\bibinfo  {journal} {Nature Physics}\ }\textbf {\bibinfo {volume} {12}},\
  \bibinfo {pages} {907} (\bibinfo {year} {2016})}\BibitemShut {NoStop}%
\bibitem [{\citenamefont {Roushan}\ \emph {et~al.}(2017)\citenamefont
  {Roushan}, \citenamefont {Neill}, \citenamefont {Tangpanitanon},
  \citenamefont {Bastidas}, \citenamefont {Megrant}, \citenamefont {Barends},
  \citenamefont {Chen}, \citenamefont {Chen}, \citenamefont {Chiaro},
  \citenamefont {Dunsworth}, \citenamefont {Fowler}, \citenamefont {Foxen},
  \citenamefont {Giustina}, \citenamefont {Jeffrey}, \citenamefont {Kelly},
  \citenamefont {Lucero}, \citenamefont {Mutus}, \citenamefont {Neeley},
  \citenamefont {Quintana}, \citenamefont {Sank}, \citenamefont {Vainsencher},
  \citenamefont {Wenner}, \citenamefont {White}, \citenamefont {Neven},
  \citenamefont {Angelakis},\ and\ \citenamefont
  {Martinis}}]{roushan2017spectroscopic}%
  \BibitemOpen
  \bibfield  {author} {\bibinfo {author} {\bibfnamefont {P.}~\bibnamefont
  {Roushan}}, \bibinfo {author} {\bibfnamefont {C.}~\bibnamefont {Neill}},
  \bibinfo {author} {\bibfnamefont {J.}~\bibnamefont {Tangpanitanon}}, \bibinfo
  {author} {\bibfnamefont {V.~M.}\ \bibnamefont {Bastidas}}, \bibinfo {author}
  {\bibfnamefont {A.}~\bibnamefont {Megrant}}, \bibinfo {author} {\bibfnamefont
  {R.}~\bibnamefont {Barends}}, \bibinfo {author} {\bibfnamefont
  {Y.}~\bibnamefont {Chen}}, \bibinfo {author} {\bibfnamefont {Z.}~\bibnamefont
  {Chen}}, \bibinfo {author} {\bibfnamefont {B.}~\bibnamefont {Chiaro}},
  \bibinfo {author} {\bibfnamefont {A.}~\bibnamefont {Dunsworth}}, \bibinfo
  {author} {\bibfnamefont {A.}~\bibnamefont {Fowler}}, \bibinfo {author}
  {\bibfnamefont {B.}~\bibnamefont {Foxen}}, \bibinfo {author} {\bibfnamefont
  {M.}~\bibnamefont {Giustina}}, \bibinfo {author} {\bibfnamefont
  {E.}~\bibnamefont {Jeffrey}}, \bibinfo {author} {\bibfnamefont
  {J.}~\bibnamefont {Kelly}}, \bibinfo {author} {\bibfnamefont
  {E.}~\bibnamefont {Lucero}}, \bibinfo {author} {\bibfnamefont
  {J.}~\bibnamefont {Mutus}}, \bibinfo {author} {\bibfnamefont
  {M.}~\bibnamefont {Neeley}}, \bibinfo {author} {\bibfnamefont
  {C.}~\bibnamefont {Quintana}}, \bibinfo {author} {\bibfnamefont
  {D.}~\bibnamefont {Sank}}, \bibinfo {author} {\bibfnamefont {A.}~\bibnamefont
  {Vainsencher}}, \bibinfo {author} {\bibfnamefont {J.}~\bibnamefont {Wenner}},
  \bibinfo {author} {\bibfnamefont {T.}~\bibnamefont {White}}, \bibinfo
  {author} {\bibfnamefont {H.}~\bibnamefont {Neven}}, \bibinfo {author}
  {\bibfnamefont {D.~G.}\ \bibnamefont {Angelakis}},\ and\ \bibinfo {author}
  {\bibfnamefont {J.}~\bibnamefont {Martinis}},\ }\bibfield  {title} {\bibinfo
  {title} {Spectroscopic signatures of localization with interacting photons in
  superconducting qubits},\ }\href@noop {} {\bibfield  {journal} {\bibinfo
  {journal} {Science}\ }\textbf {\bibinfo {volume} {358}},\ \bibinfo {pages}
  {1175} (\bibinfo {year} {2017})}\BibitemShut {NoStop}%
\bibitem [{\citenamefont {Xu}\ \emph {et~al.}(2018)\citenamefont {Xu},
  \citenamefont {Chen}, \citenamefont {Zeng}, \citenamefont {Zhang},
  \citenamefont {Song}, \citenamefont {Liu}, \citenamefont {Guo}, \citenamefont
  {Zhang}, \citenamefont {Xu}, \citenamefont {Deng}, \citenamefont {Huang},
  \citenamefont {Wang}, \citenamefont {Zhu}, \citenamefont {Zheng},\ and\
  \citenamefont {Fan}}]{xu2018emulating}%
  \BibitemOpen
  \bibfield  {author} {\bibinfo {author} {\bibfnamefont {K.}~\bibnamefont
  {Xu}}, \bibinfo {author} {\bibfnamefont {J.-J.}\ \bibnamefont {Chen}},
  \bibinfo {author} {\bibfnamefont {Y.}~\bibnamefont {Zeng}}, \bibinfo {author}
  {\bibfnamefont {Y.-R.}\ \bibnamefont {Zhang}}, \bibinfo {author}
  {\bibfnamefont {C.}~\bibnamefont {Song}}, \bibinfo {author} {\bibfnamefont
  {W.}~\bibnamefont {Liu}}, \bibinfo {author} {\bibfnamefont {Q.}~\bibnamefont
  {Guo}}, \bibinfo {author} {\bibfnamefont {P.}~\bibnamefont {Zhang}}, \bibinfo
  {author} {\bibfnamefont {D.}~\bibnamefont {Xu}}, \bibinfo {author}
  {\bibfnamefont {H.}~\bibnamefont {Deng}}, \bibinfo {author} {\bibfnamefont
  {K.}~\bibnamefont {Huang}}, \bibinfo {author} {\bibfnamefont
  {H.}~\bibnamefont {Wang}}, \bibinfo {author} {\bibfnamefont {X.}~\bibnamefont
  {Zhu}}, \bibinfo {author} {\bibfnamefont {D.}~\bibnamefont {Zheng}},\ and\
  \bibinfo {author} {\bibfnamefont {H.}~\bibnamefont {Fan}},\ }\bibfield
  {title} {\bibinfo {title} {Emulating many-body localization with a
  superconducting quantum processor},\ }\href
  {https://doi.org/10.1103/PhysRevLett.120.050507} {\bibfield  {journal}
  {\bibinfo  {journal} {Phys. Rev. Lett.}\ }\textbf {\bibinfo {volume} {120}},\
  \bibinfo {pages} {050507} (\bibinfo {year} {2018})}\BibitemShut {NoStop}%
\bibitem [{\citenamefont {Lukin}\ \emph {et~al.}(2019)\citenamefont {Lukin},
  \citenamefont {Rispoli}, \citenamefont {Schittko}, \citenamefont {Tai},
  \citenamefont {Kaufman}, \citenamefont {Choi}, \citenamefont {Khemani},
  \citenamefont {L�onard},\ and\ \citenamefont {Greiner}}]{lukin2019probing}%
  \BibitemOpen
  \bibfield  {author} {\bibinfo {author} {\bibfnamefont {A.}~\bibnamefont
  {Lukin}}, \bibinfo {author} {\bibfnamefont {M.}~\bibnamefont {Rispoli}},
  \bibinfo {author} {\bibfnamefont {R.}~\bibnamefont {Schittko}}, \bibinfo
  {author} {\bibfnamefont {M.~E.}\ \bibnamefont {Tai}}, \bibinfo {author}
  {\bibfnamefont {A.~M.}\ \bibnamefont {Kaufman}}, \bibinfo {author}
  {\bibfnamefont {S.}~\bibnamefont {Choi}}, \bibinfo {author} {\bibfnamefont
  {V.}~\bibnamefont {Khemani}}, \bibinfo {author} {\bibfnamefont
  {J.}~\bibnamefont {L�onard}},\ and\ \bibinfo {author} {\bibfnamefont
  {M.}~\bibnamefont {Greiner}},\ }\bibfield  {title} {\bibinfo {title} {Probing
  entanglement in a many-body\&\#x2013;localized system},\ }\href@noop {}
  {\bibfield  {journal} {\bibinfo  {journal} {Science}\ }\textbf {\bibinfo
  {volume} {364}},\ \bibinfo {pages} {256} (\bibinfo {year}
  {2019})}\BibitemShut {NoStop}%
\bibitem [{\citenamefont {Yao}\ \emph {et~al.}(2022)\citenamefont {Yao},
  \citenamefont {Xiang}, \citenamefont {Guo}, \citenamefont {Bao},
  \citenamefont {Yang}, \citenamefont {Song}, \citenamefont {Shi},
  \citenamefont {Zhu}, \citenamefont {Jin}, \citenamefont {Chen}, \citenamefont
  {Xu}, \citenamefont {Zhu}, \citenamefont {Shen}, \citenamefont {Wang},
  \citenamefont {Zhang}, \citenamefont {Wu}, \citenamefont {Zou}, \citenamefont
  {Zhang}, \citenamefont {Li}, \citenamefont {Wang}, \citenamefont {Song},
  \citenamefont {Cheng}, \citenamefont {Mondaini}, \citenamefont {Wang},
  \citenamefont {You}, \citenamefont {Zhu}, \citenamefont {Ying},\ and\
  \citenamefont {Guo}}]{yao2022observation}%
  \BibitemOpen
  \bibfield  {author} {\bibinfo {author} {\bibfnamefont {Y.}~\bibnamefont
  {Yao}}, \bibinfo {author} {\bibfnamefont {L.}~\bibnamefont {Xiang}}, \bibinfo
  {author} {\bibfnamefont {Z.}~\bibnamefont {Guo}}, \bibinfo {author}
  {\bibfnamefont {Z.}~\bibnamefont {Bao}}, \bibinfo {author} {\bibfnamefont
  {Y.-F.}\ \bibnamefont {Yang}}, \bibinfo {author} {\bibfnamefont
  {Z.}~\bibnamefont {Song}}, \bibinfo {author} {\bibfnamefont {H.}~\bibnamefont
  {Shi}}, \bibinfo {author} {\bibfnamefont {X.}~\bibnamefont {Zhu}}, \bibinfo
  {author} {\bibfnamefont {F.}~\bibnamefont {Jin}}, \bibinfo {author}
  {\bibfnamefont {J.}~\bibnamefont {Chen}}, \bibinfo {author} {\bibfnamefont
  {S.}~\bibnamefont {Xu}}, \bibinfo {author} {\bibfnamefont {Z.}~\bibnamefont
  {Zhu}}, \bibinfo {author} {\bibfnamefont {F.}~\bibnamefont {Shen}}, \bibinfo
  {author} {\bibfnamefont {N.}~\bibnamefont {Wang}}, \bibinfo {author}
  {\bibfnamefont {C.}~\bibnamefont {Zhang}}, \bibinfo {author} {\bibfnamefont
  {Y.}~\bibnamefont {Wu}}, \bibinfo {author} {\bibfnamefont {Y.}~\bibnamefont
  {Zou}}, \bibinfo {author} {\bibfnamefont {P.}~\bibnamefont {Zhang}}, \bibinfo
  {author} {\bibfnamefont {H.}~\bibnamefont {Li}}, \bibinfo {author}
  {\bibfnamefont {Z.}~\bibnamefont {Wang}}, \bibinfo {author} {\bibfnamefont
  {C.}~\bibnamefont {Song}}, \bibinfo {author} {\bibfnamefont {C.}~\bibnamefont
  {Cheng}}, \bibinfo {author} {\bibfnamefont {R.}~\bibnamefont {Mondaini}},
  \bibinfo {author} {\bibfnamefont {H.}~\bibnamefont {Wang}}, \bibinfo {author}
  {\bibfnamefont {J.~Q.}\ \bibnamefont {You}}, \bibinfo {author} {\bibfnamefont
  {S.-Y.}\ \bibnamefont {Zhu}}, \bibinfo {author} {\bibfnamefont
  {L.}~\bibnamefont {Ying}},\ and\ \bibinfo {author} {\bibfnamefont
  {Q.}~\bibnamefont {Guo}},\ }\href@noop {} {\bibinfo {title} {Observation of
  many-body fock space dynamics in two dimensions}} (\bibinfo {year} {2022}),\
  \Eprint {https://arxiv.org/abs/2211.05803} {arXiv:2211.05803 [quant-ph]}
  \BibitemShut {NoStop}%
\bibitem [{\citenamefont {Sierant}\ \emph {et~al.}(2019)\citenamefont
  {Sierant}, \citenamefont {Biedron}, \citenamefont {Morigi},\ and\
  \citenamefont {Zakrzewski}}]{Sierant2019many}%
  \BibitemOpen
  \bibfield  {author} {\bibinfo {author} {\bibfnamefont {P.}~\bibnamefont
  {Sierant}}, \bibinfo {author} {\bibfnamefont {K.}~\bibnamefont {Biedron}},
  \bibinfo {author} {\bibfnamefont {G.}~\bibnamefont {Morigi}},\ and\ \bibinfo
  {author} {\bibfnamefont {J.}~\bibnamefont {Zakrzewski}},\ }\bibfield  {title}
  {\bibinfo {title} {{Many-body localization in presence of cavity mediated
  long-range interactions}},\ }\href
  {https://doi.org/10.21468/SciPostPhys.7.1.008} {\bibfield  {journal}
  {\bibinfo  {journal} {SciPost Phys.}\ }\textbf {\bibinfo {volume} {7}},\
  \bibinfo {pages} {008} (\bibinfo {year} {2019})}\BibitemShut {NoStop}%
\bibitem [{\citenamefont {Ng}\ and\ \citenamefont
  {Kolodrubetz}(2019)}]{ng2019many}%
  \BibitemOpen
  \bibfield  {author} {\bibinfo {author} {\bibfnamefont {N.}~\bibnamefont
  {Ng}}\ and\ \bibinfo {author} {\bibfnamefont {M.}~\bibnamefont
  {Kolodrubetz}},\ }\bibfield  {title} {\bibinfo {title} {Many-body
  localization in the presence of a central qudit},\ }\href
  {https://doi.org/10.1103/PhysRevLett.122.240402} {\bibfield  {journal}
  {\bibinfo  {journal} {Phys. Rev. Lett.}\ }\textbf {\bibinfo {volume} {122}},\
  \bibinfo {pages} {240402} (\bibinfo {year} {2019})}\BibitemShut {NoStop}%
\bibitem [{\citenamefont {Koshkaki}\ and\ \citenamefont
  {Kolodrubetz}(2022)}]{koshkaki2022inverted}%
  \BibitemOpen
  \bibfield  {author} {\bibinfo {author} {\bibfnamefont {S.~R.}\ \bibnamefont
  {Koshkaki}}\ and\ \bibinfo {author} {\bibfnamefont {M.~H.}\ \bibnamefont
  {Kolodrubetz}},\ }\bibfield  {title} {\bibinfo {title} {Inverted many-body
  mobility edge in a central qudit problem},\ }\href
  {https://doi.org/10.1103/PhysRevB.105.L060303} {\bibfield  {journal}
  {\bibinfo  {journal} {Phys. Rev. B}\ }\textbf {\bibinfo {volume} {105}},\
  \bibinfo {pages} {L060303} (\bibinfo {year} {2022})}\BibitemShut {NoStop}%
\bibitem [{\citenamefont {Hetterich}\ \emph {et~al.}(2018)\citenamefont
  {Hetterich}, \citenamefont {Yao}, \citenamefont {Serbyn}, \citenamefont
  {Pollmann},\ and\ \citenamefont {Trauzettel}}]{hetterich2018detection}%
  \BibitemOpen
  \bibfield  {author} {\bibinfo {author} {\bibfnamefont {D.}~\bibnamefont
  {Hetterich}}, \bibinfo {author} {\bibfnamefont {N.~Y.}\ \bibnamefont {Yao}},
  \bibinfo {author} {\bibfnamefont {M.}~\bibnamefont {Serbyn}}, \bibinfo
  {author} {\bibfnamefont {F.}~\bibnamefont {Pollmann}},\ and\ \bibinfo
  {author} {\bibfnamefont {B.}~\bibnamefont {Trauzettel}},\ }\bibfield  {title}
  {\bibinfo {title} {Detection and characterization of many-body localization
  in central spin models},\ }\href {https://doi.org/10.1103/PhysRevB.98.161122}
  {\bibfield  {journal} {\bibinfo  {journal} {Phys. Rev. B}\ }\textbf {\bibinfo
  {volume} {98}},\ \bibinfo {pages} {161122} (\bibinfo {year}
  {2018})}\BibitemShut {NoStop}%
\end{thebibliography}%
	
\end{document}